\documentclass[11pt,a4paper]{article} 

\usepackage{adjustbox}
\usepackage{amsfonts}
\usepackage{amssymb}
\usepackage{amsmath}
\usepackage{amsthm}
\usepackage{authblk}
\usepackage{array}
\usepackage{bbm}
\usepackage{bigdelim}
\usepackage{booktabs}
\usepackage{caption}
\usepackage{subcaption}
\usepackage{dsfont}
\usepackage{enumerate}
\usepackage{framed,multirow}
\usepackage{graphics}
\usepackage{graphicx}
\usepackage{hyperref}
\usepackage{inputenc}
\usepackage{latexsym}
\usepackage{makecell}
\usepackage{mathrsfs}  
\usepackage[dvipsnames,table,xcdraw]{xcolor}
\usepackage{tikz}
\usetikzlibrary{spy}
\usepackage[normalem]{ulem}
\usepackage{multirow}
\usepackage{hhline}

\usetikzlibrary{calc,fit}

\def\OOO{\mathcal{O}}

\title{Recognising geometric primitives in 3D point clouds of mechanical CAD objects}

\author[ \thanks{These authors have contributed equally to this work.}]{Chiara Romanengo}
\author[{ }${}^\ast$]{Andrea Raffo}
\author[ \thanks{Corresponding author}]{Silvia Biasotti}
\author[ ]{Bianca Falcidieno}

\affil[ ]{Istituto di Matematica Applicata e Tecnologie Informatiche  ``E. Magenes", Consiglio Nazionale delle Ricerche, Via de Marini 6, 16149 Genova, Italy.}

\date{}                     
\newcolumntype{a}{>{\columncolor{blue!12}}m}
\newcolumntype{z}{>{\columncolor{teal!25}}m}

\begin{document}
\maketitle

\begin{abstract}
The problem  faced in this paper concerns the recognition of simple and complex geometric primitives in point clouds resulting from scans of mechanical CAD objects.
A large number of points, the presence of noise, outliers, missing or redundant parts and uneven distribution are the main problems to be addressed to meet this need.
In this article we propose a solution, based on the Hough transform, that can recognize simple and complex geometric primitives and is robust to noise, outliers, and missing parts. Additionally, we can  extract a series of geometric descriptors that uniquely characterize a primitive and,  based on them, aggregate the output into maximal or compound primitives, thus reducing oversegmentation. 
The results presented in the paper demonstrate the robustness of the method and its competitiveness with respect to other solutions proposed in the literature. 
\end{abstract}

\section{Introduction}
The increasing availability of affordable digital scanning devices -- such as close-range photogrammetry and laser scanners -- has made point clouds one of the most common ways of representing the surface of an object. In Computer-Aided Design, this fact results in the need to extract information from measured data points using higher-level geometric primitives. To give an example, it is highly convenient to recognise and reconstruct a digital model so that it can be interpreted in terms of some basic components and easily manipulated by CAD systems. 
A large number of points, the presence of noise and outliers, the occurrence of missing or redundant parts and the non-uniform distribution of the data severely limit the use of tessellations (e.g., meshes) as a means to ease the analysis and reconstruction of shapes; rather, they make it more convenient to analyze the point cloud directly \cite{POUX2022}.

Many approaches for detecting primitives lack robustness to noise and outliers or deal only with mesh models. Most of them are able to extract only a few classes of simple geometric primitives, being planes and cylinders the most commonly recognised, and do not consider more complex basic shapes, such as 
surfaces of revolution or generalized cylinders.
Moreover, many methods have a tendency to oversegment, thus producing very fragmented output without aggregating the non-contiguous subsets of points that lie on the same primitive. The detection of maximal or compound primitives (e.g. cylinders composed of non-adjacent parts, or patterns) is another open problem, mostly addressed only for planes or cylinders \cite{Li2011,Monzpart2015}. For example, if a pipe is interrupted by another part, as in the block model in Figure  \ref{fig:various_SGPs}(b), traditional methods recognise the two parts of the pipe as two distinct primitives.  This is particularly important, for instance, when decomposing patterns that correspond to scans of assembly CAD models, because it concurs with the recognition of compound primitives and patterns \cite{LUPINETTI2019}. When it comes to memory usage and computational complexity, learning approaches require onerous training and large annotated databases; on the other hand, the direct detection of mathematical primitives in a general space embedding typically faces the computational limitation of dealing with a number of free parameters that rapidly grows as the complexity of the primitives increases.

In this paper, we deal with the recognition of simple and complex geometric primitives in point clouds originating from scans of mechanical CAD objects, where recognition means the detection of the primitives in a given point cloud and the extraction/computation of the best  (geometric) parameters associated with those primitives. We adopt an approach based on the Hough transform (HT) to do this. The HT meets the needs of being able to recognize multiple instances of primitive functions and, through a voting procedure, to be robust to noise and outliers. Originally limited to straight lines in images \cite{c1962method}, it has been generalized and extended in multiple directions  to handle other shapes, also using families of non-parametric templates in images  \cite{ballard1981generalizing} or point clouds from CAD objects 
\cite{Woodford2014}; recently, the extended version proposed in \cite{beltrametti2012algebraic} has opened up the HT to a wide range of algebraic primitives.

Here we generalise the HT to surface primitives -- not necessarily algebraic -- represented in parametric form. Our solution is tailored to deal with point clouds and is able to deal with simple geometric primitives and some complex ones (generalized cylinders and cones, surfaces of revolution, helical surfaces, etc.), including maximal and compound primitives. To this end, we aggregate the primitives found on the basis of their size and space embedding, whenever possible.  Our strategy also reduces the oversegmentation of the output and we exploit the strategy proposed in \cite{Raffo:2022} to decrease the computational complexity of methods based on the HT thanks to an opportune preprocessing of the point cloud and its subparts.
Our experiments  confirm the robustness and completeness of the method, and the comparisons made exhibit its competitiveness with respect to the methods proposed so far in the literature.
To sum up, our main contributions are:
\begin{itemize}
    \item The introduction of a geometric primitive recognition method that is particularly robust to noise and outliers and is able to recognize multiple instances of the same primitive.
    \item The extraction and recognition of complex primitives, in addition to the most common ones (planes, cylinders, cones, spheres, tori).
    \item The aggregation of primitives on the basis of their shape, size and position to detect maximal and compound primitives.
    \item An extensive experimentation on different datasets and  comparison with state-of-the-art methods.
\end{itemize}

The rest of the paper is organised as follows. Section \ref{sec:previous} overviews previous work in geometric primitives detection and HT-based recognition methods. Section \ref{sec:recHT} briefly recalls the basic concepts of HT. Section \ref{sec:complexPrim} introduces and lists the geometric primitives that can be identified by our method, with emphasis on complex primitives such as general cylinders and cones, surfaces of revolutions, helical surfaces, and convex combinations of curves. Section \ref{sec:algorithmHT} describes our approach for the identification of primitives in point clouds based on the HT,
and the search for maximal primitives and possible relationships between primitives through a clustering technique.
Section \ref{sec:results} provides experimental results of our method for the identification of simple and complex primitives and shows a qualitative and quantitative comparative analysis. Concluding remarks end the paper.

\section{Previous work}
\label{sec:previous}

Representing an object with a set of geometric components is a long-standing problem in computer vision, computer graphics and CAD. As we are interested in recognising (simple or complex) geometric primitives in the manufacturing domain, here we focus our attention on those approaches that share the same goal.

A considerable variety of algorithms have been devised to decompose digitalized point clouds or meshes representing CAD objects into regions approximated by primitives belonging to some given sets. According to \cite{Kaiser2019}, these approaches can be grouped into four families: stochastic, parameter space, clustering and learning techniques. The first group includes the RANSAC method \cite{Schnabel2007} and its optimizations. The second family includes Hough-like voting methods and parameter space clustering, e.g., \cite{LIMBERGER2015}. The third class gathers all the other clustering techniques, and can be classified into three main types: primitive-driven region growing, e.g., \cite{Attene2010}; automatic clustering  and Lloyd-based algorithms, e.g., \cite{YAN2012}; primitive-oblivious segmentation, e.g., \cite{Le2017}. Finally, with  the growing popularity of deep learning techniques, supervised fitting methods have been proposed even for multi-class primitives \cite{LiSDYG19,uy2022point2cyl}. The reader is referred to \cite{Kaiser2019} for a comprehensive historical taxonomy of methods for simple primitive detection, which is beyond the scope of this paper. 

Despite a large number of available solutions, the problem is far from being solved; novel approaches have been proposed in the very last few years to address the shortcomings of existing paradigms or to propose novel directions to move in. A recent approach to deal with the recognition of geometric primitives is described in \cite{Qie:2021}. It consists of a curvature-based partitioning method that decomposes an input triangle mesh into maximal surface portions; the decomposition is performed so that each segment corresponds to one of the following seven invariance classes of surfaces: plane, sphere, cylinder, surface of revolution, prismatic, helix, and complex surface. The method was adapted to handle point clouds in \cite{Romanengo:2022}.
In \cite{markovic2021}, a method for the identification and fitting of planes and cylinders from unstructured point clouds in manufacturing is presented. It consists of three subsequent phases: point cloud segmentation; merging of oversegmented regions and estimation of surface parameters; extraction of cylinders and planes. Being the method able to handle only two primitive types, its applicability is however restricted. To handle the increasing availability of acquired data, \cite{POUX2022} introduces a region-growing-based system for the segmentation of large point clouds in planar regions. Other approaches, devised to detect only specific types of primitives, are: \cite{Birdal:2020}, which deals with quadric surfaces; \cite{Liu:2019} which fits surfaces of revolution; and \cite{BERGAMASCO2020107443}, which extracts  cylindrical shapes from non-oriented point clouds.
In the field of deep learning, two methods have shown promising results in the problems of segmentation and recognition: ParSeNet \cite{Sharma:2020} and HPNet \cite{Yan:2021}. Beyond their performance, outstanding merit is their capability of handling -- together with the more classical simple geometric primitives -- open and closed spline surfaces. Another  learning-based method lately proposed for fitting primitives is PriFit \cite{PriFit:2022}, which learns to decompose a point cloud of various 3D shape categories into a set of geometric primitives, such as ellipsoids and cuboids, or alternatively deformed planes. However, the natural flexibility of supervised learning approaches in identifying geometric primitives comes with a considerable cost: the need of having gigantic labelled training sets, whose difficulty of gathering limits their current application.

As for the methods that use the HT to identify geometric primitives, they were  limited, until recently,  to planes \cite{LIMBERGER2015,Borrmann2011}, combinations of linear subspaces \cite{FERNANDES2012}, spheres \cite{Camurri2014} and circular cylinders \cite{RABBANI2005}; in the case of cylinders, however, the rotational axis needs to be known before the application of the Hough transform. 
The recent advances in the use of the algebraic functions \cite{beltrametti2012algebraic} are paving the road to a larger use of the HT for recognising more complex families of geometric primitives. To the best of our knowledge, the sole approaches that exploit this idea were proposed in \cite{BELTRAMETTI2020}, for the recognition of an ellipsoid in a free-form model; and in \cite{Raffo:2022}, for the recognition of spheres, (circular) cylinders, (circular) cones and tori in presegmented point clouds, as a way to post process faulty segmentations. The work in \cite{Raffo:2022} also addresses the problem of reducing the number of the parameters of the primitive representation when the point cloud corresponds to a patch of a primitive: in this case, the point cloud is opportunely rototranslated in a space embedding so that it can be fit by a primitive in a standard form. The method has been recently  compared with other direct and learning-based approaches, see \cite{Romanengo:SHREC:2022}. However, the point clouds used for this comparison did neither originate from scans of mechanical CAD objects nor can be assumed to represent presegmented data, making the benchmark itself not of interest to this paper.

\section{Basic concepts on the Hough transform}\label{sec:recHT}

We denote by $\mathbb{A}^3_\mathbf{x}(\mathbb{R})$ and $\mathbb{A}^n_\mathbf{A}(\mathbb{R})$, respectively, the $3$-dimensional and the $n$-dimensional affine spaces over $\mathbb{R}$, 
where $\mathbf{x}:=(x,y,z)$ and $\mathbf{A}:=(A_1, \ldots, A_n)$ vectors of indeterminates. 
Given a surface $\mathcal{S}$ defined as the zero locus of a function $f$, a parameter-dependent family of surfaces can be described by functions $f_\mathbf{a}$ as $\mathcal{F} =\{ \mathcal{S}_\mathbf{a} : f_\mathbf{a}(x) = 0\ | \ \mathbf{a} \in \mathcal{U}\}$, where $\mathcal{U}$ is an open set of the parameter space $\mathbb{A}^n_\mathbf{A}(\mathbb{R})$ and  $\mathbf{a}:=(a_1,...,a_n)$ is the parameter vector. Then, given a point $P\in\mathbb{A}^3_\mathbf{x}(\mathbb{R})$, the HT of the point $P$ with respect to the family $\mathcal{F}$ is $\Gamma_P= \{f_\mathbf{a}(P)=0\}\subset\mathbb{A}^n_\mathbf{A}(\mathbb{R})$. For sufficiently general points $P\in\mathbb{A}^3_\mathbf{x}(\mathbb{R})$, it turns out that $\Gamma_P$ are hypersurfaces in $\mathbb{A}^n_\mathbf{A}(\mathbb{R})$.
If the set of hypersurfaces $\Gamma_{P}$ generated by varying $P$ on a given surface meets in one and only one point $\bar{\mathbf{a}} \in \mathcal{U}$, the family of surfaces $\mathcal{F}$ is called \textit{Hough-regular}; the intersection point $\bar{\mathbf{a}}$ of the Hough transforms $\Gamma_{P}$ uniquely identifies the surface $\mathcal{S}_{\bar{\mathbf{a}}}$ from which the points $P\in\mathbb{A}^3_\mathbf{x}(\mathbb{R})$ were sampled in the first place. If the HT regularity is not guaranteed, the solution to the intersection problem might be non-unique and a solution must be selected among more potential parameters solutions.

In the discrete scenario, the HT framework deals with the problem of finding a surface -- within a family $\mathcal{F}$ of surfaces -- that best approximates a particular shape, given in the form of a data set of $N$ points $\mathcal{P}$, where $N>>n$. The common strategy to identify the solution (or a solution), introduced in \cite{beltrametti2012algebraic}, is based on the so-called \emph{accumulator function}; it consists in a voting system whereby each point in $\mathcal{P}$
votes a $n$-uple $\mathbf{a}=(a_1, \ldots, a_n)$; the most voted $n$-uple corresponds to the most representative surface for the profile. 

While the HT traditionally deals with one shape at a time, the framework we propose in this paper can be directly applied to point clouds containing multiple shapes.
The family of surfaces we consider is represented in parametric form and we assume that the system defining $\mathcal{S}_\mathbf{a}$ with respect to the Cartesian coordinates $x$, $y$, and $z$ can be analytically solved with respect to the parameter vector $\mathbf{a}=(a_1, \ldots, a_n)$. A more detailed description of how this works in our case is given in Section \ref{sec:algorithmHT}.

\section{Definition of complex geometric primitives}
\label{sec:complexPrim}
\begin{figure}[t]
     \begin{center}
     \includegraphics[scale=0.35,trim={0cm 1cm 0cm 0.65cm}, clip]{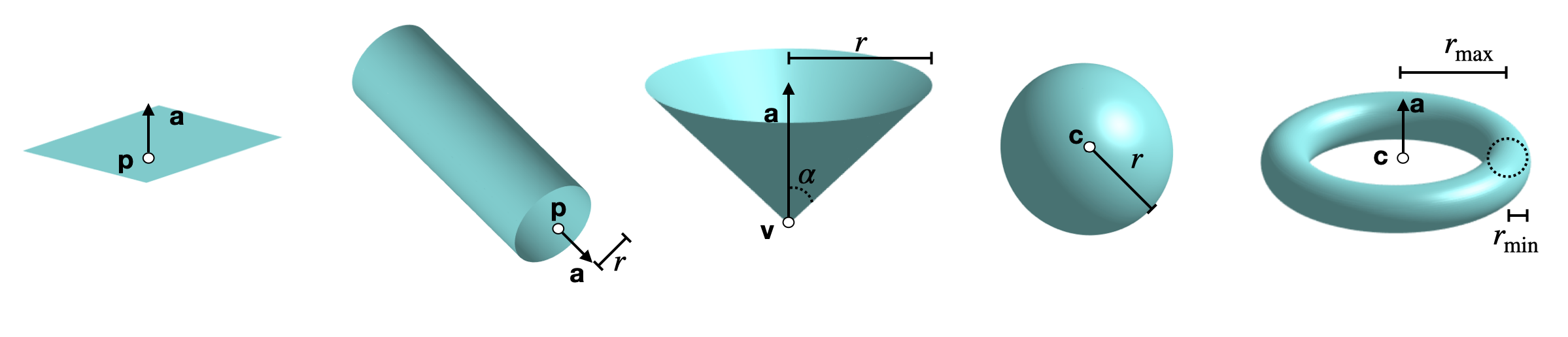}
     \end{center}
     \caption{Simple geometric primitives: plane, cylinder, cone, sphere and torus respectively, along with their attributes (geometric descriptors).}
     \label{fig:basic_primitives}
 \end{figure}

In addition to the simple geometric primitives of Figure \ref{fig:basic_primitives} (plane, cylinder, cone, sphere, and torus) -- see   \cite{Raffo:2022} for their recognition in presegmented point clouds -- in this section, we introduce a set of complex geometric primitives that  to the best of our knowledge have never been used before for recognition by HT.

In this paper we express surfaces parametrically with respect to the variables $u$ and $v$. For the sake of simplicity, some of the surfaces are here presented in their standard form or with respect to some specific axes; nevertheless, one can easily generalise these equations by applying a generic transformation of the special orthogonal group $\text{SO}(3)$.

\begin{itemize}
    \item \textit{General cylinders}.
    A cylinder is a surface traced by a straight line of fixed direction, the \emph{generatrix}, while moving along a curve, the \emph{directrix}. Given a curve $(x(u), y(u), z(u)):=(f_1(\mathbf{a},u),f_2(\mathbf{a},u),f_3(\mathbf{a},u))$ and a direction $(l,m,n)$, the parametric representation of the corresponding cylinder is given by:
    \begin{equation*}
    \begin{cases}
      x=f_1(\mathbf{a},u)+lv \\ y=f_2(\mathbf{a},u)+mv \\ 
      z=f_3(\mathbf{a},u)+nv
   \end{cases}.
   \end{equation*}
    Note that a general cylinder depends on the parameters defining generatrix and directrix. The dictionary of curves to be considered as directrix is extremely rich, see \cite{shikin1995handbook}.    Table \ref{tab:primitiveComplesse}(a) considers a $5-$convexity curve as directrix, whose equation can be found in \cite{shikin1995handbook}. 
     
    \item \textit{General cones}. A cone is a surface traced by a straight line, the \emph{generatrix}, while gliding along a curve, the \emph{directrix}, and passing through a fixed point, the \emph{vertex}.
    Given a parametrized curve $(x(u), y(u), z(u)):=(f_1(\mathbf{a},u),f_2(\mathbf{a},u),f_3(\mathbf{a},u))$ and a point $V=(x_V,y_V,z_V)$, the parametric representation of the corresponding cone is given by:
    \begin{equation*}
    \begin{cases}
    x = x_V + (f_1(\mathbf{a},u)-x_V)v
    \\
    y = y_V + (f_2(\mathbf{a},u)-y_V)v
    \\
    z = z_V + (f_3(\mathbf{a},u)-z_V)v
    \end{cases}
    .
    \end{equation*}
    
    As in the case of cylinders, we can exploit the dictionary of plane curves to create families of cones. Then, a general cone depends on the parameters that define the curve and the components of the vertex.
Table \ref{tab:primitiveComplesse}(b) shows a cone generated by a $5-$convexity curve.
        
    \item \textit{Surfaces of revolution}. A family of surfaces of revolution can be created by rotating a family of curves around an axis of rotation. For example, given the family of plane curves $(x(u), y(u), z(u)):=(f_1(\mathbf{a},u),0,f_2(\mathbf{a},u))$ and the $z-$axis, we obtain the parametric equations 
    \begin{equation*}
    \begin{cases}
      x=f_1(\mathbf{a},u)\cos v \\ y= f_1(\mathbf{a},u)\sin v\\ z=f_2(\mathbf{a},u)
   \end{cases}.
   \end{equation*}
   One of the most known examples is the \emph{torus}; another example is given by ellipsoids, which are obtained by rotating an ellipse with respect to one of its principal axes. Table \ref{tab:primitiveComplesse}(c) shows the case of a surface obtained by rotating the curve $(x(u), y(u), z(u)):=(au,0,b/u^5)$ around the $z-$axis. 
      
    \item \textit{Helical surfaces}. Table \ref{tab:primitiveComplesse}(d) presents a family of equations obtained by modifying the parametrisation of a circular cylinder. Precisely, the radius $R$ here varies between $[R_1,R_2]$, where $R_1>0$, by a cosine function; when  radii are fixed, the slope of the output surface is controlled by the parameters in $z(u,v)$. Note that the radius variation can be adapted to other shapes (e.g., the triangle wave function). 
    \item \textit{Convex combination of curves}. It is possible to define surface primitives by considering the convex combination of a pair of parametrised curves $(f_1(\mathbf{a},u),f_2(\mathbf{a},u),f_3(\mathbf{a},u))$ and $(g_1(\mathbf{b},u),g_2(\mathbf{b},u),g_3(\mathbf{b},u))$. This family has the following parametric equations:
    \begin{equation*}
     \begin{cases}
      x=vf_1(\mathbf{a},u)+(1-v)g_1(\mathbf{b},u) 
      \\ 
      y= vf_2(\mathbf{a},u)+(1-v)g_2(\mathbf{b},u)
      \\ 
      z=vf_3(\mathbf{a},u)+(1-v)g_3(\mathbf{b},u)
   \end{cases},
   \end{equation*}
   where $v\in[0,1]$. Note that the primitive parametrisation depends on the same parameters which define the pair of curves, i.e., $\mathbf{a}$ and $\mathbf{b}$.
   A planar example is given by the \emph{annulus}, i.e., the region bounded by two concentric circles. A helical strip can be obtained by cutting and bending an annular strip; this corresponds to considering a convex combination of two helices of an equal slope but different radii. An example of a helical strip is provided in Table \ref{tab:primitiveComplesse}(e).
\end{itemize}

The use of the standard form allows us to diminish the number of unknown parameters in the recognition process. To give a practical example, the family 
\begin{equation*}
    \begin{cases}
      x = a_1\cos(u)+b_1\sin(u)+c_1v+d_1 \\
      y = a_2\cos(u)+b_2\sin(u)+c_2v+d_2 \\
      z = a_3\cos(u)+b_3\sin(u)+c_3v+d_3
    \end{cases},
\end{equation*}
 contains (circular) cylinders; it has $12$ unknown parameters, which make the representation computationally and memory-wise unmanageable when used in a voting procedure. By further assuming that the cylinder can be rototranslated so that it has the rotational axis aligned with the $z-$axis and it is centred in the origin, the number of parameters reduces to  $1$ in the case of a circular cylinder, i.e., the sole radius:
\begin{equation*}
    \begin{cases}
      x = r\cos(u)\\
      y = r\sin(u)\\
      z = v
    \end{cases}.
\end{equation*}
A strategy to obtain the standard form in an automatic way starting from a point cloud is provided in \cite{Raffo:2022}, and it can be naturally extended to complex primitives.

\begin{table*}[h!]
    \caption{Parametric representation of some complex geometric primitives.}
      \centering
\resizebox{1.0\hsize}{!}{
    \begin{tabular}{|c|c|c|c|c|}
    \hline
           \cellcolor{ForestGreen!15}(a)  & \cellcolor{ForestGreen!15}(b) & \cellcolor{ForestGreen!15}(c) & \cellcolor{ForestGreen!15}(d) & \cellcolor{ForestGreen!15}(e)\\
           \cellcolor{ForestGreen!15}generalized & \cellcolor{ForestGreen!15}generalized & \cellcolor{ForestGreen!15}surface of & \cellcolor{ForestGreen!15}helical & \cellcolor{ForestGreen!15}helical \\
           \cellcolor{ForestGreen!15}cylinder & \cellcolor{ForestGreen!15}cone & \cellcolor{ForestGreen!15}revolution & \cellcolor{ForestGreen!15}surface & \cellcolor{ForestGreen!15}strip \\
           \hline
           & & & & \\
           \includegraphics[width=2.2cm,trim={2cm 1cm 0.75cm 1.2cm}, clip]{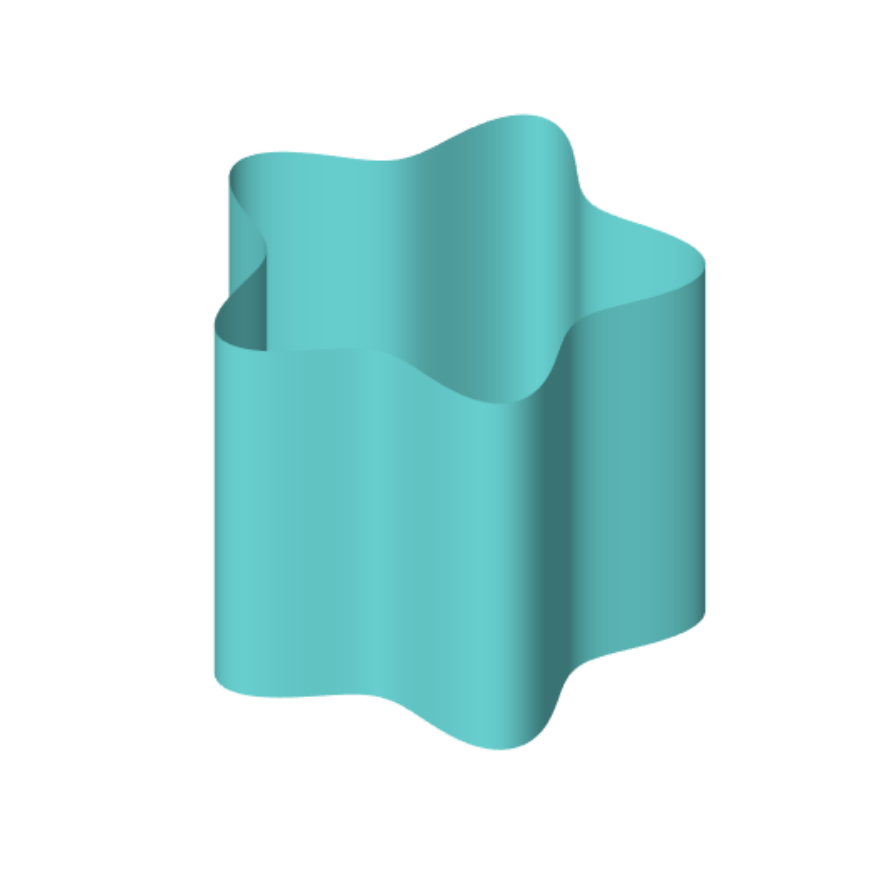}
           &
           \includegraphics[width=2.6cm,trim={2cm 2.35cm 1cm 1.5cm}, clip]{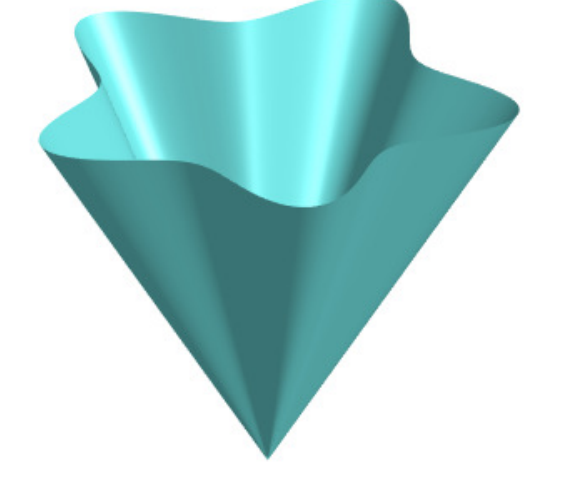}
           &
           \includegraphics[width=2.5cm,trim={1.3cm 1cm 1.1cm 1.7cm}, clip]{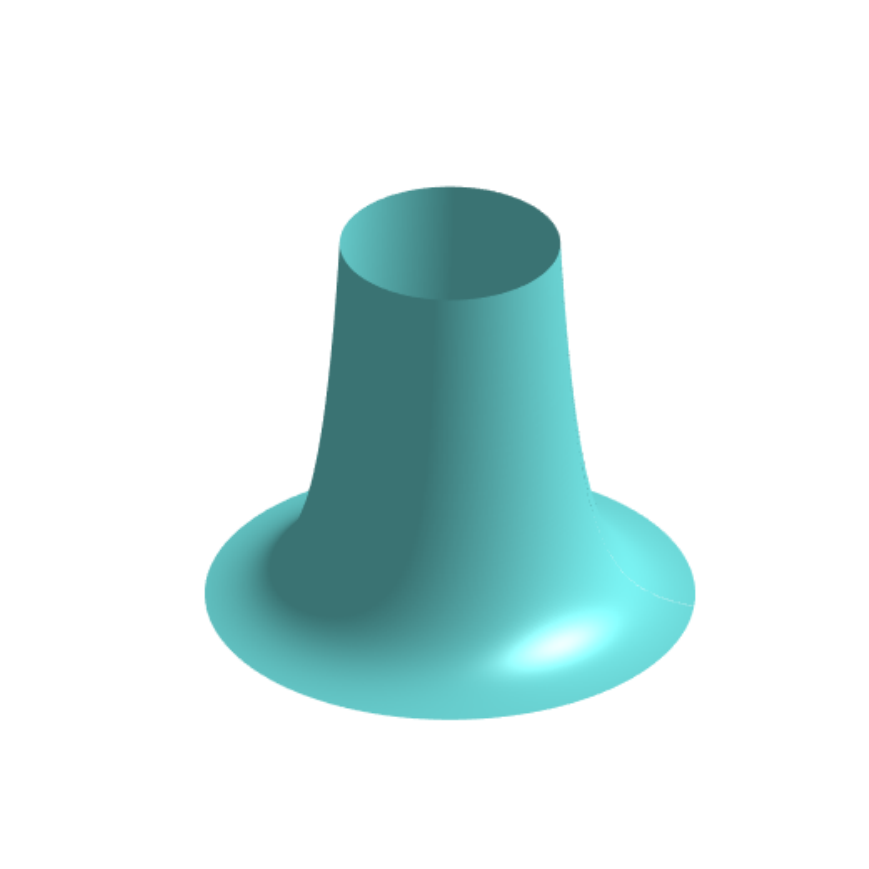}
           & 
           \includegraphics[width=1.9cm,trim={2cm 1.25cm 2cm 2cm}, clip]{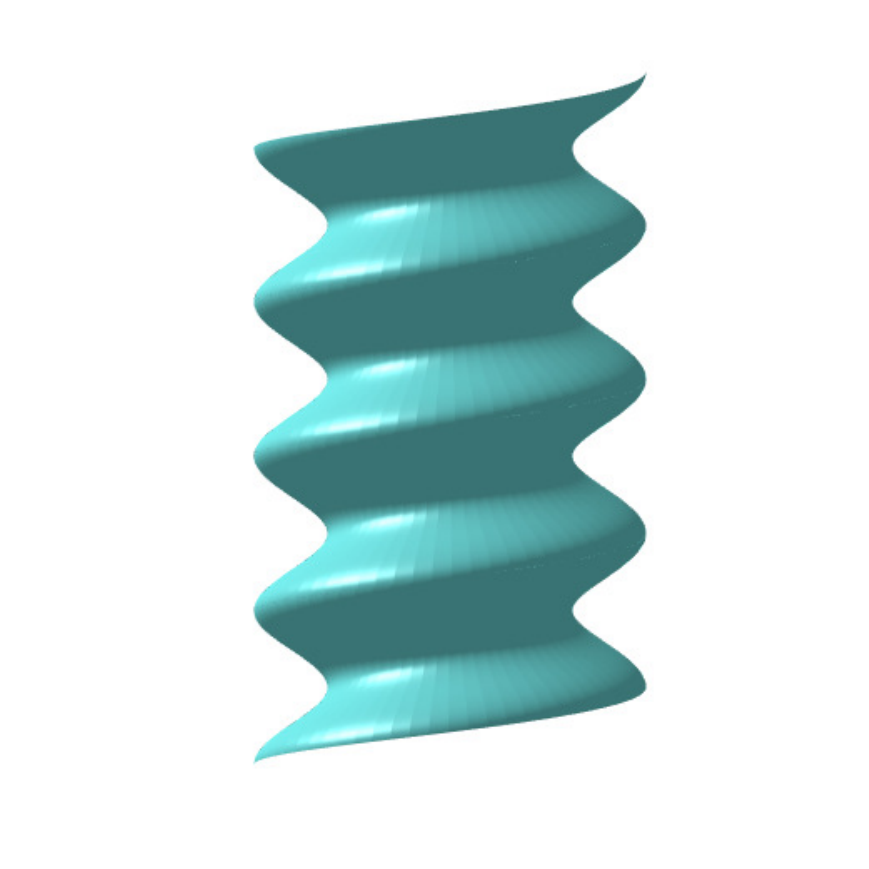}
           &
           \includegraphics[width=1.8cm,trim={2cm 1.2cm 2cm 1.2cm}, clip]{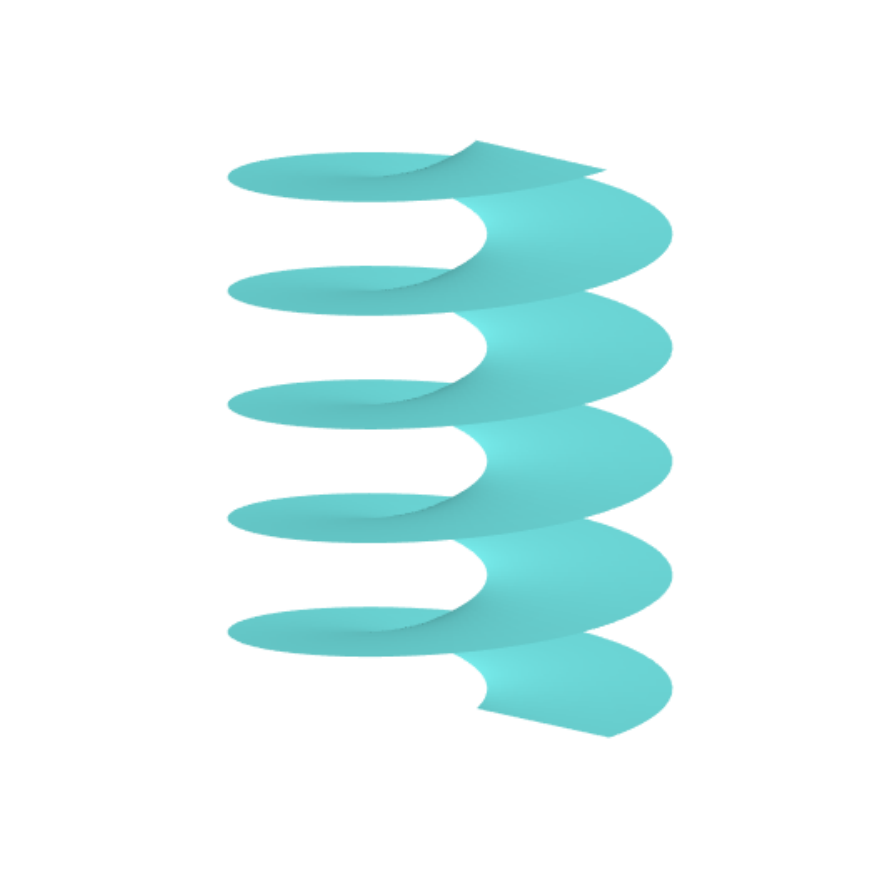}
           \\
           & & & & \\
    $
    \begin{cases}
      x=\frac{a\cos u}{1+b\cos(5u)} 
      \\ 
      y=\frac{a\sin u}{1+b\cos(5u)} 
      \\ 
      z=v
   \end{cases}
   $ 
   &
   $\begin{cases}
      x=\frac{av\cos u}{1+b\cos(5u)} 
      \\ 
      y=\frac{av\sin u}{1+b\cos(5u)}
      \\ 
      z=Av+B
   \end{cases}$ 
   &
   \scalebox{0.75}{$
   \begin{cases}
      x=au\cos v 
      \\ 
      y=au\sin v 
      \\ 
      z=\frac{b}{u^5} 
   \end{cases}
   $}
   &
   \makecell{\scalebox{0.75}{$
    \begin{cases}
    x=R(u)\cos v 
    \\ 
    y=R(u)\sin v 
    \\ 
    z=k_1(u+nv)+k_2
   \end{cases},
   $} \\ 
   \scalebox{0.75}{where} 
   \\ \scalebox{0.75}{$R(u):=R_1+\dfrac{R_2-R_1}{2}(\cos u+1) $,}
   \\ \scalebox{0.75}{$n\in\mathbb{Z}$}}
   &
    \makecell{\scalebox{0.75}{$
    \begin{cases}
    x=R(u)\cos v 
    \\ 
    y=R(u)\sin v 
    \\ 
    z=v
   \end{cases},
   $} \\ \scalebox{0.75}{where} \\ \scalebox{0.75}{R(u):=au+(1-u)b }} \\
           
           \hline
    \end{tabular}
}
        \label{tab:primitiveComplesse}
    \end{table*}

\section{Recognising geometric primitives using Hough transforms}\label{sec:algorithmHT}
In this section, we describe a method to recognise an input point cloud $\mathcal{P}$ with geometric primitives via the HT technique. It consists of the following main steps: an initial  point cloud preprocessing followed by the iteration of a recognition step and 
a splitting phase, and a final 
step which applies a clustering technique  to discover geometric relationships between primitives or parts of them. A graphical illustration of its pipeline is given in Figure \ref{fig:pipeline}.

\subsection{Point cloud preprocessing}
First, the input point cloud $\mathcal{P}$ is translated to place its barycenter at the origin of the Cartesian coordinate system. Then, the normals at the input points are estimated. In the literature, there are several ways to estimate normals; in general, we are interested in using an approach suitable for handling noisy input.

For our purposes the orientation of the normal vector is not relevant, as we are only interested in aligning the most voted normal direction to the z-axis. We can therefore use the approach introduced in \cite{Raffo:2022}, which estimates the normal vectors through a voting procedure.
For convenience, when $\mathcal{P}$ is clean we can also consider the method presented in \cite{pcnormals}, which is available in MATLAB with the \texttt{pcnormals} function; for each input point in $\mathcal{P}$, the method selects the $k$ points of $\mathcal{P}$ closest to that input point and then uses principal component analysis (PCA) to estimate the normal vector. Note that this does not necessarily generate normal vectors that are coherent in their orientation; however, this is not a problem for our pipeline since we are not interested in the normal orientation but rather in the direction of the (estimated) normal vectors.

The point cloud is then rotated so that the most voted direction among the (just-estimated) normal vectors coincides with the $z-$axis. This is an essential step for multiple reasons. Firstly, our dictionary contains primitives that -- in their standard form -- are given with the axis coincident with the $z-$axis. In addition, this process is necessary to test the existence -- as a first check -- of primitives in their standard form (i.e., with centres or vertices in the origin of the Cartesian axes and with the normal or the principal axis aligned to the $z$-axis), thus reducing the number of parameters that will need to be estimated by the HT.

Finally, $\mathcal{P}$ is scaled into a unit cube, in order to restrict the range of variation for most parameters.

\subsection{Recognition step\label{sec:recognition_step}}

After being preprocessed, $\mathcal{P}$ becomes the input of the HT-based recognition step. Since $\mathcal{P}$ can contain different instances of the same primitive type and primitives of different types, the standard HT procedure must be adjusted to allow such cases. This step can be summarized as follows:
    \begin{itemize}
        \item \textit{Selection of one or more families of primitives}. The user can select the families of primitives to be used for recognition; in its default configuration, the algorithm tests  the presence of simple geometric primitives -- one family at a time. By studying the geometric properties of $\mathcal{P}$ and by relating them to the geometric characteristics of the selected family (e.g., bounding box, radius, diagonal length), it is possible to initialise a region \textit{T} of the parameter space. The region \textit{T} is discretized into cells, which are uniquely identified by the coordinates of their centre. Then,  an accumulator function $H$, in the discretized form of a matrix, is initialized. The matrix entries are in a one-to-one correspondence with the cells of the discretization performed in the previous step.
        \item \textit{Estimation of the accumulator function}. An entry of the accumulator function $H$ is increased by $1$ each time the Hough transform $\Gamma_P$ of a point $P$ intersects the corresponding cell. The way we check if and where a Hough transform intersects some cells depends on whether the family of surfaces is described in implicit or parametric form.
        In our case,  surfaces are expressed in parametric form;  we adapt to surfaces the method devised for curves in \cite{stagLoghi2020}. Specifically, if the system of equations $f_\mathbf{a}$ defining the family can be solved analytically with respect to the unknown parameters $\mathbf{a}$, we automatically calculate  a sample of $\Gamma_P$ by exploiting the Moore-Penrose pseudo-inverse \cite{penrose_1956} of the matrix that defines the coefficients of $\mathbf{a}$. The evaluation of the intersection is translated into an inequality between the components of the sample points of $\Gamma_P$ and the coordinates of the cell endpoints. 
        \item \textit{Selection of potential fitting primitives.} The cells corresponding to the peak values of the accumulator function $H$ have to be identified. When the set of points is assumed to represent a single surface profile, the traditional HT formulation aims at finding the maximum of the accumulator function $H$; when the family of surfaces is not Hough-regular, there might exist several maxima. On the other hand, if the point cloud is composed of different primitives, different peaks of $H$ identify potential primitives that might fit different parts.
          To select these peaks, we observe that peaks of the accumulator function corresponding to primitive shapes rise distinctly with respect to their neighbours and are well-characterised as isolated peaks. Formally said, we proceed by identifying peaks that have high  \emph{topological persistence}\footnote{The notion of topological persistence was introduced in \cite{EdelZomo2002} for encoding and simplifying the points of a filtration $f$ by classifying them as either a feature or noise depending on its lifetime or persistence within the filtration. In practice, given a pair of points $p$ and $q$, their persistence is defined as $f(p)-f(q)$. Pairing is defined in terms of the topological connection between the points, for details on topological persistence and saliency, we refer to \cite{EdelZomo2002,DORAISWAMY2013}. In our case, the domain is represented by the grid of \textit{T}, the role of filtration is played by the accumulator function $H$, and we are interested only in the peaks of $H$.}. In our implementation,  peaks that correspond to primitives are automatically recognised by keeping the local maxima having a persistence higher than 10\% of the maximum value of $H$, by using the algorithm for persistent maxima proposed in \cite{biasotti2016tracking}.
          The coordinates of the cell centres of the maxima  correspond to the parameters of potentially recognised surface primitives.
             \item \textit{Evaluation of the approximation accuracy}. 
             To measure the recognition accuracy of a specific primitive, we select the set of input points $\mathcal{X}$ closer to such a primitive than a given threshold and study its density via the $k$-Nearest Neighbor algorithm (see, for example, \cite{Friedman:1977}). If $\mathcal{X}$ is sparse, the recognised primitive is considered a false positive; otherwise, for each dense subset $\mathcal{X}_i\subset{\mathcal{X}}$ we define the \emph{Mean Fitting Error} (MFE) as:
             \begin{equation}
            E(\mathcal{X}_i,\mathcal{P}):=\dfrac{1}{|\mathcal{X}_i|}\sum_{\mathbf{x}\in\mathcal{X}_i}d(\mathbf{x},\mathcal{P})/l_i,
             \label{eqn:MFE}
             \end{equation}
             where $\mathcal{P}$ is the current primitive, $d$ is the Euclidean distance, and $l_i$ is the diagonal of the axis-aligned bounding box containing $\mathcal{X}_i$.
             What can  happen when the selected family of primitives does not fit any part of the point cloud? There are two possibilities:
             \begin{itemize}
                 \item The accumulator function is identically zero, with the result that its persistence is zero; therefore the selection of potential fitting primitives returns the empty solution.
                 \item The accumulator function $H$ does not present predominant peaks, resulting in false positives which are identifiable by studying the sparsity of the fitted points.
             \end{itemize}
            
            Given a set of dense points $\mathcal{X}_i$ and two candidate primitives $\mathcal{P}_{i,1}$ and $\mathcal{P}_{i,2}$, we first calculate the fitting errors $E(\mathcal{X}_i,\mathcal{P}_{i,1})$ and $E_i(\mathcal{X}_i,\mathcal{P}_{i,2})$ between each primitive and $\mathcal{X}_i$; the primitive having lowest error is kept.  
                         
            Each time a primitive is recognized, the points of $\mathcal{P}$ close to the recognised primitive less than a threshold $\varepsilon$ are discarded from $\mathcal{P}$. The value of $\varepsilon$ typically ranges from $1\%$ to $3\%$ of the diagonal of the minimum bounding box of the $\mathcal{P}$, according to the type of primitive.             
   \end{itemize}

\begin{figure*}[h!]
    \centering
    \includegraphics[width=\textwidth,trim={0.3cm 2cm 0.3cm 2.25cm}, clip]{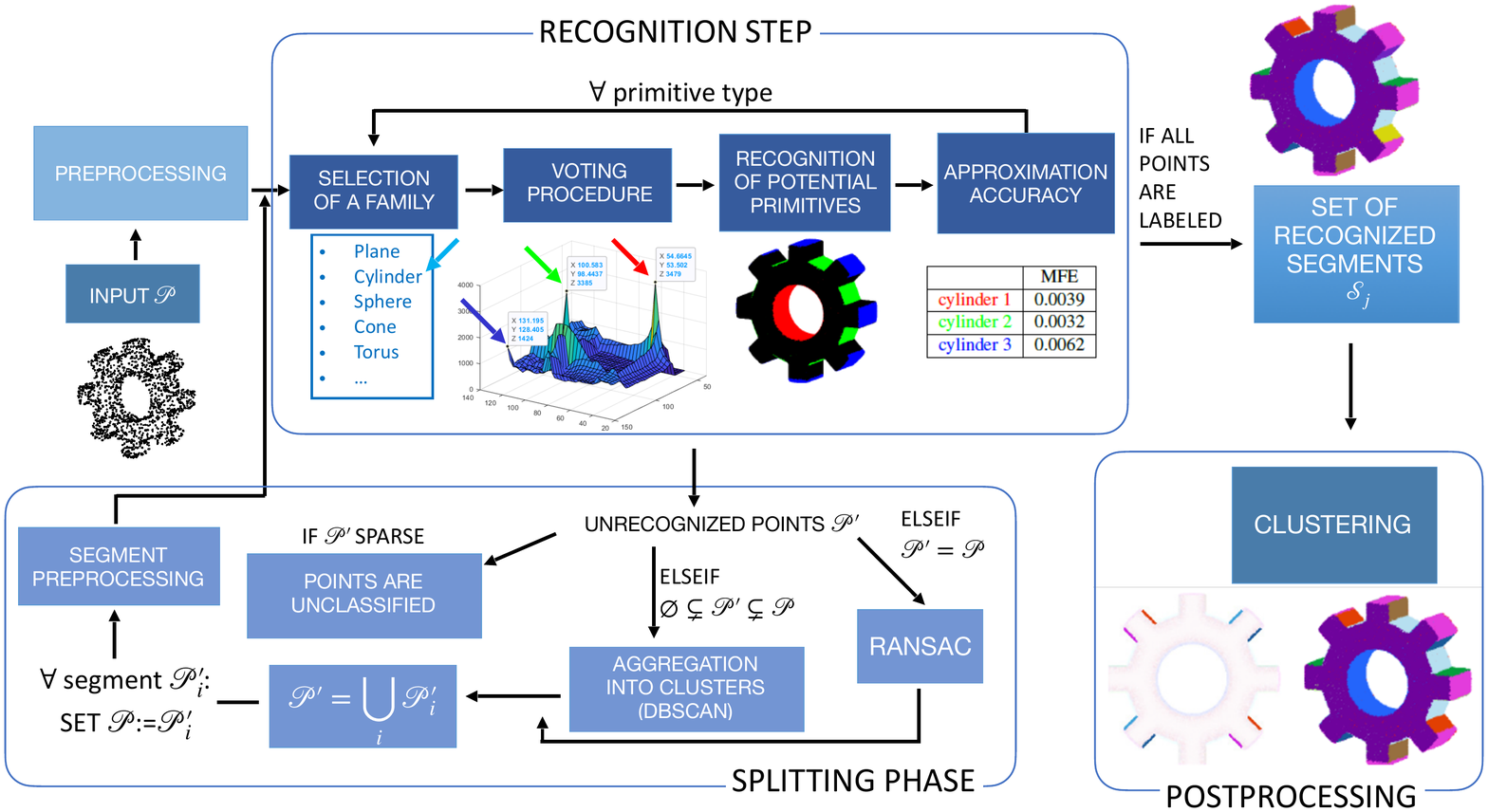}
    \caption{Pipeline of the method.}
    \label{fig:pipeline}
\end{figure*}

The algorithm returns the parameters of the geometric primitives and the corresponding points fitted by them, as well as the set of points that were not fitted by any primitive -- denoted by $\mathcal{P}'$. Note that, if more geometric primitives potentially fit the same region of the point cloud, we select the one with the minimum fitting accuracy (i.e., the lowest MFE).

\subsection{Splitting phase \label{sec:splitting_phase}}
The algorithm chooses the next step according to the resulting $\mathcal{P}'$:

\begin{itemize}
    \item If $\mathcal{P}'$ is sparse, then its points are returned as unclassified. 
    \item If $\emptyset\subsetneq{}\mathcal{P}'\subsetneq{}\mathcal{P}$, i.e., if $\mathcal{P}'$ is a proper nonempty subset of $\mathcal{P}$,  we proceed by aggregating points in $\mathcal{P}'$ into clusters $\mathcal{P}'_j$, with $j=1,\dots,N_{\text{clust}}$, by adopting the \textit{Density-Based Spatial Clustering of Applications with Noise} (DBSCAN) method \cite{EKS*96}, which groups together nearby points and marks as outliers isolated points in low-density regions. It requires two parameters: the first is a threshold used as the radius of the density region, while the second represents the minimum number of points required to form a dense region. Then, the recognition step from Section \ref{sec:recognition_step} is iterated over each cluster $\mathcal{P}'_j$ as long as some geometric primitives are recognised.
    Before proceeding with a new recognition round, we preprocess each cluster $\mathcal{P}'_j$; the preprocessing adopted here differs from that applied to the entire point cloud $\mathcal{P}$: we exploit the strategy presented in \cite{Raffo:2022} to estimate its standard form, thus reducing the number of parameters that have to be estimated in the new iteration of the recognition step. Specifically, it computes the normals of the points of the cluster $\mathcal{P}'_j$ through a voting procedure and it estimates the position of vertices or centres and the direction of axes of symmetry, exploiting different geometric rules specialised for each type of primitive. To reduce the dimension of the parameter space, vertices and centres are translated to the origin of the coordinate system, while axes are aligned to the $z-axis$. Then, it automatically centres and orients a (spherical, cylindrical, conical or toric) cluster so that it can be fitted with a primitive in standard form.
    In case the recognition step involves complex primitives, the strategy proposed in \cite{Raffo:2022} can be naturally extended since they are characterized by the presence of symmetry axes that can be estimated in a similar way through the use of the normals of points.
    \item It is possible that, in some steps, no primitive is recognized. This can happen in two cases: (i) when applying our algorithm to an input point cloud that has no primitives in their standard form, or (ii) when the estimation of the standard position of some cluster $\mathcal{P}'_j$ fails -- because $\mathcal{P}'_j$ contains more than one primitive. In both cases, we proceed by oversegmenting the model: in our experiments, we use the RANSAC algorithm proposed in \cite{Schnabel2007}, because of its efficiency and its tendency to oversegment point clouds (see, for example, \cite{Le2017}); however, we proceed by estimating primitive types and geometric parameters with a new round of the recognition step, as voting procedures have shown greater robustness to point cloud artifacts. 
\end{itemize}

This step ends by transforming the parameters found by the Hough transform with respect to the inverse translations, rotations and scaling applied in the previous steps of the algorithm. The result of this procedure is the decomposition of an input point cloud $\mathcal{P}$ into several subsets, called \emph{surface segments} or simply \emph{segments}, in such a way that points of the same segment are well approximated by the same primitive. We denote these final segments by $\mathcal{S}_j$.

\subsection{Segment association based on geometric relationships}
After decomposing the input point cloud into the segments $\mathcal{S}_j$, a clustering technique is applied to unveil geometric relationships. The goal of this step is to find maximal primitives that are not automatically detected in the HT-based recognition process, to detect patterns of primitives or to relate primitives (or parts of them) even if they do not belong to the same primitive. More specifically, we relate primitives on the basis of their positions, orientation and dimension, as described in \cite{Raffo:2022} for simple geometric primitives: segments are aggregated only if they are part of the same primitives; otherwise, we only identify  relationships that  may be of interest, for example in the processing phases (e.g., process planning, machining). To give some examples:

\begin{itemize}
    \item Segments sampled from circular cylinders can be associated 
    with respect to their radii and rotational axes. For instance, it is possible to check whether such segments: originate from the exact same primitive shape, i.e., if they are all sampled from a cylinder of a given radius and rotational axis (such as for the red segments in Figure \ref{fig:various_SGPs}); are characterized by the same radius and have parallel rotational axes (such as for the black segments in Figure \ref{fig:NuGear}); share the radius but not the rotational axis, not even in terms of parallelism.
    \item Segments sampled from helical surfaces can be associated 
    with respect to their parameters $R_1$, $R_2$ and $n$, $k_1$, $k_2$ and the direction -- or a combination of such geometric descriptors.
    \item Segments sampled from $n$-convexity cylinders can be associated 
    with respect to their parameters $a$ and $b$, the number of convexities $n$, and the direction of the generatrix -- or a combination of such geometric descriptors.
\end{itemize}
To perform segment associations, 
 a well-known (hierarchical) clustering strategy is considered -- the complete linkage -- as it penalizes chaining effects.

\subsection{An illustrative example}
Figure \ref{fig:pipeline} provides a graphical illustration of the pipeline of our method. The main steps of the algorithm are associated with an example of a point cloud representing a gear. 

After preprocessing the point cloud, we start with the first round of the recognition step -- which aims at recognizing the presence of primitives in their canonical position. For the sake of clarity, however, the graphical illustration only focuses on the recognition of cylinders. Specifically, once the family of cylinders is selected, the corresponding accumulator function is computed;
it exhibits three peaks, which indicate the presence of three potential solutions: the three cylinders highlighted in the colours red, green, and blue. The approximation accuracy for this type of primitive is less than $1\%$. 

Being some segments (the non-axis-aligned planar segments) non in their canonical position, their points are not recognized in the first round of the recognition step. Instead, these points are collected in $\mathcal{P}'$ and aggregated into dense clusters in the splitting phase; each of such clusters is individually studied by a new round of the recognition step and recognized as planar segments. However, being these segments studied separately, we lose some geometric information. We then apply clustering to recognise that some planar segments lie on the same parametric plane, by exploiting the geometric descriptors provided by the HT.

The final result is a segmentation of $17$ segments. It is worth mentioning that the method is able to group some of the non-adjacent parts, which belong to the same primitive in canonical position -- as in the example of the two external cylinders, and of the axis-aligned planes that form four couples. However, for primitives not in their canonical position, postprocessing based on clustering is required. 

\subsection{Computational complexity}
In the preprocessing step, our method includes the estimation of the normal vectors and the individuation of the most voted normal, which is implemented as explained in \cite{pcnormals} -- in the case of clean point clouds -- and in \cite{Raffo:2022} -- in the presence of point cloud artifacts. For a thorough analysis of the computational complexities of this estimation, the reader is referred to the corresponding papers.

The formulation of the HT for surface primitives embedded in $\mathbb{R}^3$ naturally extends that for plane curves in  $\mathbb{R}^2$. In the pipeline presented in this paper, for each point cloud $\mathcal{P}$ (both in the case of the initial point cloud and of single clusters at subsequent iterations) we apply the HT by considering the different types of geometric primitives one at a time; thus, the dimension of the parameter space changes accordingly to the type of primitive considered. The quantization of a region of interest and the dimension of the parameter space determines the size of the accumulator function, and dominate both the memory usage and the computational complexity of the Hough transform: it is, therefore, necessary to balance the samples for each parameter and their number. To reduce the computational cost,   primitives are put in their (estimated) standard positions, in this way the number of parameters does not exceed $3$, as explained in \cite{Raffo:2022}; complementarily, adaptive approaches can be used to further speed up the search (e.g., \cite{HTadaptive}). 
The overall computational complexity of the HT-based recognition step is $\OOO(ML)$, where $M$ denotes the number of cells of the parameter space and  $L$ represents the number of points (in the initial point cloud or in a cluster obtained at a subsequent iteration) at which we evaluate the HT accumulator function. More precisely:
\begin{itemize}
    \item At the first iteration the HT is applied to the whole input point cloud. Supposing that the number of families of primitives to be used for the recognition is $K$ and the number of points in $\mathcal{P}$ is $N$, the overall complexity of the first iteration is $\OOO(N\sum_{k=1}^K M_k)$, where $M_k$ is the dimension of the parameter space for the $k$-th primitive type.
    \item In the subsequent steps, the HT-based recognition method is applied to each cluster $\mathcal{P}'_j$ and then the computational cost corresponds to $\OOO(\sum_{j=1}^{N_{\text{clust}}}N_j\sum_{k=1}^K M_k)$, where $N_j$ denotes the number of points in  cluster $\mathcal{P}'_j$ while $N_{\text{clust}}$ is the number of clusters.
\end{itemize}
Notice that, being the recognition step performed independently w.r.t. the primitive type, the task is \emph{embarrassingly parallel}. The same applies to the clusters of points obtained in the same iteration.

Finally, agglomerative hierarchical clustering approaches require, in their naive implementation, $\OOO(N_\text{seg}^3)$ operations, where $N_\text{seg}$ denotes the number of segments in the output segmentation; see, for example, \cite{Day:1984}. When it comes to complete-linkage clustering, one can consider more efficient implementations, such as the one proposed in \cite{Defays:1977}, which costs $\OOO(N_\text{seg}^2)$. Although the dissimilarity-matrix assembly costs $\OOO(N_\text{seg}^2)$, one may note that each entry is computed independently; again, the task is therefore embarrassingly parallel.

\section{Experimental results}
\label{sec:results}

In this section, we show the results obtained with our method. Specifically, Section \ref{sec:simplePrim} presents some point clouds of mechanical objects containing only simple geometric primitives, while Section \ref{sec:comlexPrim} provides some examples with complex primitives. In Section \ref{sec:robustness}, we exhibit some noisy point clouds to demonstrate the robustness of our pipeline. In all cases, whenever possible, we show the maximal primitives of the selected point cloud. Finally, a qualitative and quantitative comparative analysis is provided in Section \ref{sec:comparative}.

\subsection{Overall analysis}
\subsubsection{Simple geometric primitives}\label{sec:simplePrim}

\begin{figure*}[h!]
    \centering
    \begin{tabular}{ccc}
    \begin{minipage}[c]{0.1\linewidth}
    \hspace*{-5cm}
    \vspace*{0cm}
    \includegraphics[width=2.5cm]{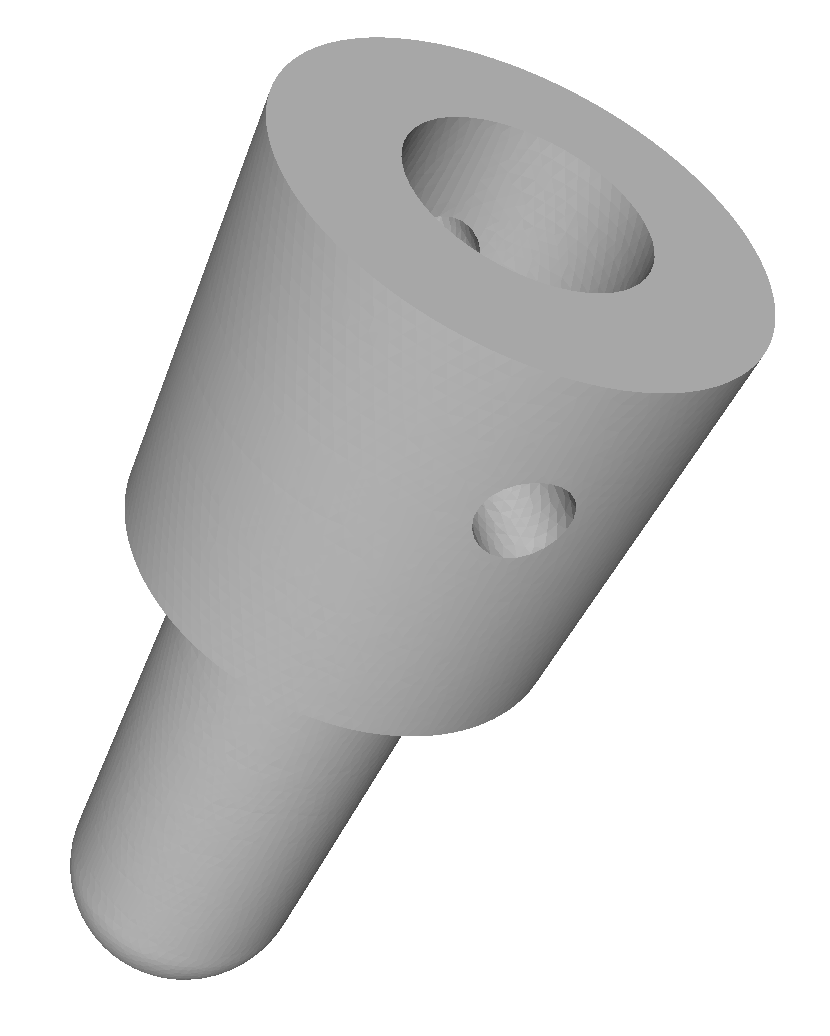}\\ 
    \end{minipage}
    &
    \begin{minipage}[c]{0.1\linewidth}
    \hspace*{-4.4cm}
    \vspace*{0cm}
    \includegraphics[width=4.75cm]{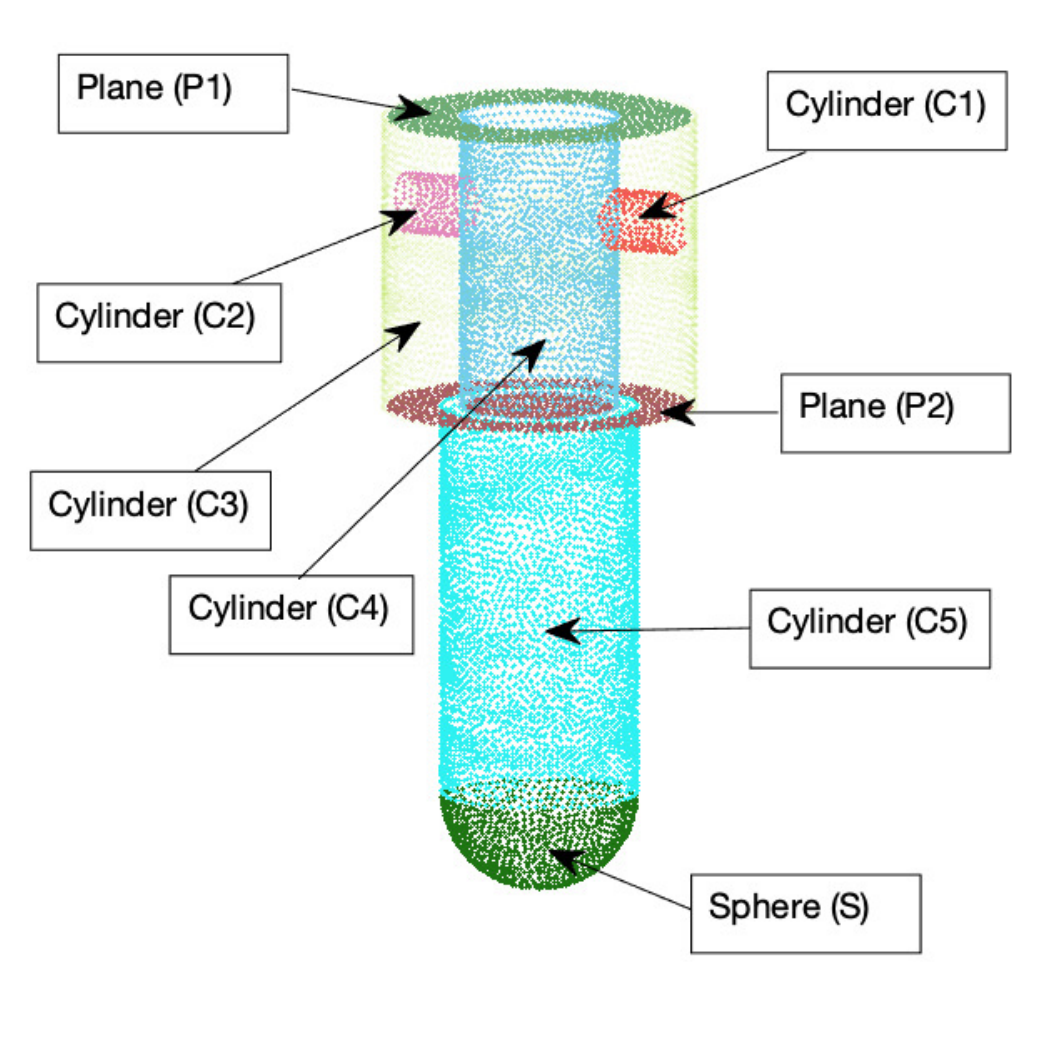}
    \end{minipage}
     &
     \begin{minipage}[c]{0.1\linewidth}
    \hspace*{-1.5cm}
    \vspace*{-0.9cm}
   \resizebox{5\textwidth }{!}{  
    \begin{tabular}{|c|c|c|c|}
    \hline
         \cellcolor{ForestGreen!15}equation & \cellcolor{ForestGreen!15}code & \cellcolor{ForestGreen!15}ground truth & \cellcolor{ForestGreen!15}HT parameters \\
         \hline
        \multirow{5}{*}{$x^2+y^2=r^2$} & C1 & $r=1.50$ & $r=1.50$ \\
        & C2 & $r=8.00$ & $r=8.00$\\
         & C3 & $r=4.00$ & $r=4.00$ \\
         & C4 & $r=1.50$ & $r=1.50$ \\
        & C5 & $r=5.00$ & $r=5.00$ \\
        \hline
        \multirow{2}{*}{$z=k$} &  P1 & $k=20.00$ & $k=20.00$ \\
        & P2 & $k=35.00$ & $k=35.00$ \\
        \hline
        $x^2+y^2+z^2=r^2$ & S & $r=5.00$ & $r=5.05$ \\
        \hline
        \end{tabular}}
        \par\vspace{1.2cm}
   \end{minipage}
  \\
        \begin{minipage}[c]{0.1\linewidth}
        \hspace*{-4.5cm}
        (a)
        \end{minipage} 
        & 
        \begin{minipage}[c]{0.1\linewidth}
        \hspace*{-2.25cm}
        (b)
        \end{minipage} 
        & 
        \begin{minipage}[c]{0.1\linewidth}
        \hspace*{2.5cm}
        (c)
        \end{minipage}\\
    \end{tabular}
        
        \caption{
        In (a) a mechanical CAD model from the  benchmark in \cite{Koch_2019_CVPR}; in (b) the vertices of its triangle mesh decomposed into $8$ surface segments; in (c), for each primitive, the HT parameters are compared with those  provided by the database}
    \label{fig:modelliABC1}
\end{figure*}

\begin{figure*}[h!]
\centering
    \begin{tabular}{ccc}
    \begin{minipage}[c]{0.1\linewidth}
    \hspace*{-5.25cm}
    \vspace*{0cm}
    \includegraphics[width=2.5cm]{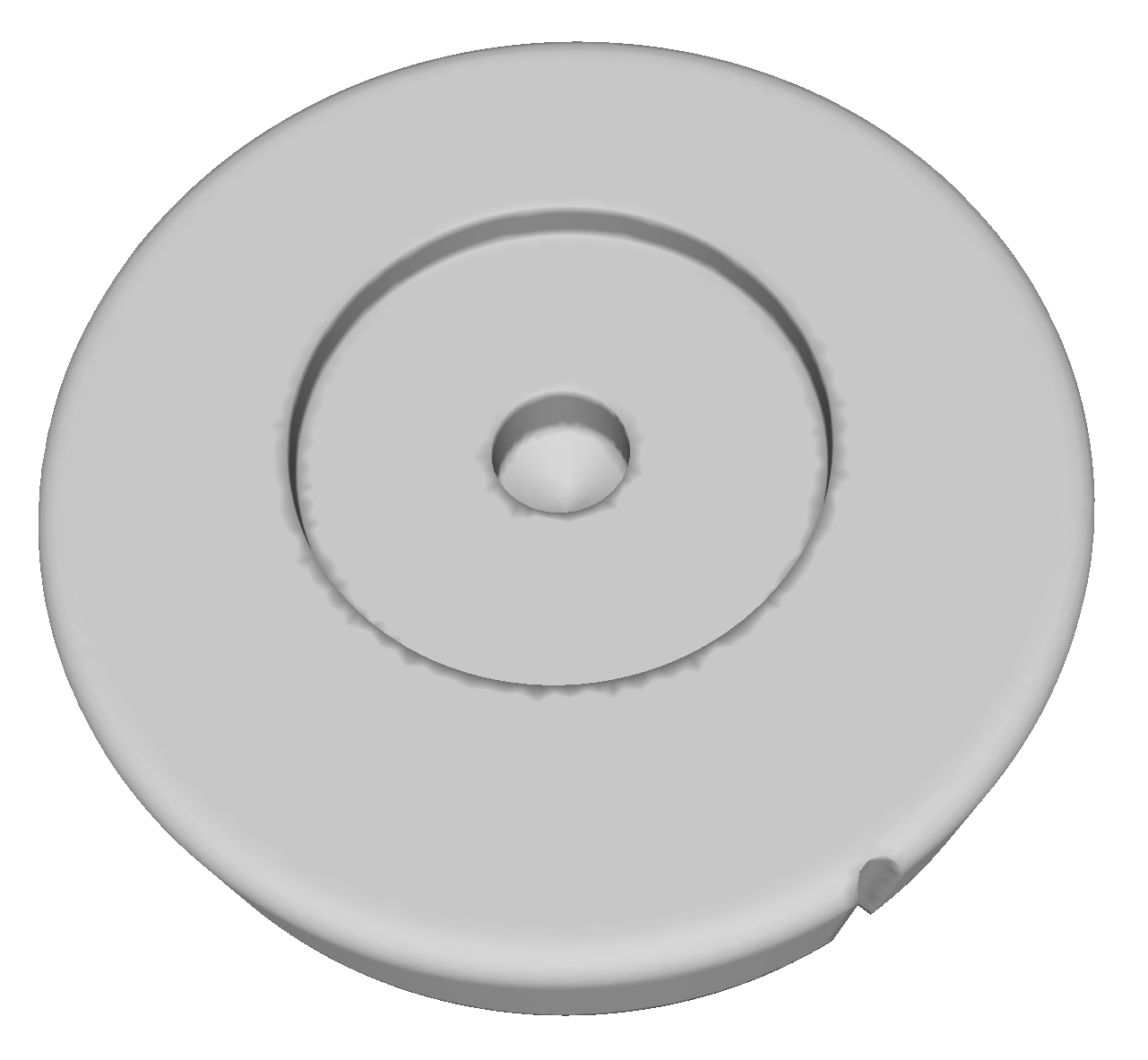}\\ 
    \end{minipage}
    &
    \begin{minipage}[c]{0.1\linewidth}
    \hspace*{-4.75cm}
    \vspace*{0cm}
    \includegraphics[width=4.4cm,trim=1cm 0cm 0cm 0cm, clip]{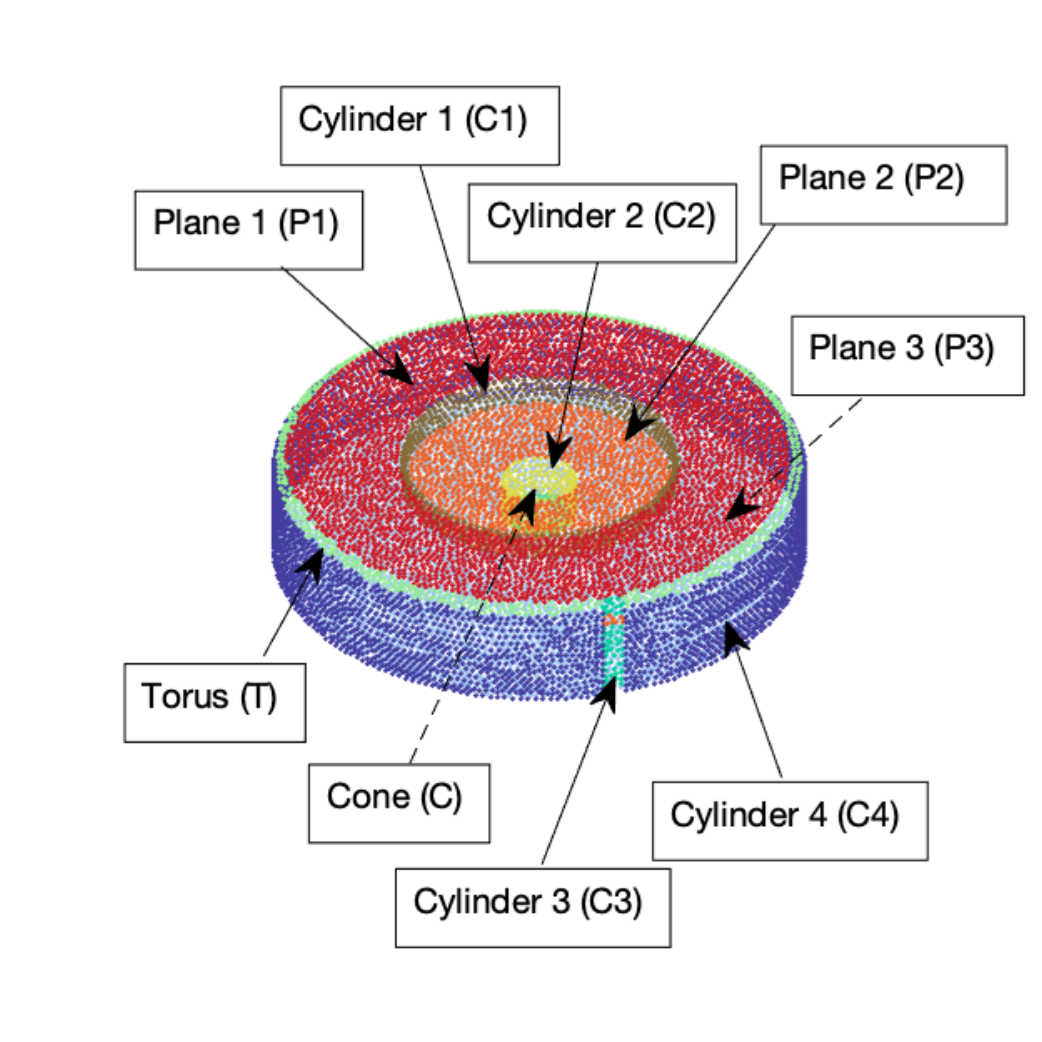}
    \end{minipage}
     &
         \begin{minipage}[c]{0.1\linewidth}
    \hspace*{-2.1cm}
    \vspace*{-0.9cm}
   \resizebox{5.6\textwidth }{!}{   \begin{tabular}{|c|c|c|c|}
    \hline
         \cellcolor{ForestGreen!15}equation & \cellcolor{ForestGreen!15}code & \cellcolor{ForestGreen!15}ground truth & \cellcolor{ForestGreen!15}HT parameters \\
         \hline
          \multirow{4}{*}{$x^2+y^2=r^2$} & C1 & $r=15.01$ & $r=15.01$ \\
        & C2 & $r=3.97$ & $r=3.97$ \\
        & C3 & $r=1.28$ &  $r=1.26$ \\
         & C4 & $r=29.25$ & $r=29.25$\\
        \hline
         \multirow{3}{*}{$z=k$} & P1 & $k=12.00$ & $k=12.00$ \\
       &  P2 & $k=9.08$ & $k=9.08$ \\
        & P3  & $k=0.00$ & $k=0.00$ \\
        \hline
        \multirow{2}{*}{ $(R-\sqrt{x^2+y^2})^2+z^2-r^2=0$}& T  & $R=27.25$ & $R=27.25$ \\
       & & $r=2.00$ & $r=2.00$ \\
        \hline
        \multirow{2}{*}{$x^2+y^2+a(z-b)^2=0$} &C & $a= -2.77$ & $a=-2.81$\\
        & & $b=2.19$ & $b=2.22$\\
        \hline
        \end{tabular}}
        \par\vspace{1.2cm}
        \end{minipage}
        
        \\
        \begin{minipage}[c]{0.1\linewidth}
        \hspace*{-4.25cm}
        (a)
        \end{minipage} 
        & 
        \begin{minipage}[c]{0.1\linewidth}
        \hspace*{-2.85cm}
        (b)
        \end{minipage} 
        & 
        \begin{minipage}[c]{0.1\linewidth}
        \hspace*{2.25cm}
        (c)
        \end{minipage}\\
        
    \end{tabular}
    \\
        \caption{
        In (a) a mechanical CAD model from the  benchmark in \cite{Koch_2019_CVPR}; in (b) the vertices of its triangle mesh decomposed into $9$ surface segments; in (c), for each primitive,  the HT parameters are compared with those  provided by the database}
    \label{fig:modelliABC2}
\end{figure*}

As a sanity check, we first apply our method to two models from \cite{Koch_2019_CVPR}, which were selected because containing all simple geometric primitives -- plane, sphere, cylinder, cone and torus -- and because of the availability of ground truth. A first example is provided in Figure \ref{fig:modelliABC1}. The point cloud, corresponding to the set of vertices of the original triangle mesh, is decomposed in $8$ surface segments: $5$ cylinders, $2$ planes and $1$ sphere. The high accuracy of our method is proved by comparing the parameters from the HT with those in the database. In the second example, presented in Figure \ref{fig:modelliABC2}, the corresponding point cloud is subdivided into $9$ surface primitives: $4$ cylinders, $1$ torus, $3$ axis-aligned planes and $1$ cone. Again, we are able to recognise all primitives up to a small error in the parameters with respect to those provided in the dataset. 

\begin{figure*}[h!]
    \centering
    \begin{tabular}{|cc|c|}
    \hline
    \multicolumn{2}{|c|}{\cellcolor{ForestGreen!15}Segments (a)} & \cellcolor{ForestGreen!15}Segments (b) \\
    \hline
    & &
    \\
    \includegraphics[scale=0.5, trim=1.5cm 1cm 1cm 0.5cm, clip]{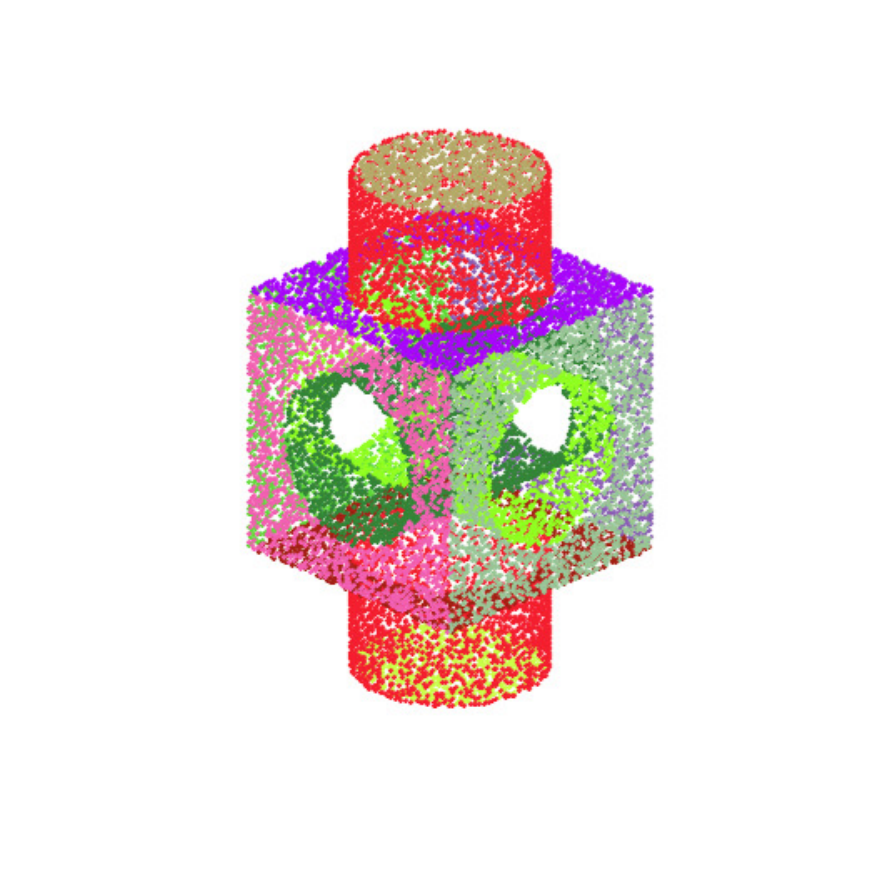}
    &
    \includegraphics[scale=0.5, trim=1.5cm 1cm 0.5cm 0.5cm, clip]{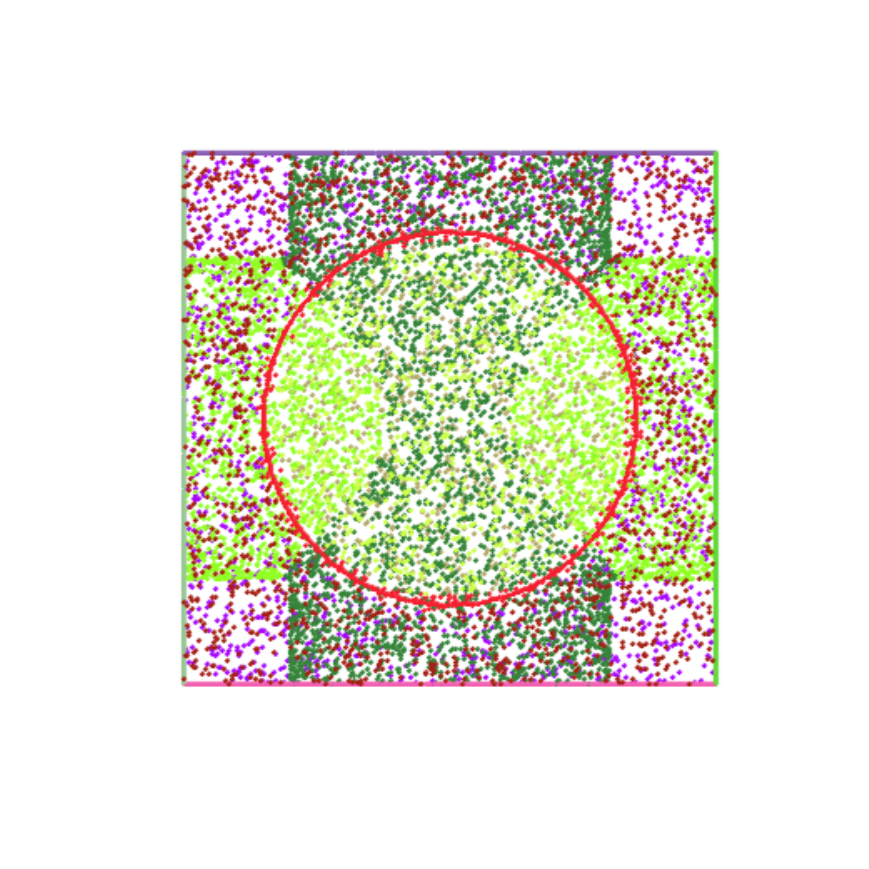}
    &
   \includegraphics[scale=0.275, trim=0cm 5.5cm 0cm 5cm, clip]{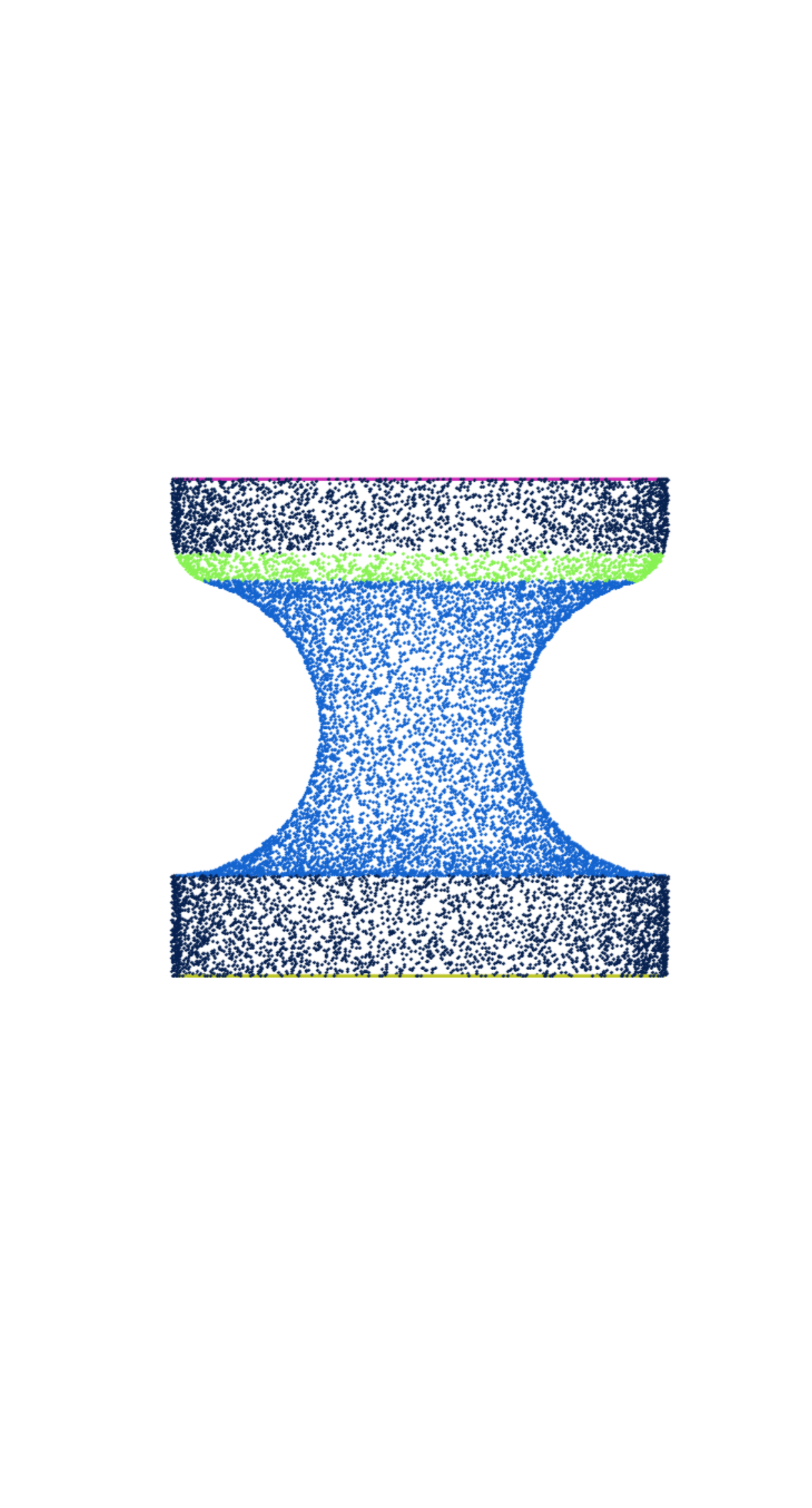}
   \\[10pt]
   \hline
   \end{tabular}
   \caption{Recognition of CAD point clouds containing only simple geometric primitives. The identification of maximal segments does not require, in these cases, the application of any clustering algorithm.}
    \label{fig:various_SGPs}
\end{figure*}

Figure \ref{fig:various_SGPs}(a-b) shows point clouds that can be segmented into complete geometric primitives without the need for a further step of aggregation. The first example, shown in Figure \ref{fig:various_SGPs}(a), consists of $11$ segments --  $3$ cylinders and $8$ planes -- and it highlights the robustness of our  method when it comes to detecting intersecting cylinders, here arranged similarly to a Steinmetz solid. Complete geometric primitives, each of which is represented by a specific colour, are automatically detected. Another point cloud, displayed in Figure \ref{fig:various_SGPs}(b), contains $5$ segments: $2$ tori, $1$ cylinder and $2$ axis-aligned planes. As for the previous case, the HT-based recognition  successfully detects all the primitives, even those made up of non-adjacent parts.

\begin{figure}[h!]
    \centering
    \begin{tabular}{|c|c|c|}
    \hline
    \cellcolor{ForestGreen!15}(a) Model & \cellcolor{ForestGreen!15}(b) Segments & \cellcolor{ForestGreen!15}(c) Clustering \\
    \hline
    \begin{tikzpicture}[spy using outlines={circle,black,magnification=3.8,size=2.75cm, connect spies}]
    \node {\pgfimage[height=4cm]{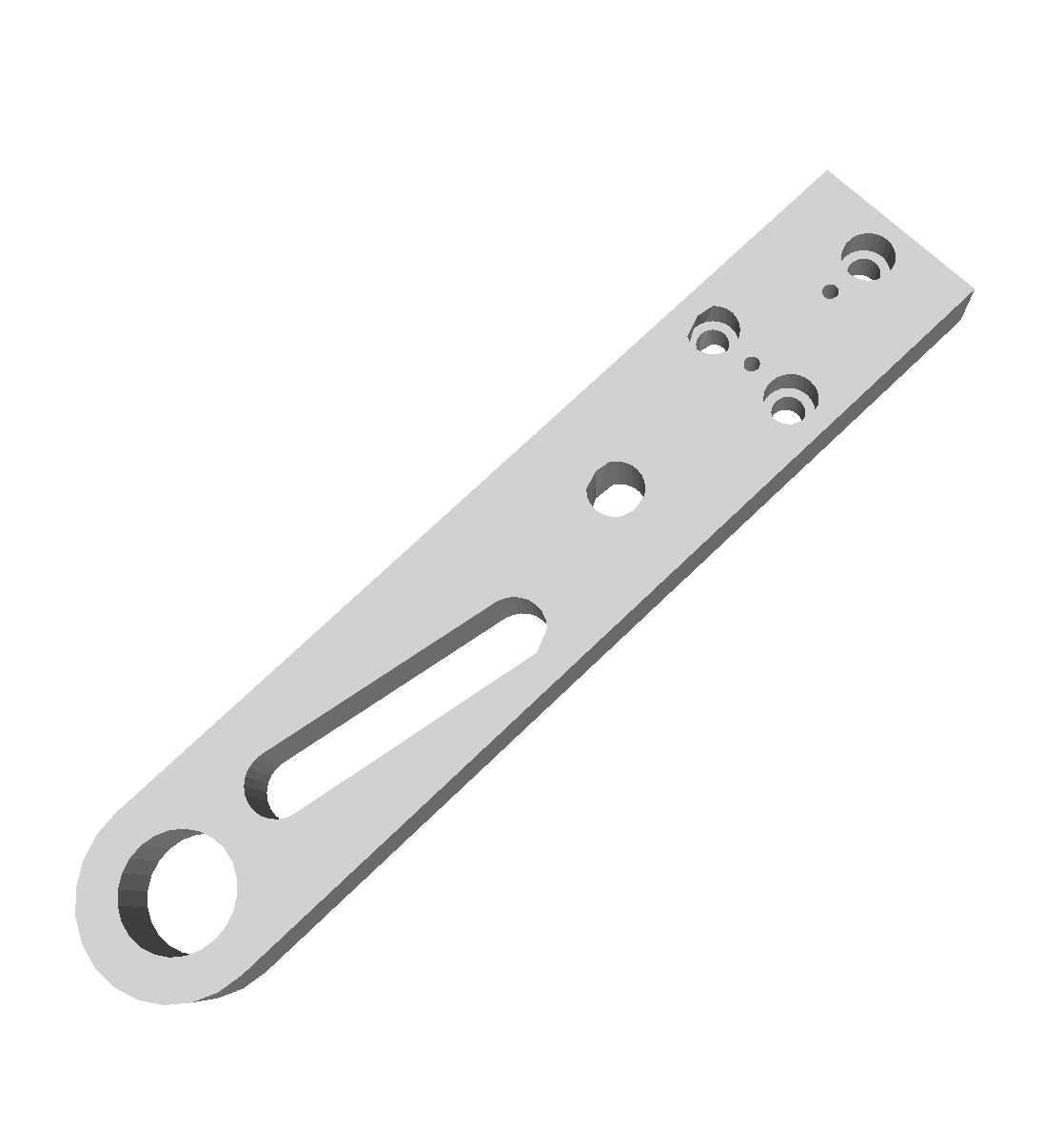}};
    \spy on (0.7,0.7) in node [left] at (3,-1.0);
    \end{tikzpicture}
    &
    \includegraphics[scale=0.25, trim=2cm 0cm 2cm 0cm, clip]{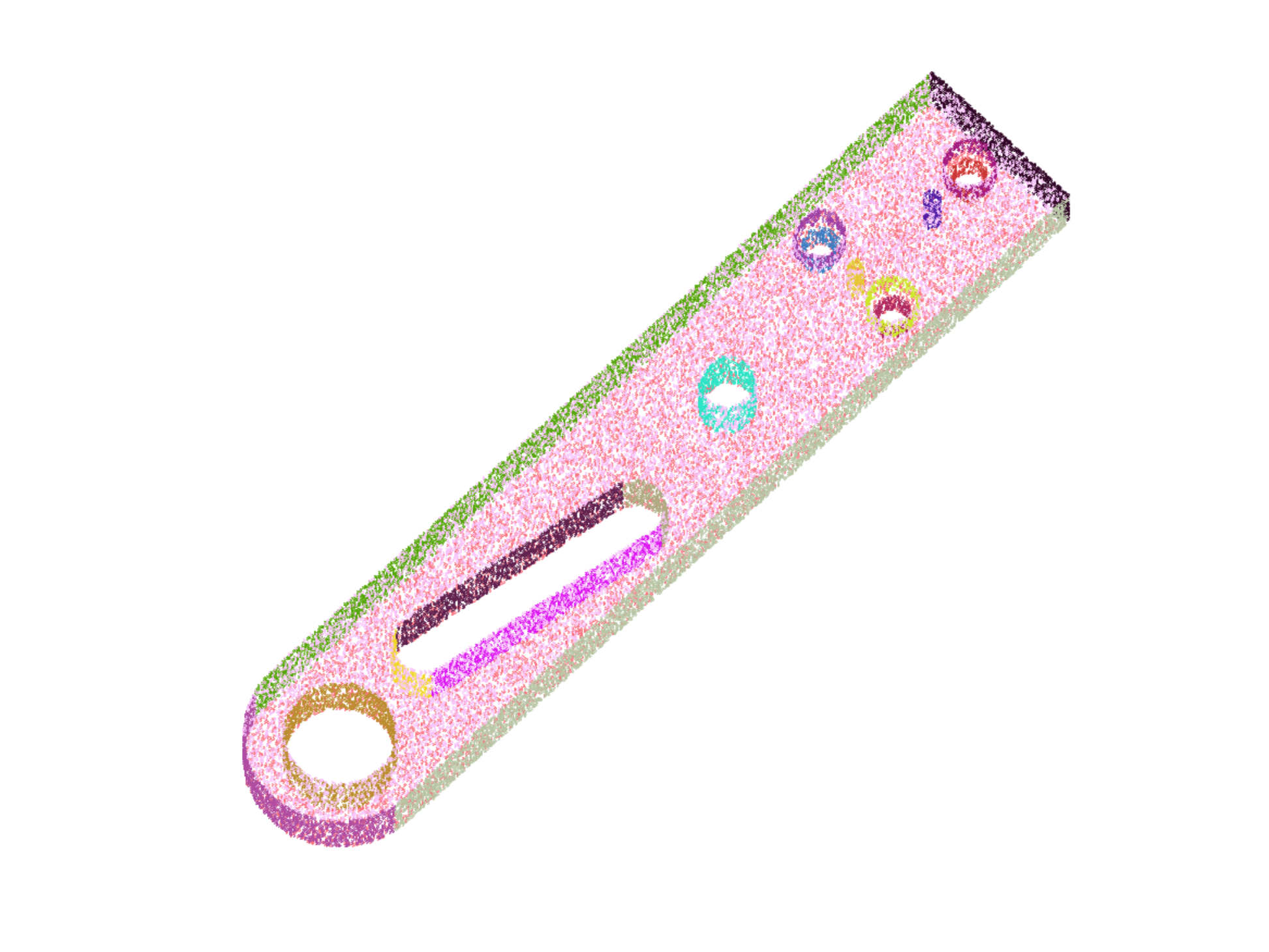}
    &
    \includegraphics[scale=0.25, trim=2cm 0cm 2cm 0cm, clip]{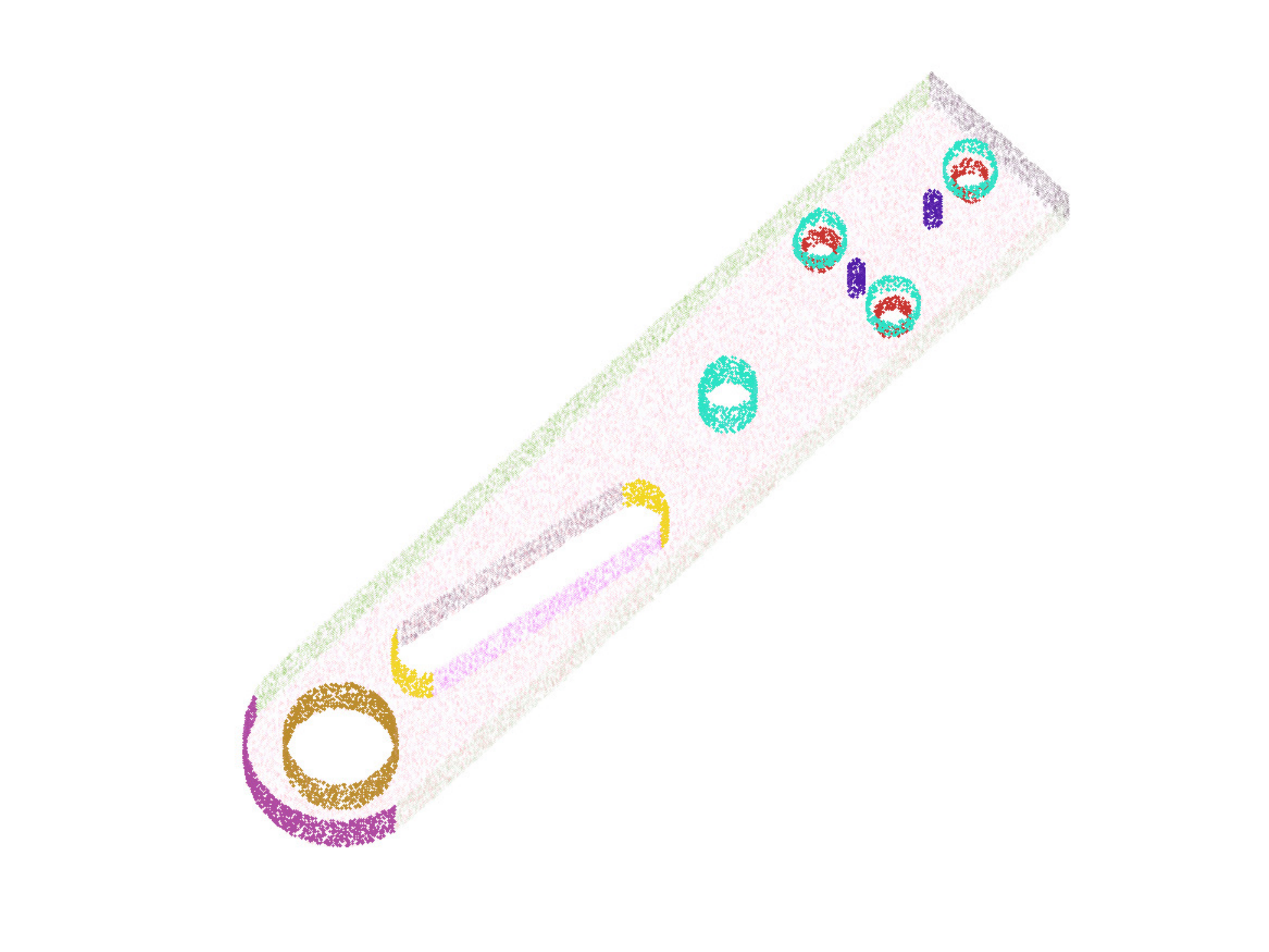}\\
    \hline
    \end{tabular}
    \caption{A linkage arm. In (a), the original model is shown, together with a magnification revealing some imperfections. The segments identified by the HT are shown in (b) in different colours. (c) draws  attention to cylinders, among which we can recognize segments lying on the same primitive, up to a translation.}
    \label{fig:linkage}
\end{figure}

\begin{figure*}[h!]
    \centering
    \begin{tabular}{|c|c|c|}
    \hline
    \cellcolor{ForestGreen!15}(a) Model & \cellcolor{ForestGreen!15}(b) Segments & \cellcolor{ForestGreen!15}(c) Clustering \\
    \hline
    & & \\
    \includegraphics[scale=0.135, trim=0cm 0cm 0cm 0cm, clip]{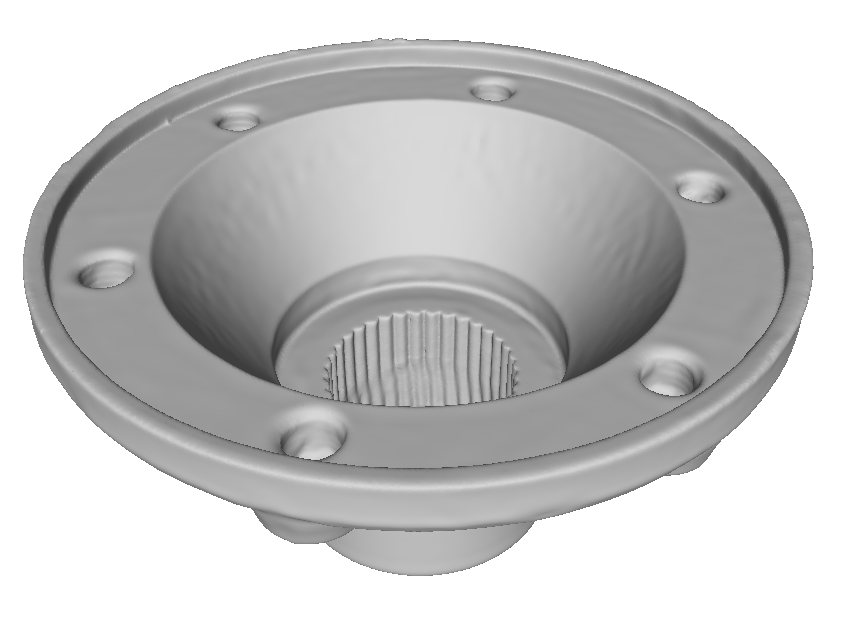}
    &
    \includegraphics[scale=0.50, trim=2cm 4cm 2cm 2.2cm, clip]{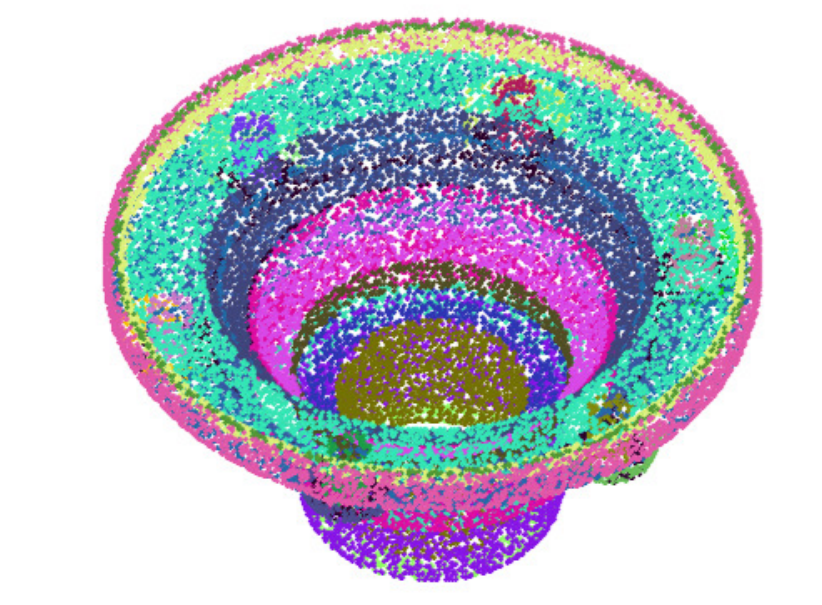}
    &
    \includegraphics[scale=0.50, trim=2cm 4cm 2cm 2.2cm, clip]{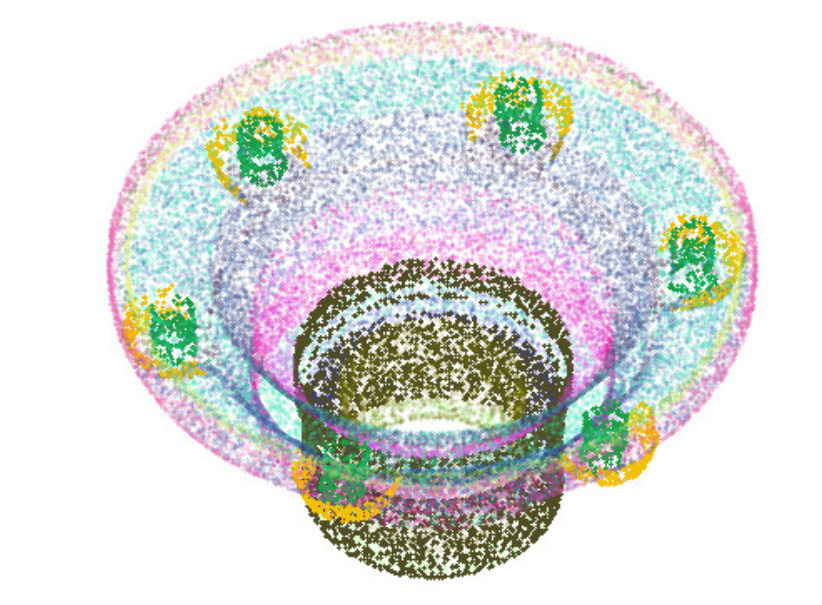}
    \\
    \begin{tikzpicture}[spy using outlines={circle,black,magnification=3.85,size=1.25cm, connect spies}]
    \node {\pgfimage[height=3.3cm]{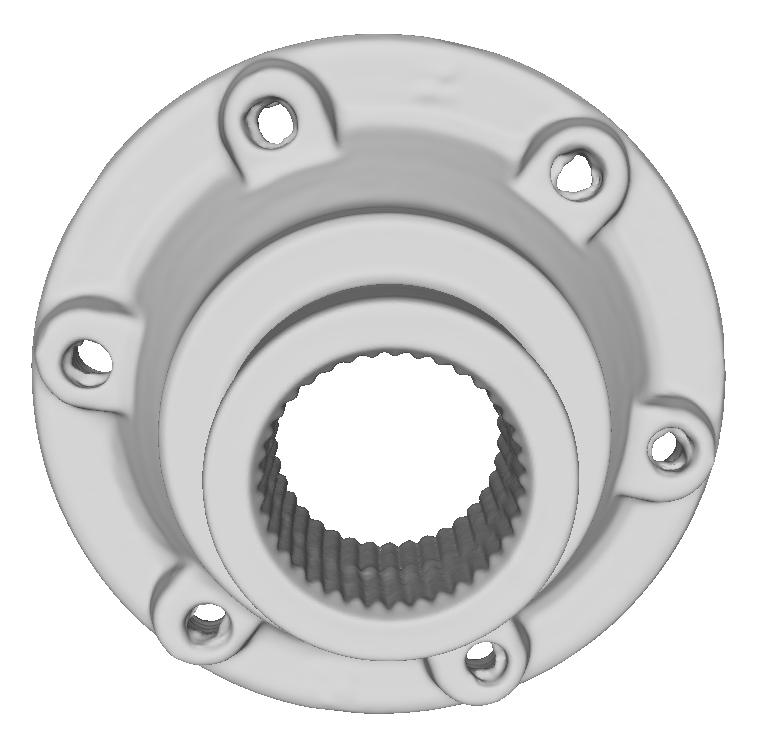}};
    \spy on (0.890,0.890) in node [left] at (2.6,-1.0);
    \end{tikzpicture}
    &
    \includegraphics[scale=0.50, trim=2cm 3.5cm 2cm 2cm, clip]{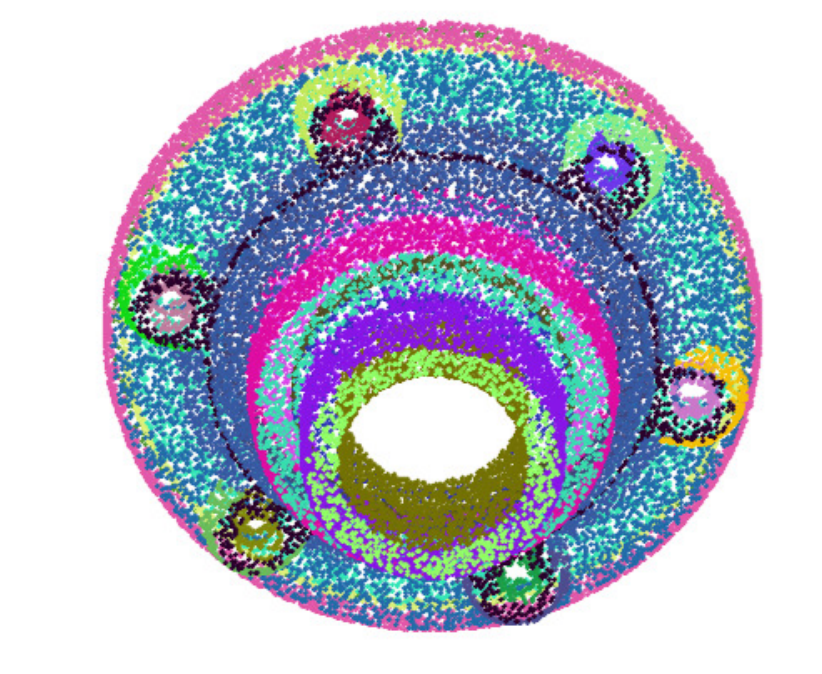}
    &
    \includegraphics[scale=0.50, trim=2cm 3.5cm 2cm 2cm, clip]{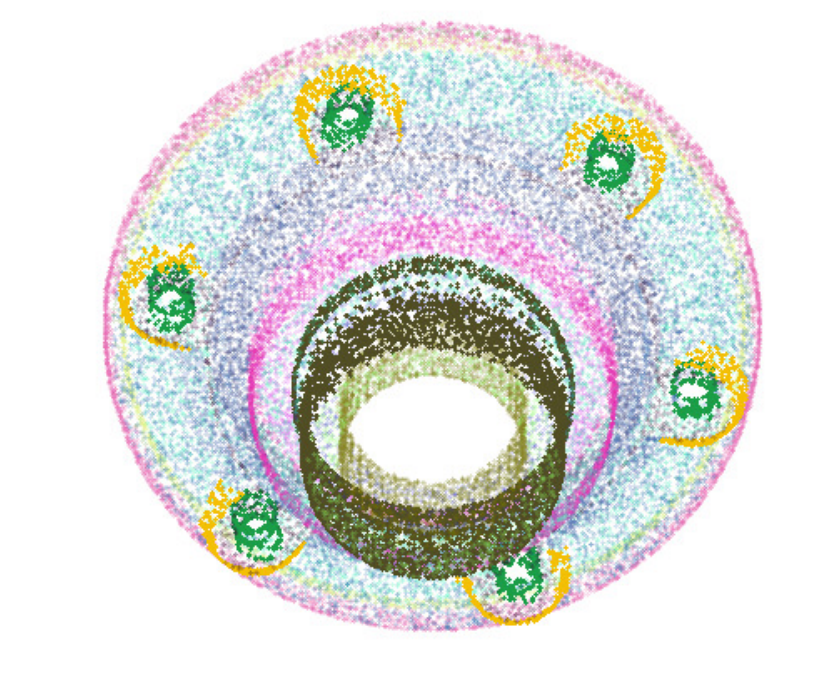}
    \\
    \hline
    \end{tabular}
    \caption{A carter. In (a), the original model is shown. The surface segments found by the HT approach are depicted in (b), while (c) shows the result of primitive  association, when one is interested in identifying the same primitive  up to a translational transformation. Different rows correspond to different points of view.}
    \label{fig:carter}
\end{figure*}

\begin{figure*}[ht!]
    \centering
    \begin{tabular}{|c|c|c|}
    \hline
    \cellcolor{ForestGreen!15}(a) Model & \cellcolor{ForestGreen!15}(b) Segments & \cellcolor{ForestGreen!15}(c) Clustering \\
    \hline
    \includegraphics[scale=0.08, trim=20cm 12cm 20cm 12cm, clip]{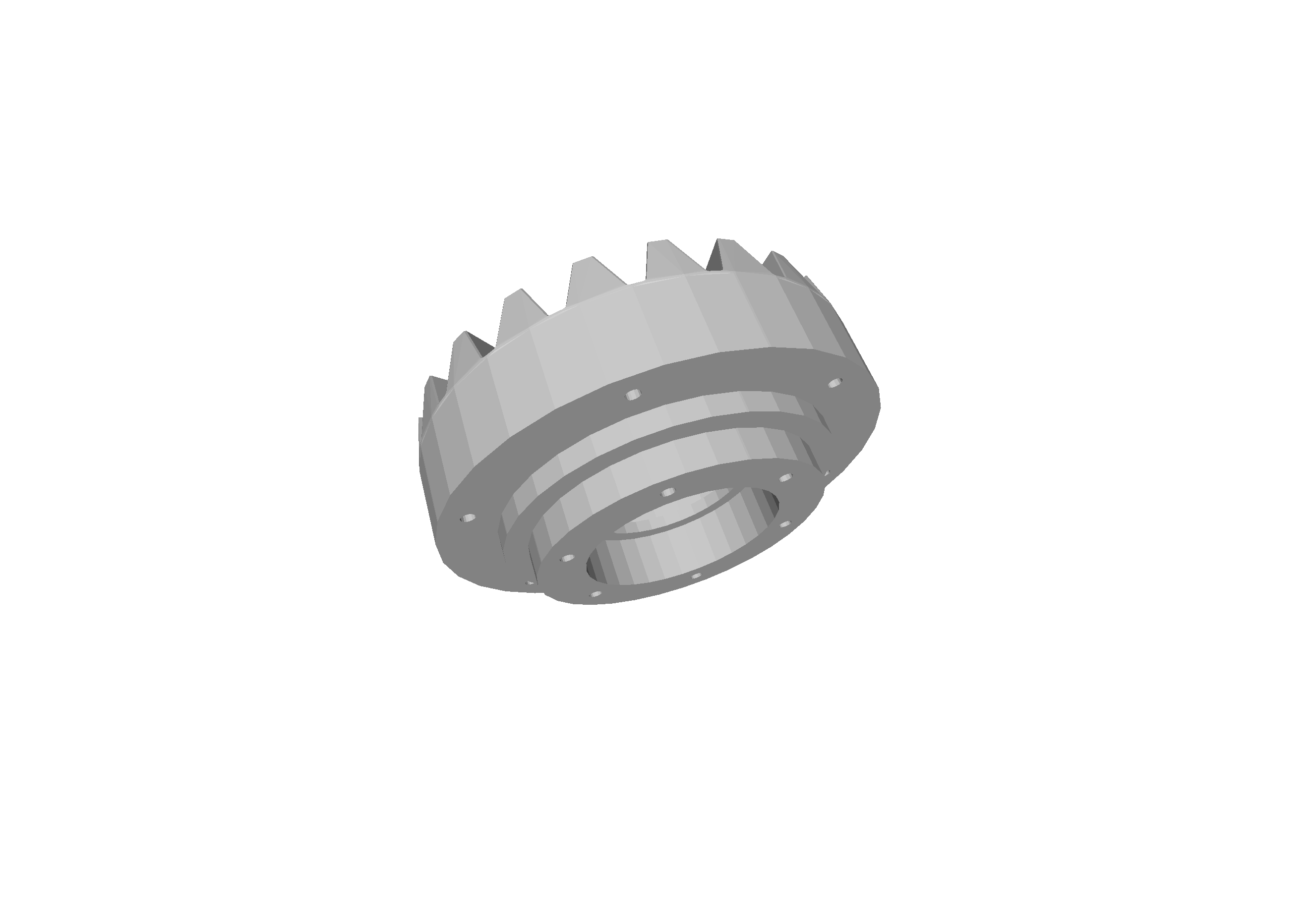}
    &
    \includegraphics[scale=0.50, trim=0cm 0.5cm 0cm 0.5cm, clip]{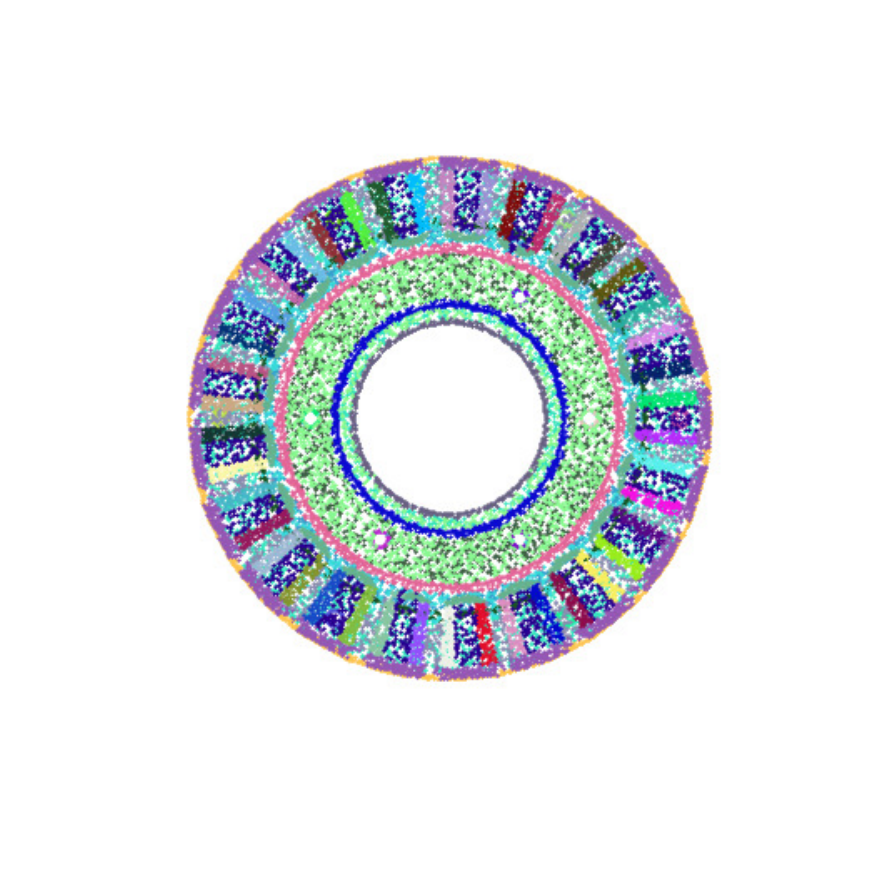}
    &
    \includegraphics[scale=0.50, trim=0cm 0.5cm 0cm 0.5cm, clip]{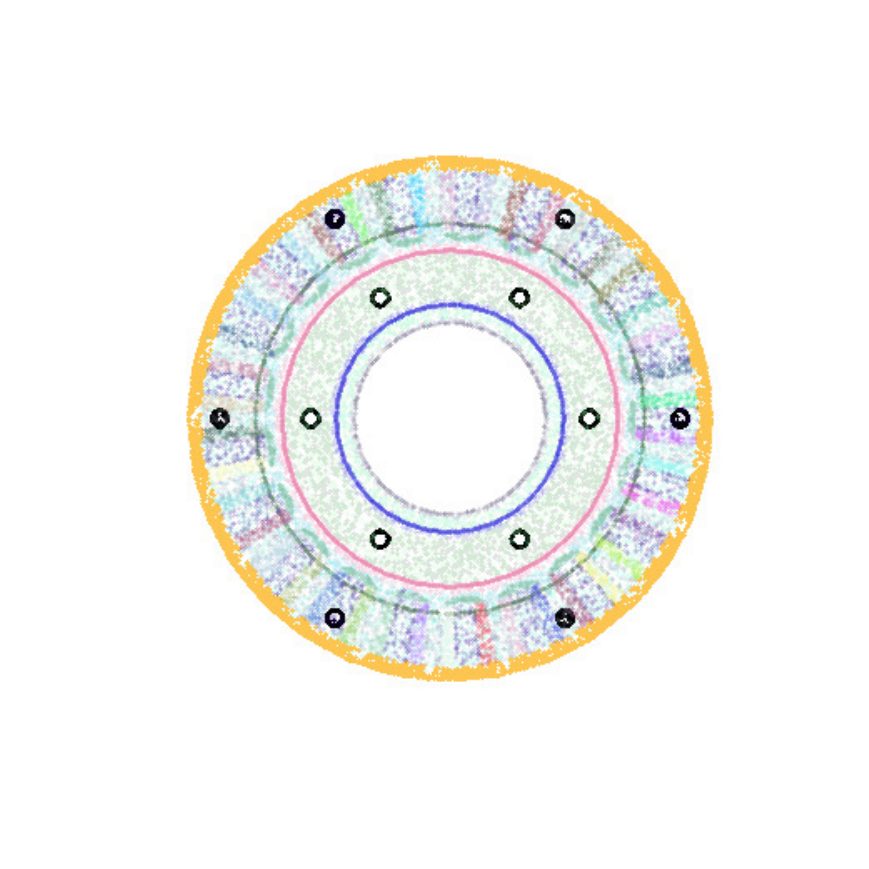}
    \\
    \includegraphics[scale=0.08, trim=20cm 7cm 20cm 12cm, clip]{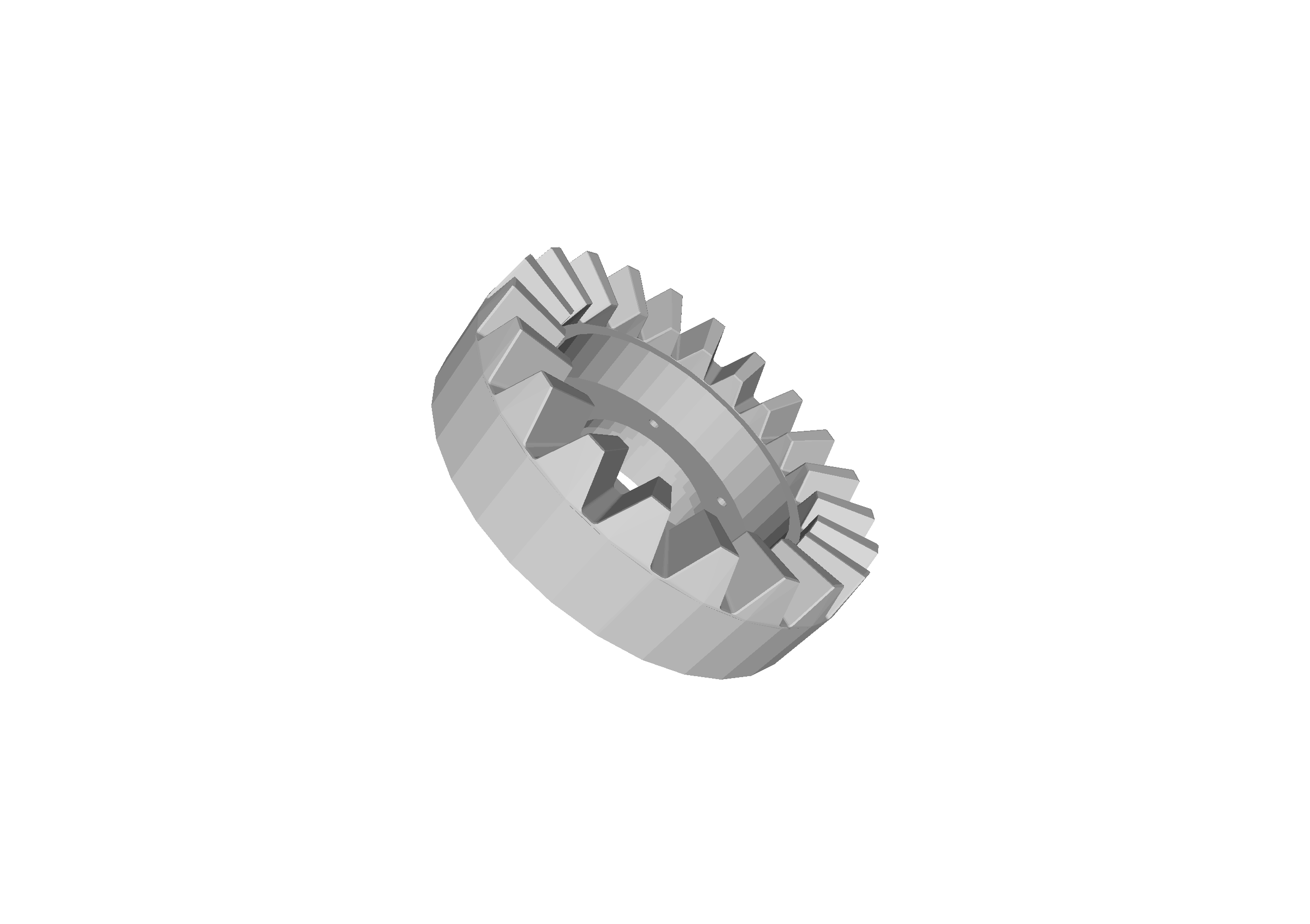}
    &
    \includegraphics[scale=0.4875, trim=0cm 0.5cm 0cm 2cm, clip]{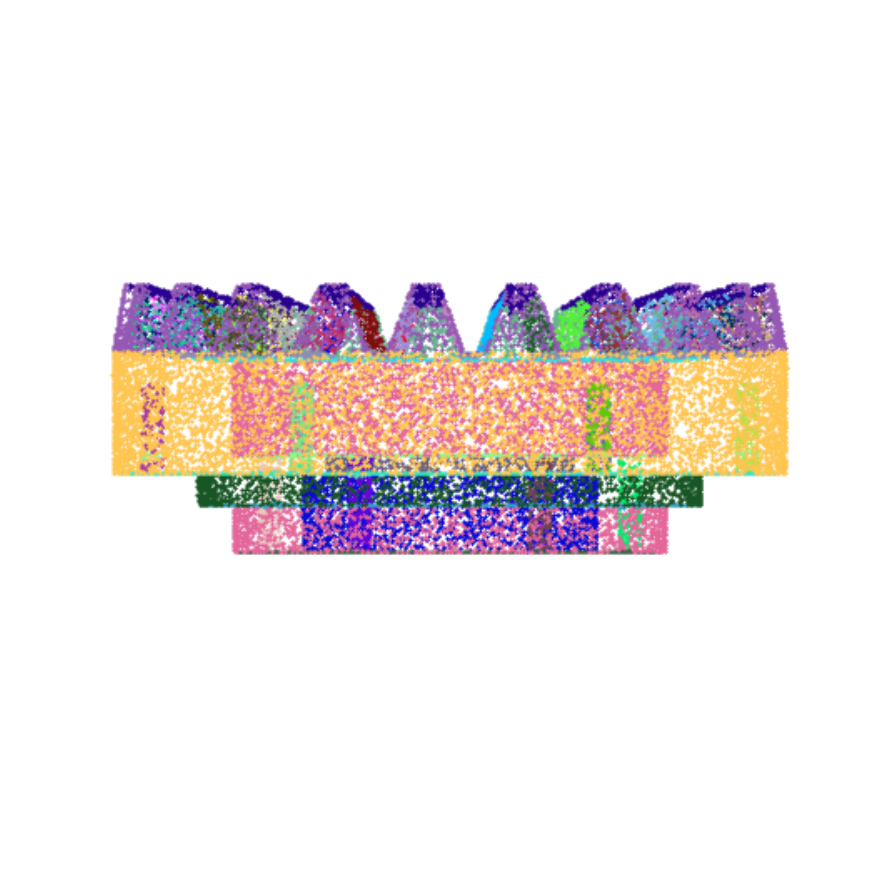}
    & 
    \includegraphics[scale=0.4875, trim=0cm 0.5cm 0cm 2cm, clip]{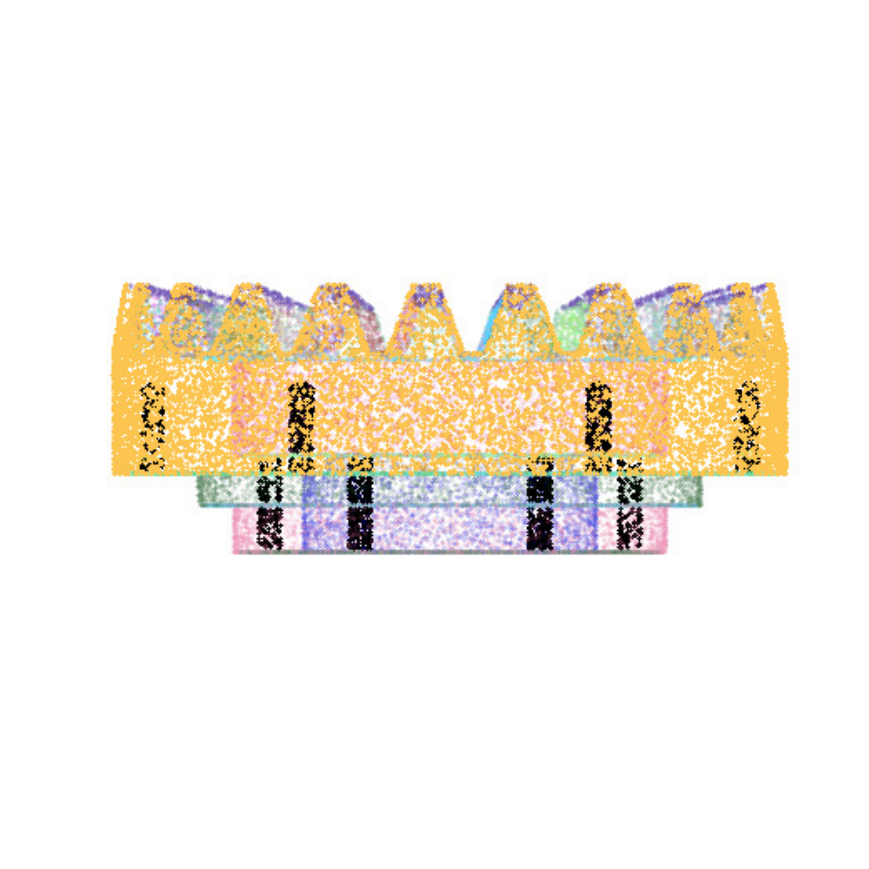}
    \\
    \hline
    \end{tabular}
    \caption{A prototype of the NuGear component, courtesy of STAM S.r.l. (Genoa, Italy). The original model is shown in (a). The decomposition in clusters of points produced by the HT approach is given in (b). The output of the additional clustering association procedure is shown in (c), which highlights the similarity between $12$ cylindrical holes (in black) and between two cylinders (in yellow).}
    \label{fig:NuGear}
\end{figure*}

\begin{figure*}[h!]
    \centering
    \begin{tabular}{|c|c|c|}
    \hline
    \cellcolor{ForestGreen!15}(a) Model & \cellcolor{ForestGreen!15}(b) Segments & \cellcolor{ForestGreen!15}(c) Clustering \\
    \hline
    & & \\
    \includegraphics[scale=0.095]{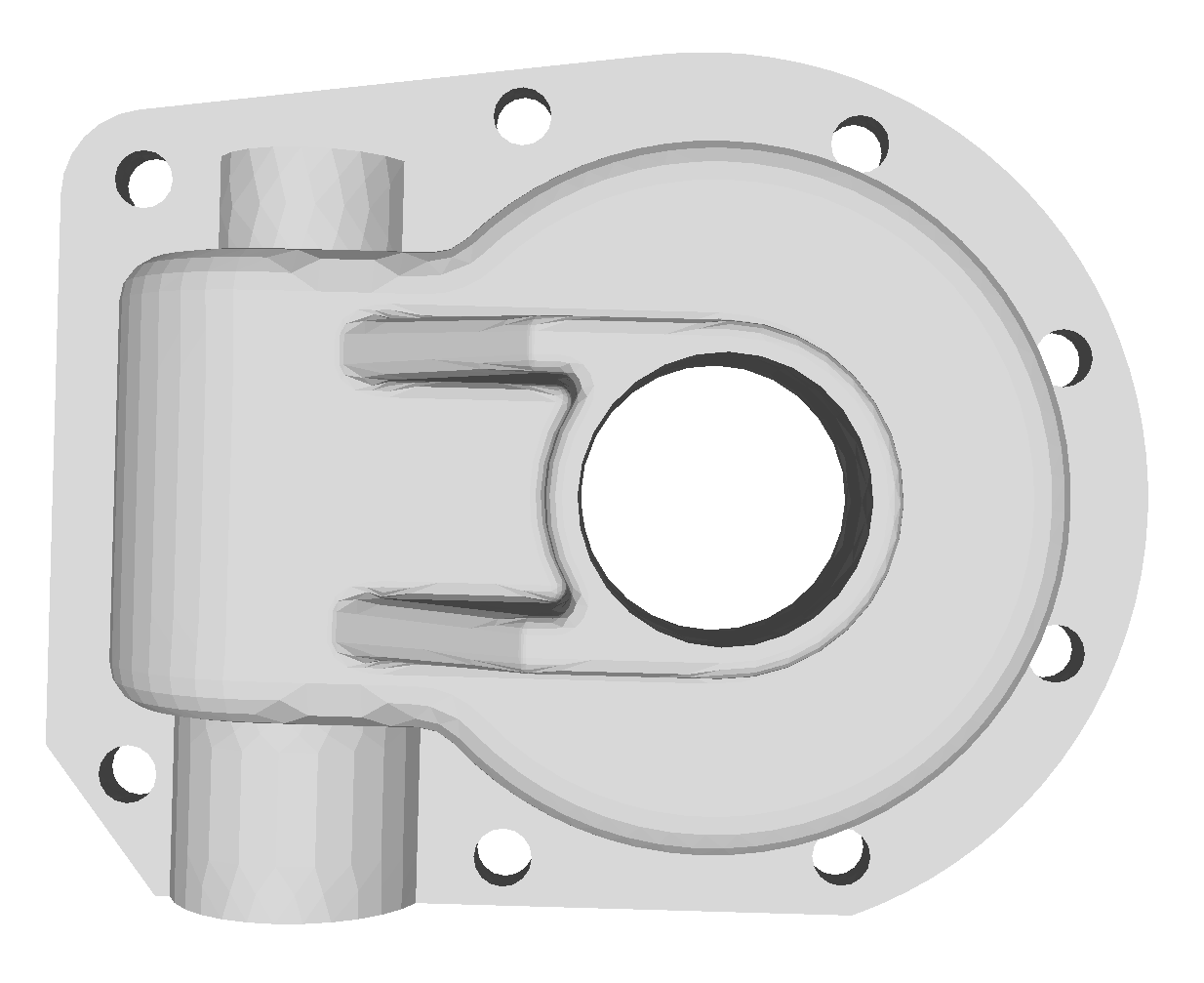}
    &
    \includegraphics[scale=0.575, trim=3cm 3.5cm 2cm 3cm, clip]{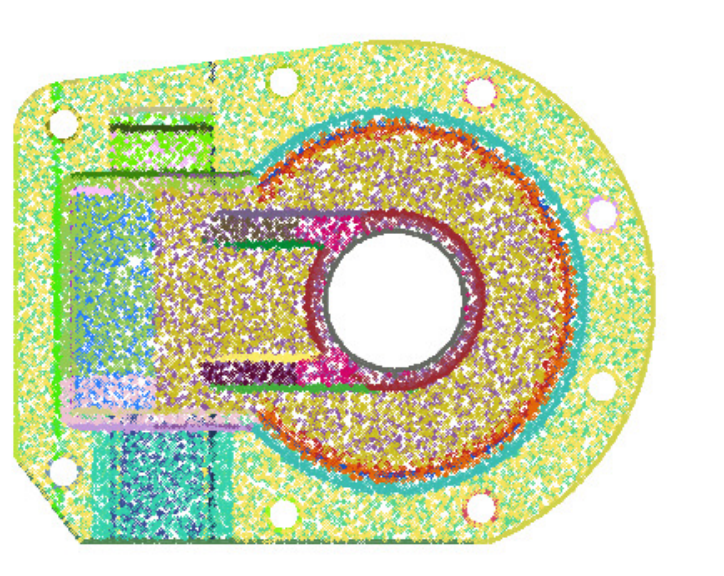}
    &
    \includegraphics[scale=0.575, trim=3cm 3.5cm 2cm 3cm, clip]{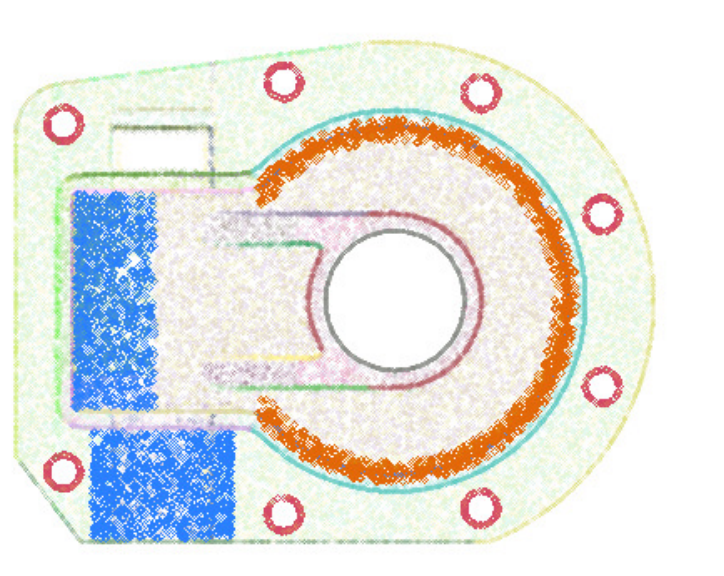}
    \\
    & & \\
    \hline
    \end{tabular}
    \caption{A mechanical part. In (a) the original model is shown, while in (b) the decomposition of the corresponding point cloud into segments produced by the HT. In (c) the result of the clustering procedure: $8$ cylindrical holes, in red, have a high similarity, up to translations; the same applies for $2$ cylindrical segments, in blue; $2$ tori, in orange, identify the same primitive, up to a rototranslation.}
    \label{fig:ModMarco}
\end{figure*}

Figures (\ref{fig:linkage}-\ref{fig:ModMarco}) show point clouds wherein the segments produced by the HT approach are post-processed by hierarchical clustering.
In Figure \ref{fig:linkage}(b), the input point cloud is first segmented into $20$ surface primitives via the HT-based recognition algorithm (cylinders and planes), which are then associated thanks to the comparison of the geometric descriptors that characterize them uniquely. As expected, no pair of primitives lying on the same surface is found. Despite the presence of imperfections in the original model (see Figure \ref{fig:linkage}(a)), we can correctly identify repeating primitives, here in the form of circular cylinders of equal radii. Figure \ref{fig:linkage}(c) highlights the similarities identified by associating 
cylinders: $4$ in light blue, $3$ in red, $2$ in purple and $2$ in yellow. 

Figure \ref{fig:carter} increases the difficulty by considering a point cloud containing a higher number of segments, some of which are low-quality as the small holes highlighted in Figure \ref{fig:carter}(a). In this example, the geometric primitives identified by the HT algorithm are cylinders, cones, and planes. The result is a segmentation in $28$ surface segments, see Figure \ref{fig:carter}(b). Note that the central hole presents some grooves, i.e., a surface detail that was not present as a geometric feature in the original CAD model; therefore, we recognise it as a cylinder and the HT is able to ignore the shallow grooves. A final application of clustering makes it possible to identify repeating primitives, up to translations. Figure \ref{fig:carter}(c) illustrates the similarity between $6$ cylindrical holes (in green) and between other $6$ cylindrical segments (in yellow).

Another example of gear is shown in Figure \ref{fig:NuGear}(a). Here, the total number of extracted geometric primitives is $68$: $19$ cylinders; $2$ cones; $47$ planes, $5$ of which are axis-aligned. The result is presented in Figure \ref{fig:NuGear}(b).
As highlighted by the original model, in this prototypical version of the NuGear component, cylinders are roughly approximated by a series of planar primitives; in this specific case, we fit a cylinder instead of many planes, since the former can describe a much larger area without significantly increasing the error.
 Clustering identifies here a similarity between the $12$ cylindrical holes -- up to translations -- and between $2$ external cylinders, while the surface segments identified by non-axis-aligned planes are not grouped in pairs; this is because the tooth inclination prevents any alignment. Figure \ref{fig:NuGear}(c) shows this result, colouring the holes in black and the external cylinders in yellow.

Finally, Figure \ref{fig:ModMarco}(a) shows another mechanical part. The HT-based decomposition of the input point cloud consists of $50$ surface segments, all of which are tori, cylinders, and planes, see Figure \ref{fig:ModMarco} (b). The clustering method can identify  the similarity between $8$ red cylindrical holes and between $2$ blue cylindrical segments, up to translations -- see Figure \ref{fig:ModMarco}(c). Finally, $2$ tori are aggregated, because they lie on the same surface primitive up to rototranslations.

Table \ref{tab:simplePrim_mfe_times} summarises the characteristics of each point cloud processed in this section and the mean fitting error for all the simple primitives recognised on it.

\begin{table}[h!]
    \centering
    \caption{Statistics of the MFES for all models of Section \ref{sec:simplePrim}. Being the MFE normalized by definition, we can conclude that the maximum error for the fitting of the simple primitives is $4.48\%$, which corresponds to the noisy holes in the carter of Figure \ref{fig:carter}.  }
    \begin{tabular}{|c|c|c|c|c|c|}
    \hline
         \cellcolor{ForestGreen!15} Model & \cellcolor{ForestGreen!15} \# points & \cellcolor{ForestGreen!15} \# segs & \cellcolor{ForestGreen!15} min($E_i$) & \cellcolor{ForestGreen!15} mean($E_i$) & \cellcolor{ForestGreen!15} max($E_i$)  \\ 
         \hline
          Fig. \ref{fig:modelliABC1}& $15,216$ &  $8$ & $0.0006$ & $0.0021$ & $0.0046$  \\
         \hline
          Fig. \ref{fig:modelliABC2}& $15,022$ & $9$ & $0.0008$ & $0.0029$ & $0.0051$ \\
         \hline
          Fig. \ref{fig:various_SGPs}(a) & $25,000$ & $11$ & $0.0009$ & $0.0024$ & $0.0056$  \\
         \hline
          Fig. \ref{fig:various_SGPs}(b) & $25,000$ & $5$ & $0.0004$ & $0.0031$ & $0.0059$  \\
         \hline
         Fig. \ref{fig:linkage}(b) & $50,000$ & $20$ & $0.0004$ & $0.0098$ & $0.0300$  \\
         \hline
          Fig. \ref{fig:carter}(b) & $50,000$ & $28$ & $0.0013$ & $0.0107$ & $0.0448$ \\
         \hline
          Fig. \ref{fig:NuGear}(b) & $50,000$ & $68$ & $0.0006$ & $0.0053$ & $0.0178$  \\
         \hline
          Fig. \ref{fig:ModMarco}(b) & $50,000$ & $50$ & $ 0.0008$ & $0.0035$ & $0.0057$ \\
         \hline
    \end{tabular}
    \label{tab:simplePrim_mfe_times}
\end{table}

\subsubsection{Complex geometric primitives}\label{sec:comlexPrim}

As anticipated, our method can recognize other primitives in addition to the simple ones shown in Section \ref{sec:simplePrim}. Here we show some of the complex primitives identified in our experiments.

\begin{figure*}[h!]
    \centering
    \begin{tabular}{|c|c|cc|}
    \hline
     \cellcolor{ForestGreen!15}Segments (a) & \cellcolor{ForestGreen!15}Segments (b) & \multicolumn{2}{|c|}{\cellcolor{ForestGreen!15}Segments (c)} \\
    \hline
    & & & 
    \\
    \includegraphics[scale=0.35, trim=1cm 0cm 1cm 1cm, clip]{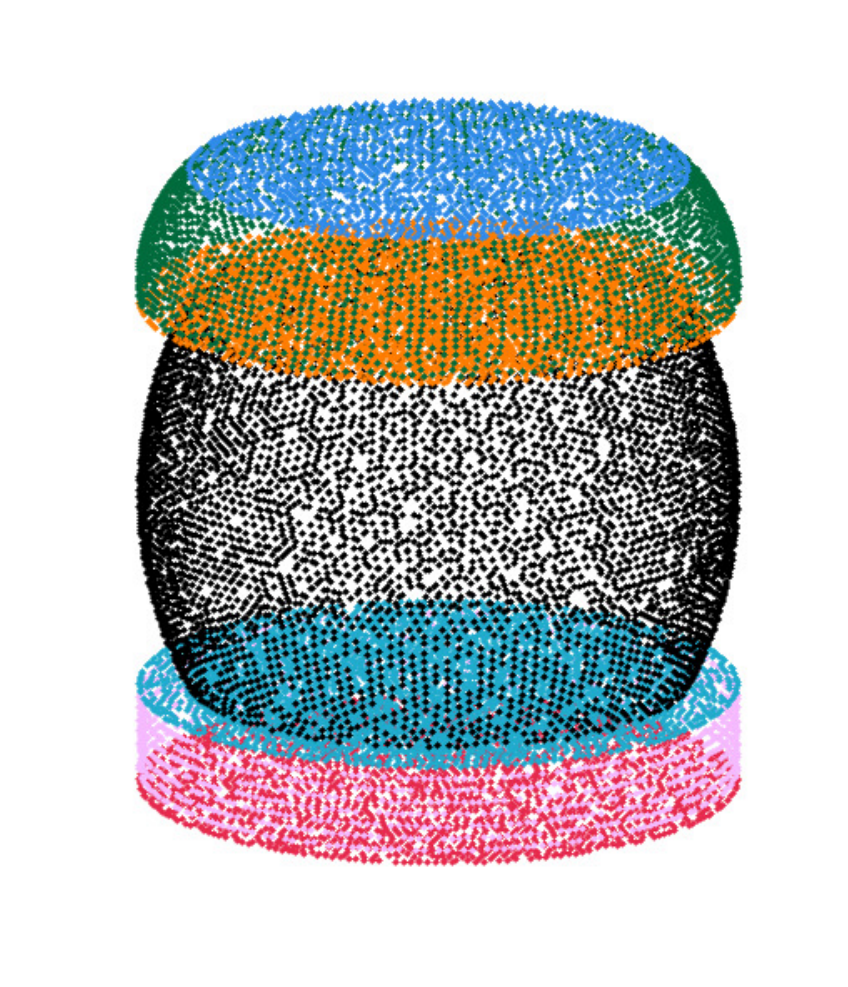}
    &
    \includegraphics[scale=0.27, trim=2cm 1cm 2cm 2cm, clip]{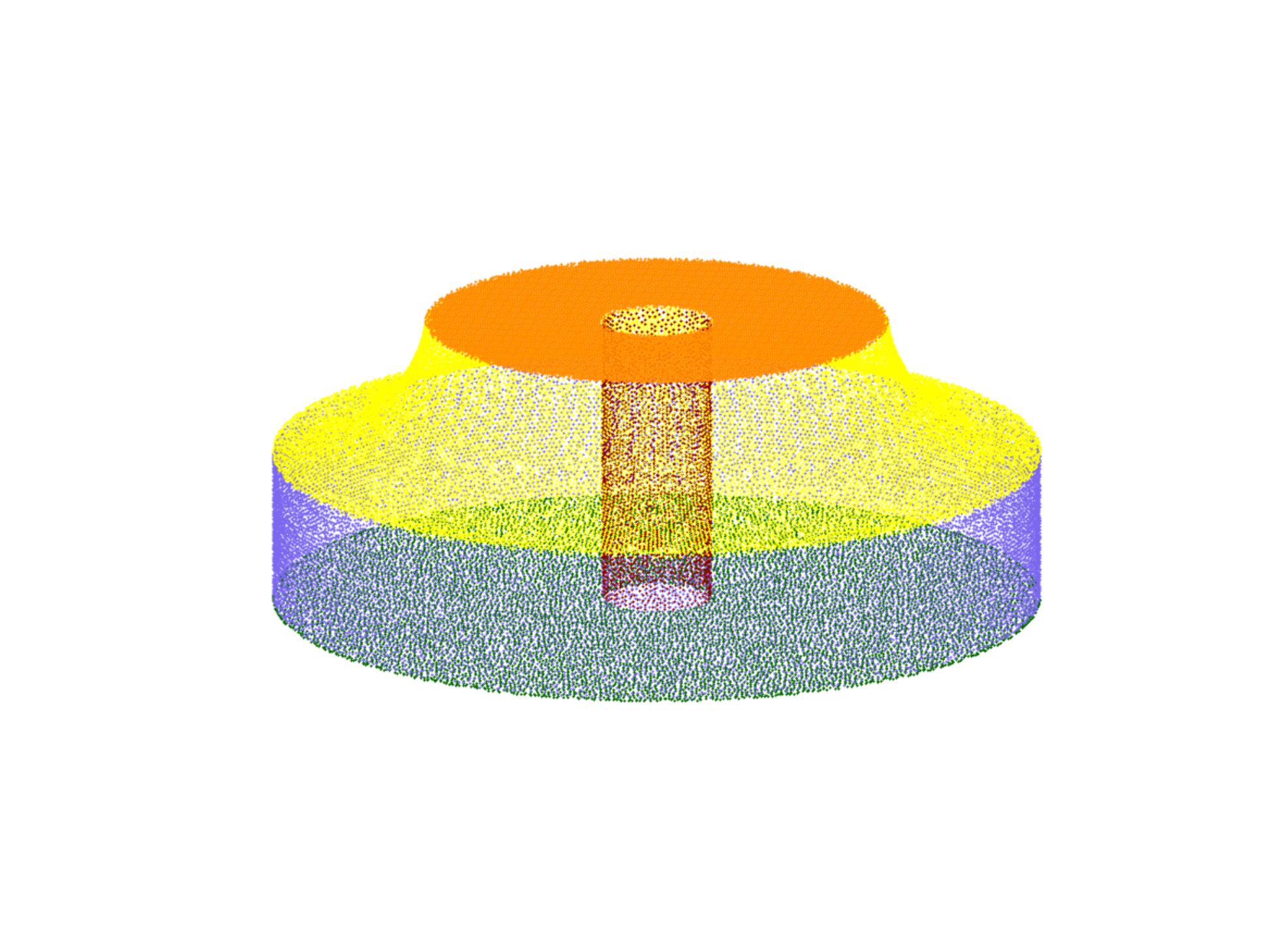}
    &
    \includegraphics[scale=0.5, trim=1cm 1cm 1cm 0.5cm, clip]{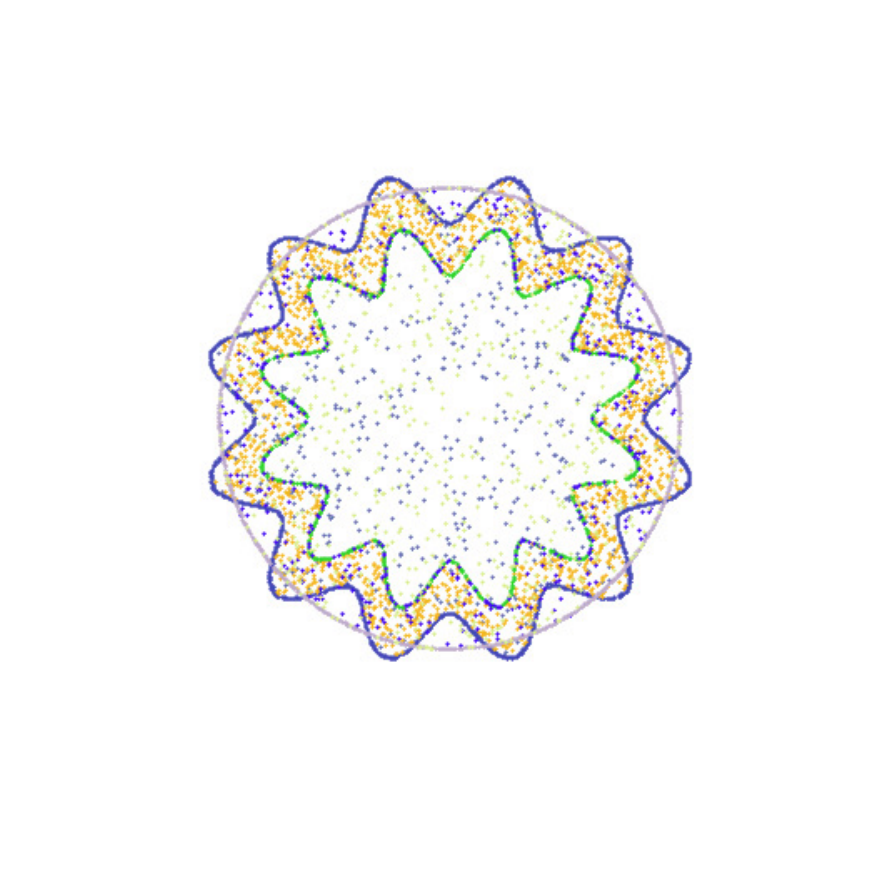}
    &
    \includegraphics[scale=0.50, trim=1cm 1cm 1cm 0.5cm, clip]{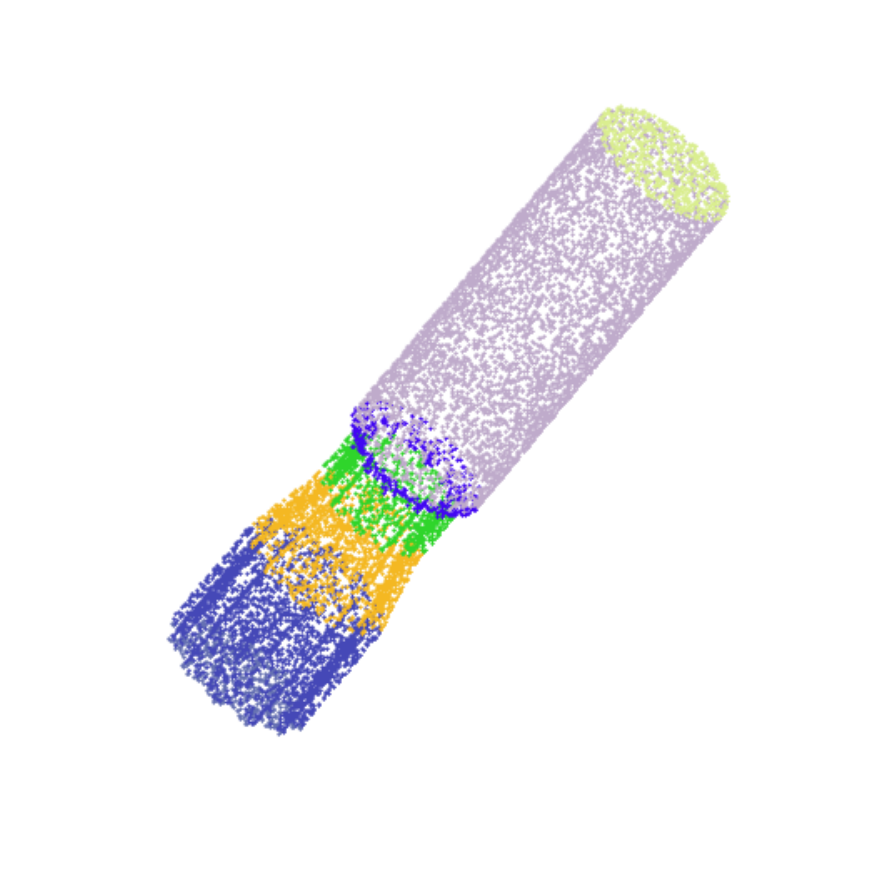}
    \\[10pt]
    \hline
    \end{tabular}
    \caption{Recognition of complex geometric primitives in 
     CAD point clouds. Their identification does not require, in these cases, the application of any clustering algorithm.}
    \label{fig:various_CGPs}
\end{figure*}

The first example -- shown in Figure \ref{fig:various_CGPs}(a) -- contains an ellipsoid, which is easily recognized by our method at the price of an additional parameter in the parameter space; additionally, the point cloud can be partitioned into $4$ planes, $1$ torus and $1$ cylinder. In the point cloud from Figure \ref{fig:various_CGPs}(b), we are able to identify correctly the yellow segment as a surface of revolution from Table \ref{tab:primitiveComplesse}(c); the remaining points are segmented into $2$ cylinders and $2$ planes. In the point cloud shown in Figure \ref{fig:various_CGPs}(c), we recognize the gold part as a generalized cone, while the blue and the green segments are fitted with generalized cylinders: all of the three have the same directrix, a $5-$convexity curve. This last point cloud has been segmented into $7$ clusters. 

\begin{figure}[h!]
    \centering
    \begin{tabular}{|c|c|c|}
    \hline
    \cellcolor{ForestGreen!15}(a) Model & \cellcolor{ForestGreen!15}(b) Segments & \cellcolor{ForestGreen!15}(c) Clustering \\
    \hline
    & & \\
    \includegraphics[scale=0.1]{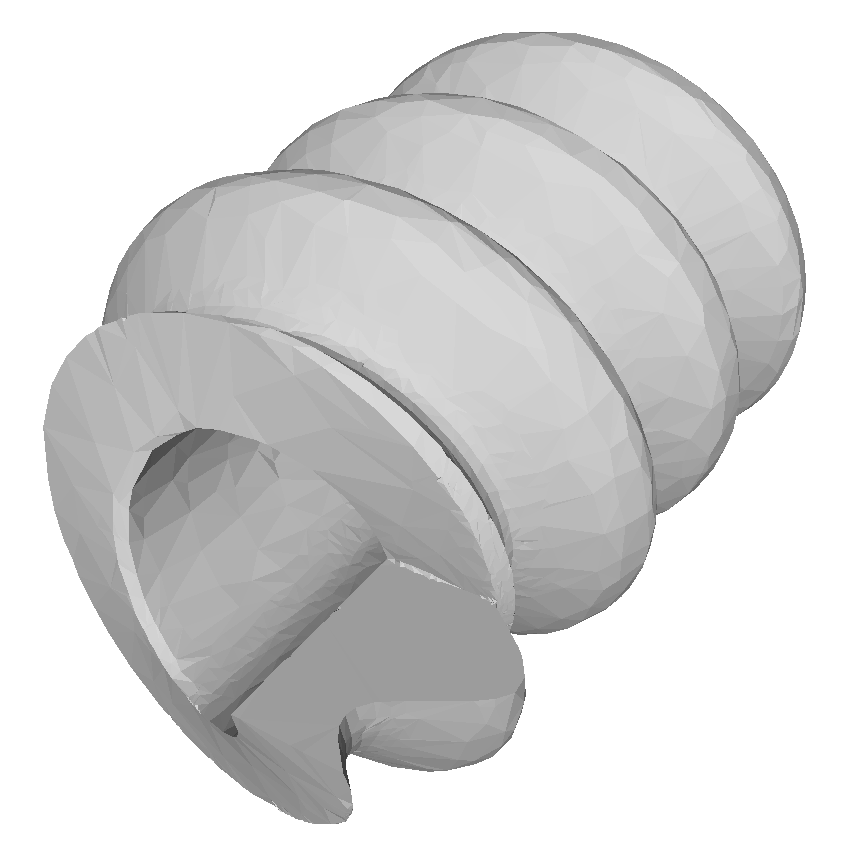}
    &
    \includegraphics[scale=0.25, trim=2cm 0cm 2cm 0cm, clip]{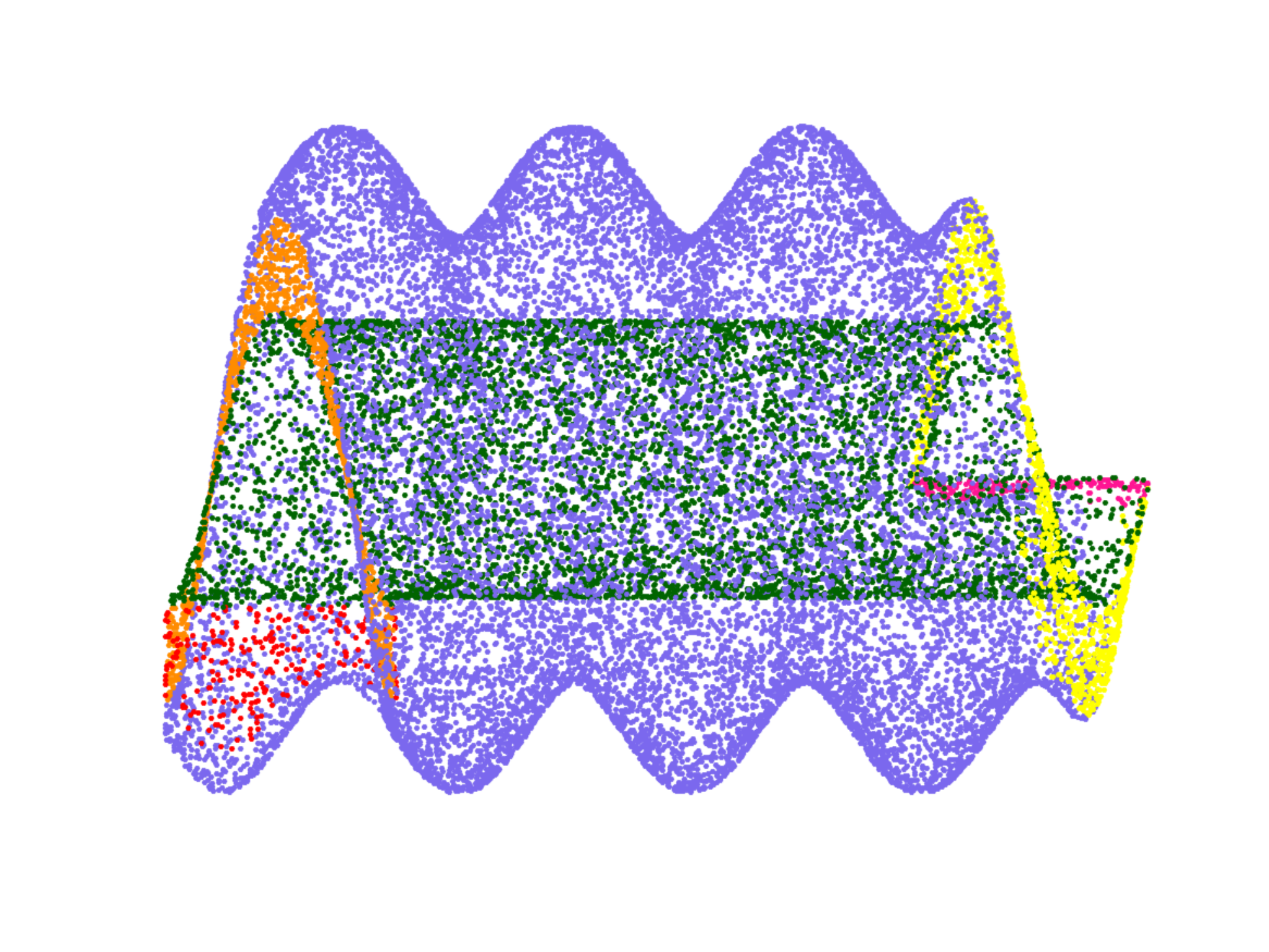}
    &
    \includegraphics[scale=0.25, trim=2cm 0cm 2cm 0cm, clip]{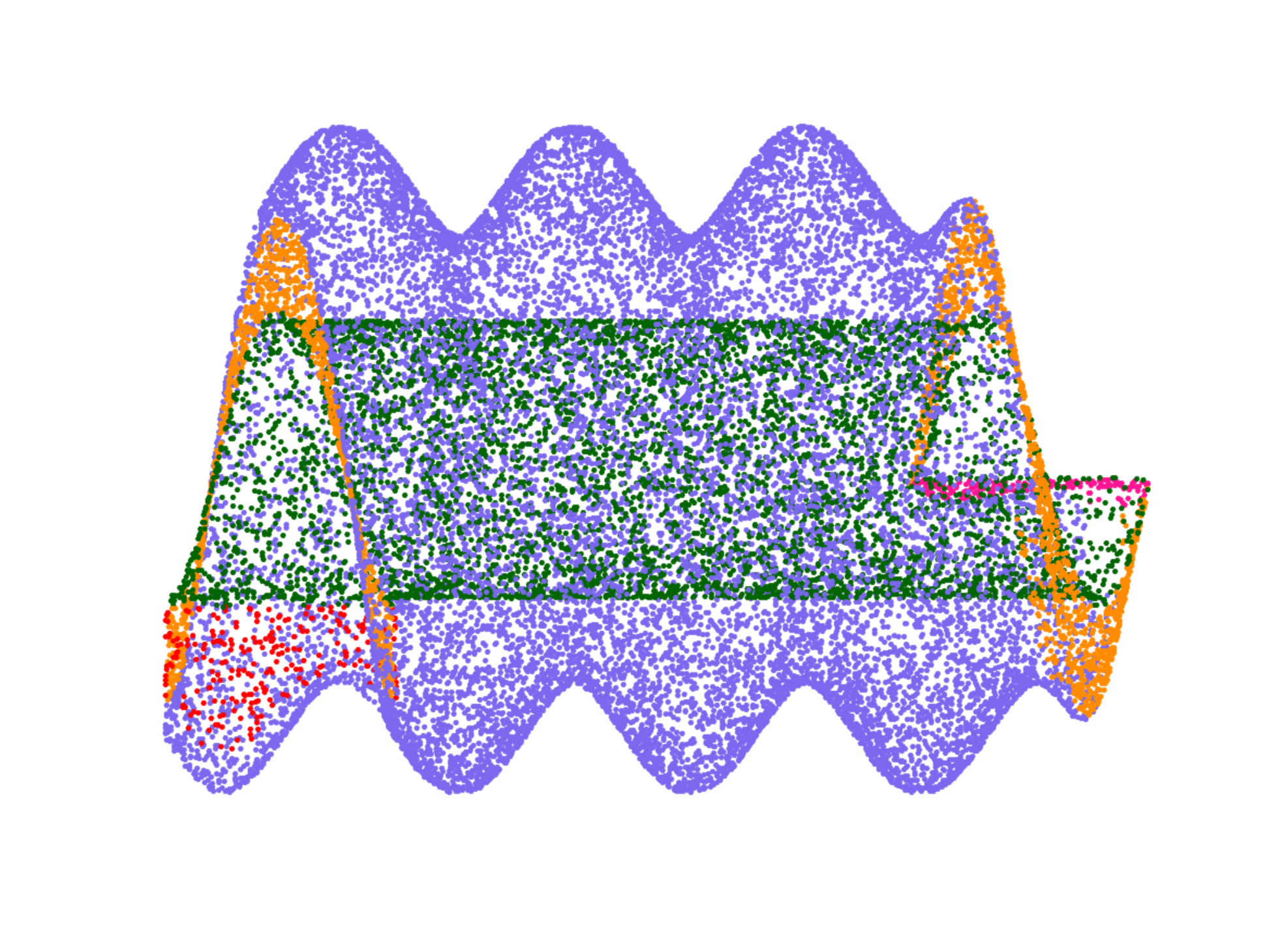}
    \\
    \hline
    \end{tabular}
    \caption{A screw-like part. The original model, (a), is sampled. The surface primitives detected via HT are shown in (b) using different colours: a helical surface (in purple), two planes (in red and magenta), and two helical strips (in orange and yellow). Although no pair of them lies on the same parametrised surface, the 2 helical strips have the same equation up to a translation, as shown in (c).}
    \label{fig:viteggiante}
\end{figure}

Figure \ref{fig:viteggiante}(a) displays a mechanical part which can be accurately described by combining a portion of a helical surface, as introduced in Table \ref{tab:primitiveComplesse}(d), with a pair of planes and a pair of convex combinations of helices, see Table \ref{tab:primitiveComplesse}(e). The result is a segmentation of the point cloud into $6$ 
primitives, Figure \ref{fig:viteggiante}(b). Two of them are then grouped by the clustering technique since the helical strips have the same equation up to a translation, as shown in Figure \ref{fig:viteggiante}(c). 

Table \ref{tab:complexPrim_mfe_times} provides the main characteristics of each point cloud processed in this section and the mean fitting error for all the simple and complex primitives recognised on it. 

\begin{table}[h!]
    \centering
        \caption{Statistics of the MFEs for all models of Section \ref{sec:comlexPrim}. Being the MFE normalized by definition, we can conclude that the maximum error for the fitting of the simple primitives is $1,54\%$, which corresponds to the helical surface of Figure \ref{fig:viteggiante}. }
    \begin{tabular}{|c|c|c|c|c|c|c|}
        \hline
         \cellcolor{ForestGreen!15} Model & \cellcolor{ForestGreen!15}\# points & \cellcolor{ForestGreen!15}\# segs & \cellcolor{ForestGreen!15}min($E_i$) & \cellcolor{ForestGreen!15}mean($E_i$) & \cellcolor{ForestGreen!15}max($E_i$) \\ 
         \hline
          Fig. \ref{fig:various_CGPs}(a) & $25,524$ & $7$ & $0.0009$ & $0.0042$ & $0.0063$  \\
         \hline
          Fig. \ref{fig:various_CGPs}(b) & $67,777$ & $5$ & $0.0006$ & $0.0031$ & $0.0071$ \\
         \hline
         Fig. \ref{fig:various_CGPs}(c) & $25,000$ & $7$ & $0.0020$ & $0.0053$ & $0.0081$  \\
         \hline
          Fig. \ref{fig:viteggiante}(b) & $25,000$ & $6$ & $0.0033$ & $0.0086$ & $0.0154$ \\
         \hline
      
    \end{tabular}
    \label{tab:complexPrim_mfe_times}
\end{table}

\subsubsection{Robustness of the pipeline}\label{sec:robustness}
The use of the HT naturally leads to a robust method for the recognition of mathematical surfaces, as suggested in the examples of Figures \ref{fig:linkage} and \ref{fig:carter} -- which were characterized by spurious parts. 

\begin{figure*}[h!]
    \centering
    \begin{tabular}{|c|c|c|}
    \hline
    \cellcolor{ForestGreen!15}(a) Model & \cellcolor{ForestGreen!15}(b) Segments & \cellcolor{ForestGreen!15}(c) Clustering \\
    \hline
    & & \\
    \includegraphics[scale=0.1]{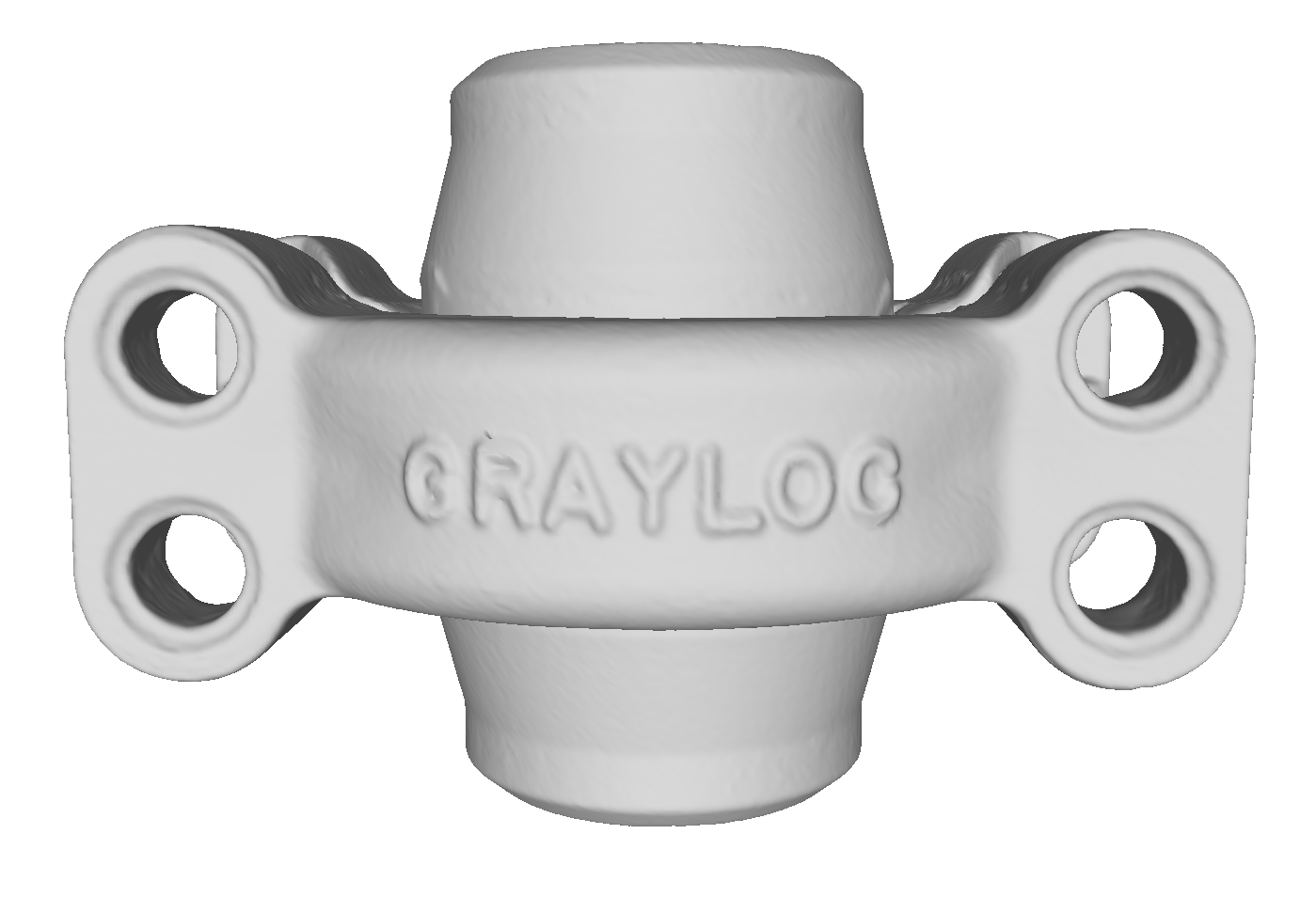}
    &
    \includegraphics[scale=0.40, trim=0cm 2cm 0cm 2cm, clip]{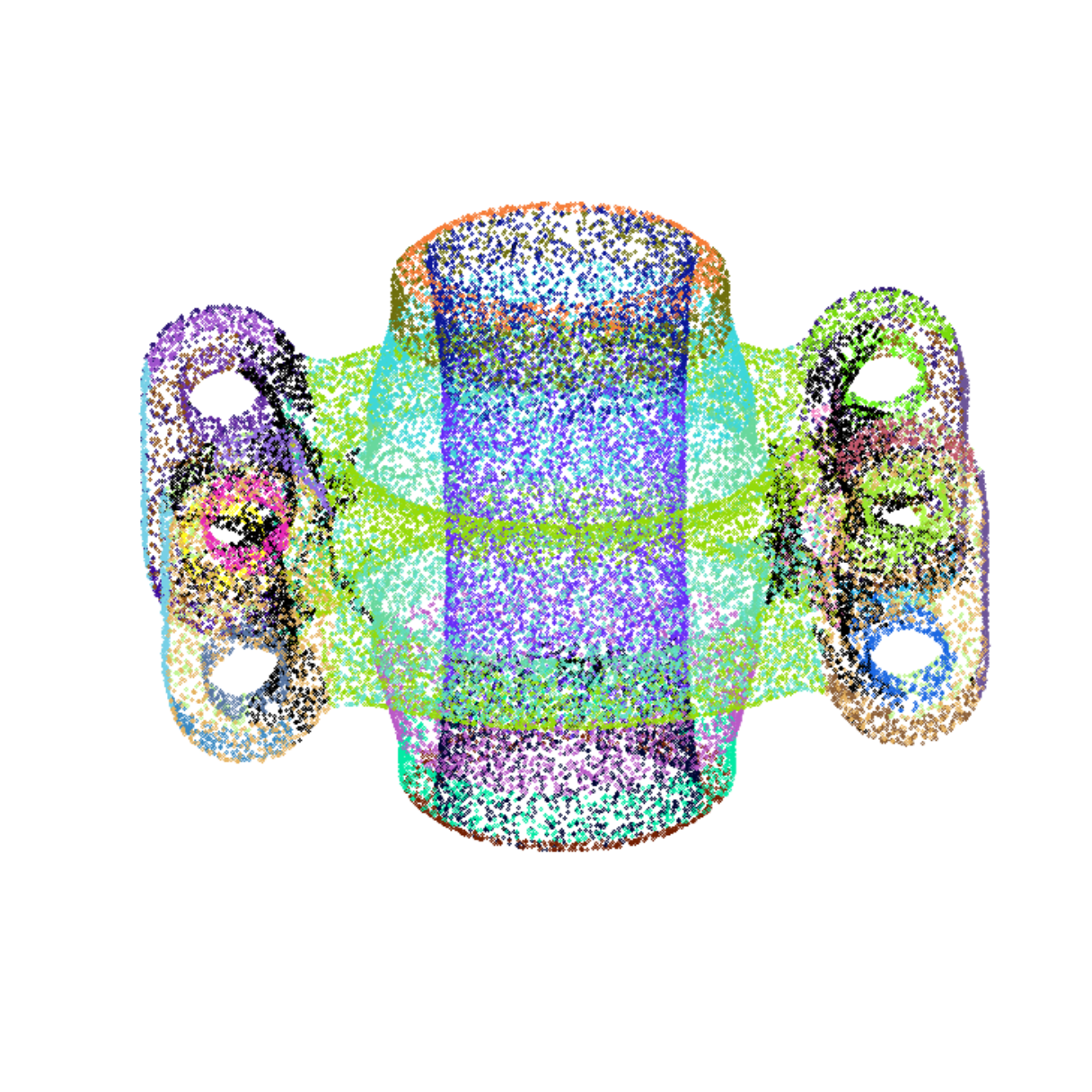}
    &
    \includegraphics[scale=0.40, trim=0cm 2cm 0cm 2cm, clip]{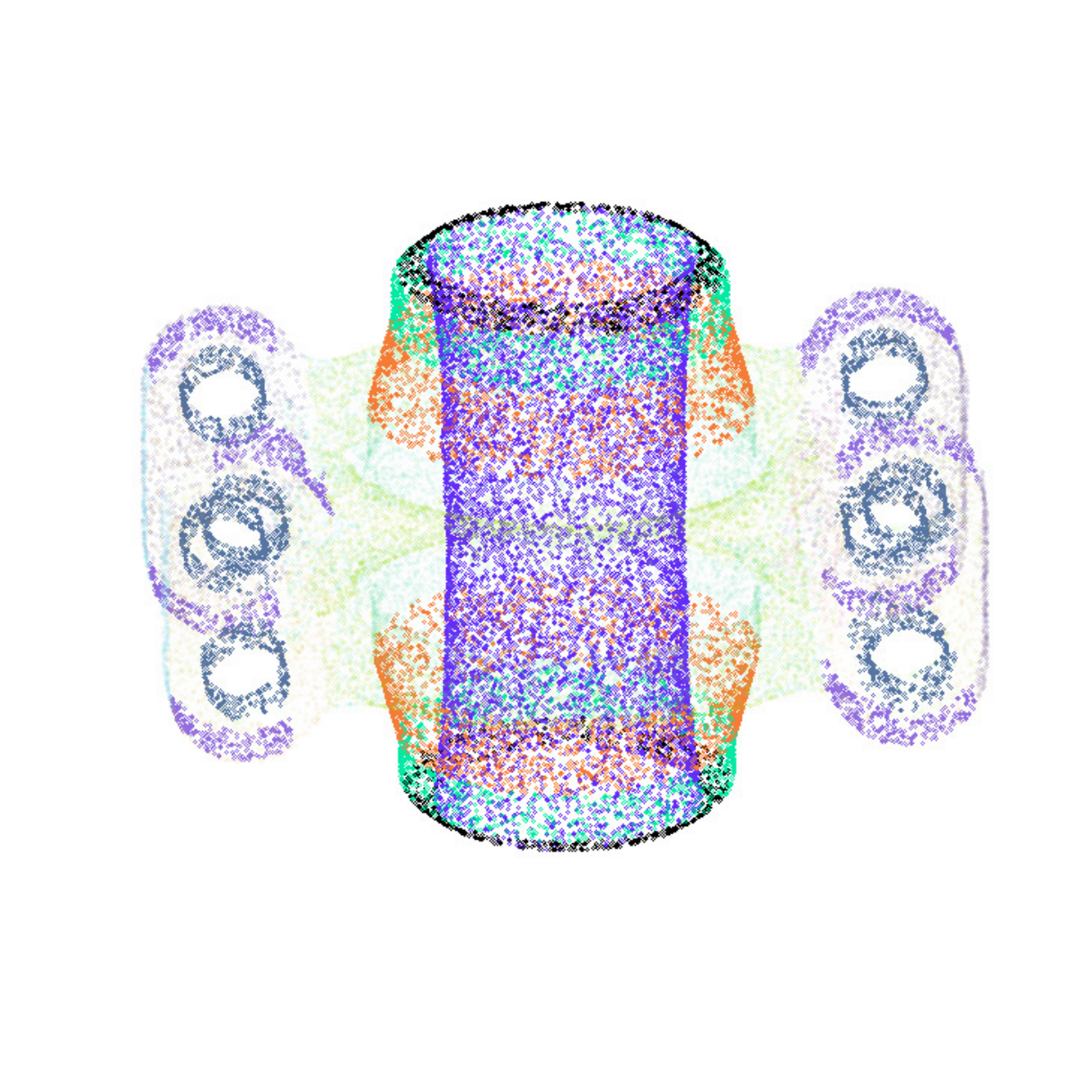}
    \\
    \hline
    \end{tabular}
    \caption {A clamp connector. In (a) the original model. In (b) the decomposition of the corresponding point cloud into $38$ segments provided by the HT procedure. In (c), the final grouping obtained by clustering, consisting of $6$ clusters of primitives (here, singletons of segments are transparent).}
    \label{fig:graylock}
\end{figure*}

In the point cloud of Figure \ref{fig:graylock}, the HT recognition correctly identifies the cylinder that fits the central part, without being negatively influenced by  the letters in relief -- see the original model in Figure \ref{fig:graylock}(a). The point cloud is decomposed into $38$ segments: $23$ cylinders with different axes, $10$ planes, $4$ cones, and $1$ torus that automatically identifies the top and bottom of the cylinder with the ``GRAYLOC" inscription (see Figure \ref{fig:graylock}(b)). The application of the hierarchical clustering technique allows us to group together: $8$ grey cylindrical holes (up to rototranslations); $8$ purple cylindrical segments; $2$ aquamarine circular cylinders; $3$ violet circular cylinders; $2$ orange circular cones (up to a reflection); $2$ black cones (up to a reflection).
 Moreover, the small imperfections of the manufacture on the central part of the body (recognised by vertical cones, cylinders and tori at the top and bottom) and on the lateral holes do not prevent the clustering technique from appropriately 
associating the corresponding segments, correctly dealing with rotations and reflections. However not everything is recognised:  the  black dots in Figure \ref{fig:graylock}(b) correspond to points that are not fitted by any of the geometric primitives at our disposal, as they originate from irregular elements that act as a connection between better-defined segments. We label such points as ``unsegmented" because of the high mean fitting error.

\begin{figure*}[h!]
    \centering
    \begin{tabular}{|c|c|c|c|}
    \hline
    \cellcolor{ForestGreen!15} $\sigma=0.01$ & \cellcolor{ForestGreen!15} $\sigma=0.05$ & \cellcolor{ForestGreen!15} $\sigma=0.10$ & \cellcolor{ForestGreen!15} $\sigma=0.20$\\
    \hline
    & & & \\
    \includegraphics[width=3cm]{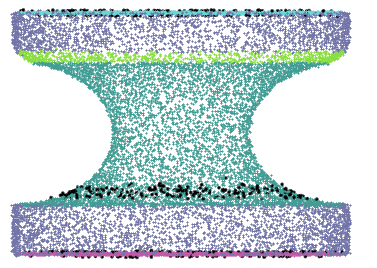}
    &
    \includegraphics[width=3cm]{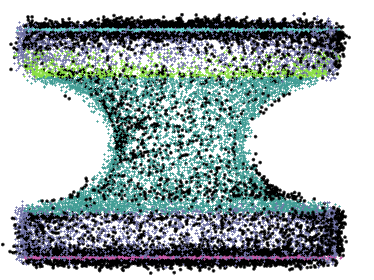}
    &
    \includegraphics[width=3cm]{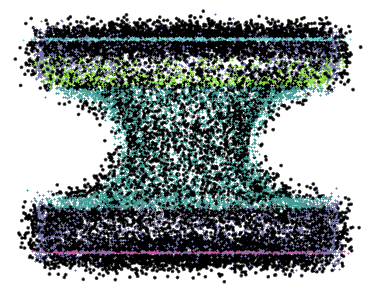}
    &
    \includegraphics[width=3cm, trim={0cm 0.3cm 0cm 0.1cm}, clip]{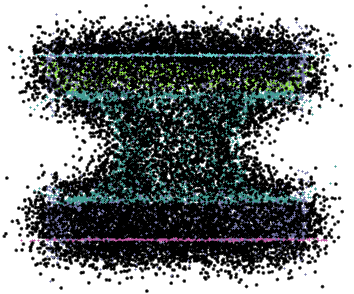}
    \\

    \includegraphics[width=2.7cm]{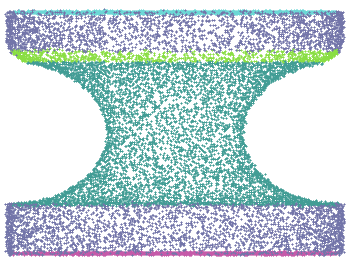}
    &
    \includegraphics[width=2.7cm]{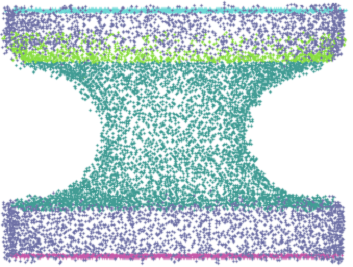}
    &
    \includegraphics[width=2.7cm]{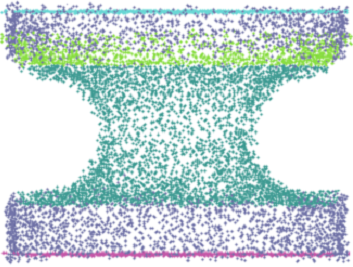}
    &
    \includegraphics[width=2.7cm]{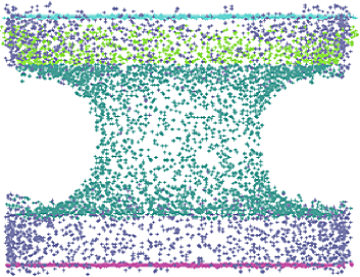}\\
    & & & \\
    \hline
\end{tabular}
\caption{The point cloud in Figure \ref{fig:various_SGPs}(b) is perturbed by adding zero-mean Gaussian noise of standard deviation: $0.01$, $0.05$, $0.10$ and $0.20$. The first row superimposes the points identified as noise (in black) to the final segments found by our method; the second row depicts the points that fit the primitives found and provides a denoised segmentation. \label{fig:noise_robustness}}
\end{figure*}

Further proof of the robustness of our method applied to raw data is presented in Figure \ref{fig:noise_robustness} and quantitatively analysed in Table \ref{tab:noise_robustness}.  In this example, we perturb the point cloud of Figure \ref{fig:various_SGPs}(b) by adding zero-mean Gaussian noise of standard deviation: $0.01$, $0.05$, $0.10$ and $0.20$. The first row shows the points classified as noise (in black) and the segments found by our method in the same image, for each level of noise. The second row focuses only on the  points that fit the identified primitives, thus providing a denoised segmentation.
The robustness to noise is quantitatively studied in  Table \ref{tab:noise_robustness}: the parameters obtained in the original point cloud are there compared with those found in the perturbed point clouds.

\begin{table}[h]
    \centering
        \caption{ Parameter comparison between the original point cloud from Figure  \ref{fig:various_SGPs}(b) and the perturbed versions from Figure \ref{fig:noise_robustness}.}
\begin{tabular}{c}
\\
\begin{tabular}{|c|c|c|c|c|c|}
        \hline
            \cellcolor{ForestGreen!15}Segment &  \cellcolor{ForestGreen!15}Original & \cellcolor{ForestGreen!15} $\sigma=0.01$ & \cellcolor{ForestGreen!15} $\sigma=0.05$ & \cellcolor{ForestGreen!15} $\sigma=0.10$ & \cellcolor{ForestGreen!15} $\sigma=0.20$ \\
            \hline
            Plane $1$ & $k=-1.28$ & $k=-1.28$ & $k=-1.29$ & $k=-1.28$ &$k=-1.31$ \\
            \hline
            Plane $2$ & $k=1.28$ & $k=1.28$ & $k=1.28$ & $k=1.28$ & $k=1.26$\\
            \hline
            Cylinder & $r=1.80$ & $r=1.79$ & $r=1.78$ & $r=1.79$ & $r=1.78$\\
            \hline
            \multirow{2}{*}{Torus $1$} & $R=1.49$ & $R=1.49$ & $R=1.47$ & $R=1.48$ & $R=1.56$\\
             & $r=0.72$ & $r=0.73$ & $r=0.70$ & $r=0.67$ & $r=0.74$\\
             \hline    
             \multirow{2}{*}{Torus $2$} & $R=1.05$ & $R=1.09$ & $R=1.02$ & $R=1.17$ & $R=1.13$\\
             & $r=0.79$ & $r=0.78$ & $r=0.78$ & $r=0.74$ & $r=0.80$\\
             \hline
        \end{tabular} \\
        \end{tabular}
    \label{tab:noise_robustness}
\end{table}

\subsection{Comparative analysis}\label{sec:comparative}
We here present two comparisons with alternative methods, all of which rely on estimating the normal vectors at the given input points.

\begin{figure*}[th]
\begin{tabular}{ccccc}
  \begin{tabular}{l}
  \centering
  \footnotesize
  RANSAC
  \end{tabular} &
  \begin{tabular}{l}
  \includegraphics[scale=0.30]{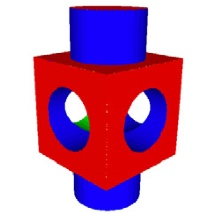}
  \end{tabular}
  &
  \begin{tabular}{l}
  \includegraphics[scale=0.35, trim={0.75cm 1cm 0.5cm 0.5cm}, clip]{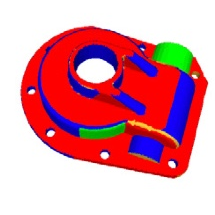}
  \end{tabular}
  &
  \begin{tabular}{l}
  \includegraphics[scale=0.40, trim={0.50cm 0.7cm 0.5cm 0.4cm}, clip]{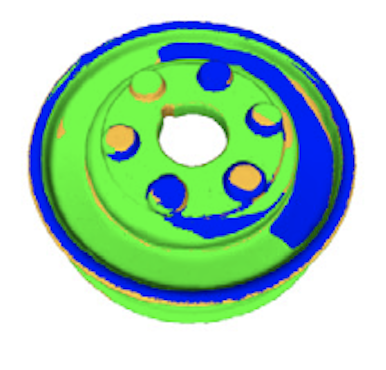}
  \end{tabular}
   &
  \begin{tabular}{l}
    \includegraphics[scale=0.35, trim={0.50cm 0.7cm 0.5cm 0.4cm}, clip]{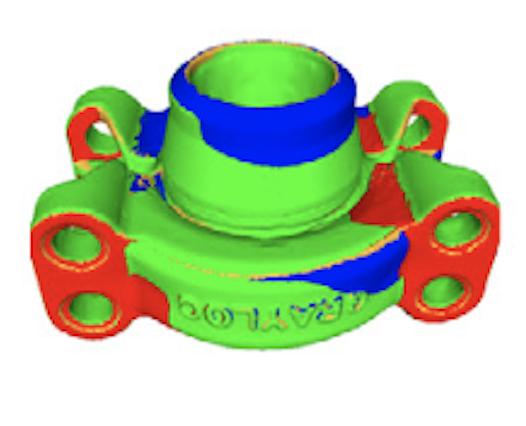}
  \end{tabular}
  \\
 
  \begin{tabular}{l}
  \centering
  \footnotesize
  Le and Duan
  \end{tabular} &
  \begin{tabular}{l}
    \includegraphics[scale=0.30]{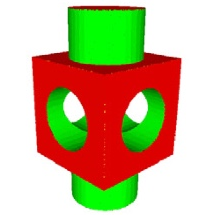}
  \end{tabular} 
  &
  \begin{tabular}{l}
    \includegraphics[scale=0.35, trim={0.75cm 1cm 0.5cm 0.5cm}, clip]{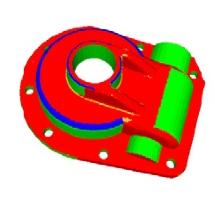}
  \end{tabular}
  &
  \begin{tabular}{l}
    \includegraphics[scale=0.34, trim={0.35cm 0.3cm 0.365cm 0.3cm}, clip]{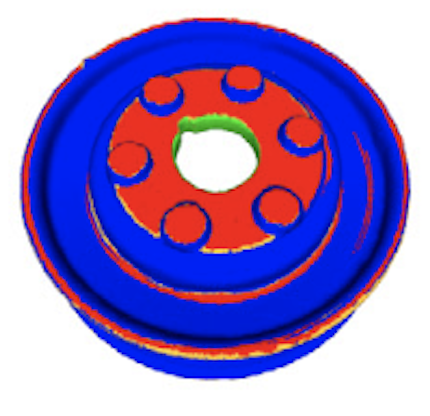}
  \end{tabular}
  &
  \begin{tabular}{l}
    \includegraphics[scale=0.35, trim={0.30cm 0.3cm 0.5cm 0.3cm}, clip]{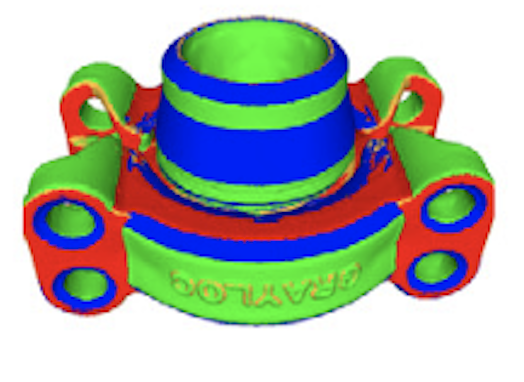}
  \end{tabular}
  \\
  
  \begin{tabular}{l}
  \centering
  \footnotesize
  ParSeNet
  \end{tabular} &
  \begin{tabular}{l}
    \includegraphics[scale=0.135]{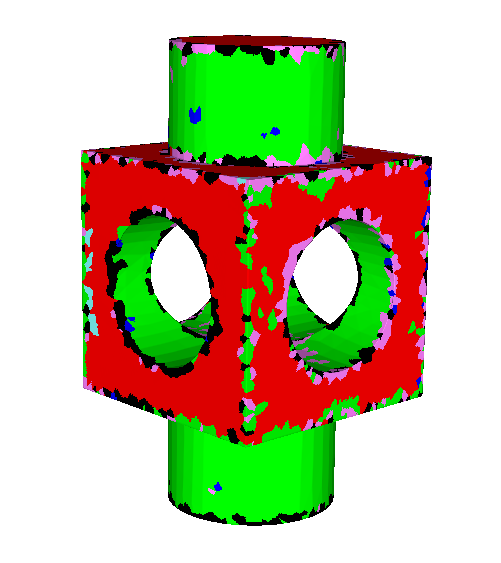}
  \end{tabular} 
  &
  \begin{tabular}{l}
    \includegraphics[scale=0.095, trim={0.25cm 0.5cm 0.25cm 0.25cm}, clip]{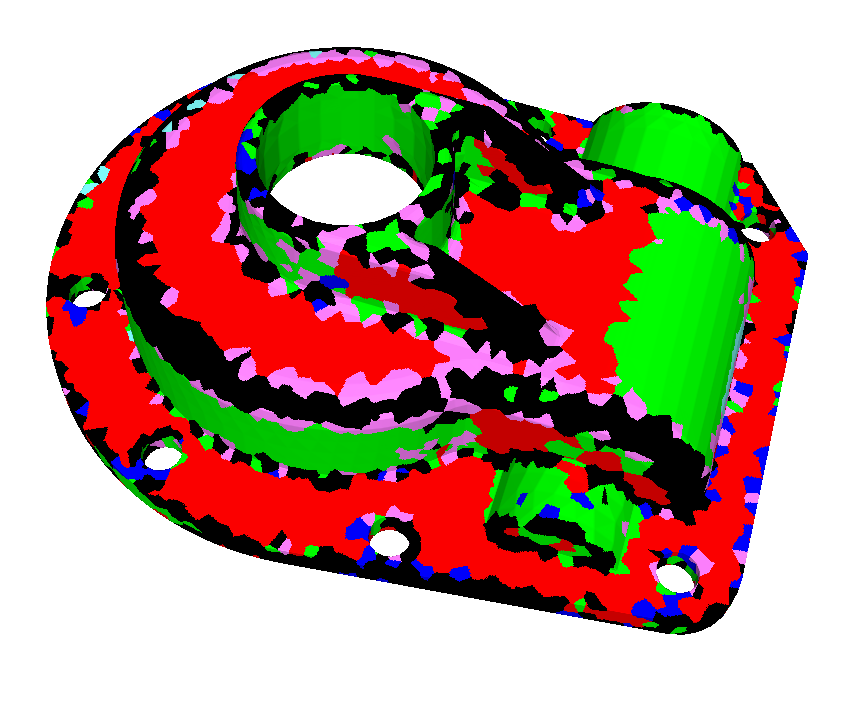}
  \end{tabular}
  &
  \begin{tabular}{l}
    \includegraphics[scale=0.185, trim={0.28cm 0.42cm 0.33cm 0.27cm}, clip]{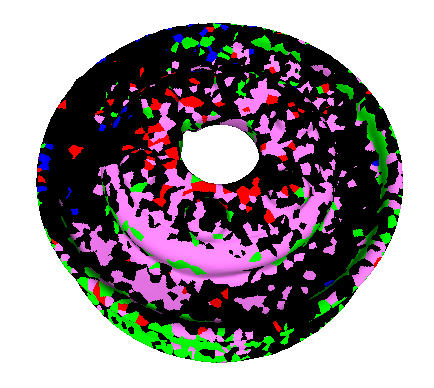}
  \end{tabular}
  &
  \begin{tabular}{l}
    \includegraphics[scale=0.15, trim={0.30cm 0.3cm 0.5cm 0.3cm}, clip]{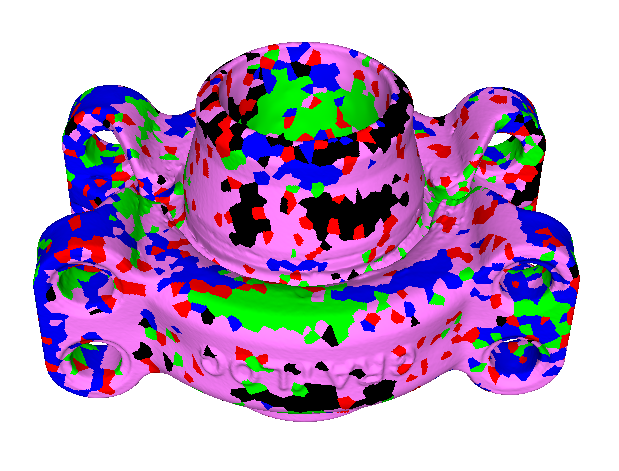}
  \end{tabular}
  \\
  
  \begin{tabular}{l}
  \centering
  \footnotesize
  HPNet
  \end{tabular} &
  \begin{tabular}{l}
    \includegraphics[scale=0.085]{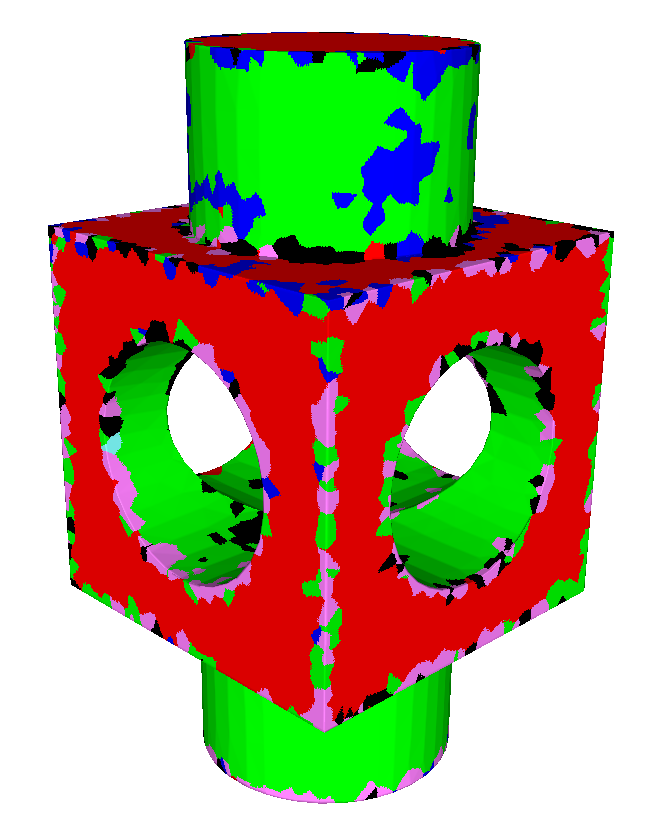}
  \end{tabular} 
  &
  \begin{tabular}{l}
    \includegraphics[scale=0.075, trim={0.25cm 0.5cm 0.25cm 0.25cm}, clip]{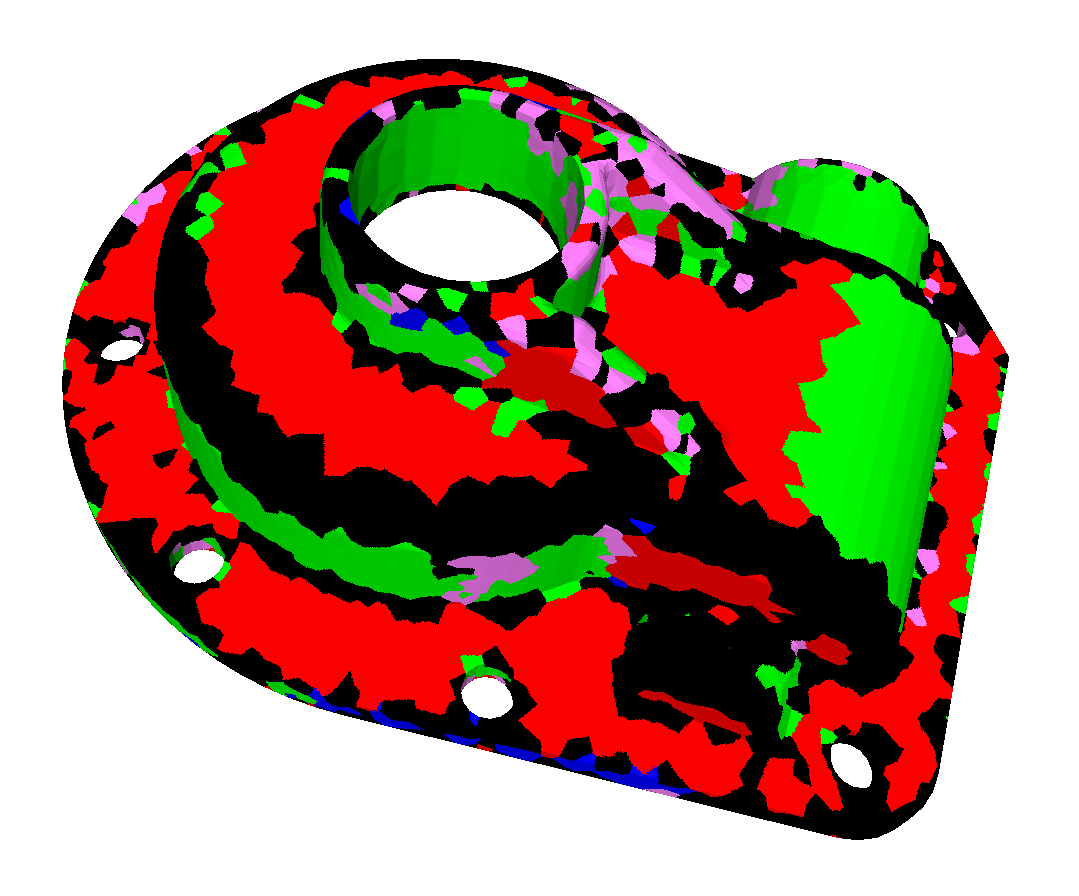}
  \end{tabular}
  &
  \begin{tabular}{l}
    \includegraphics[scale=0.075, trim={0.28cm 0.42cm 0.33cm 0.27cm}, clip]{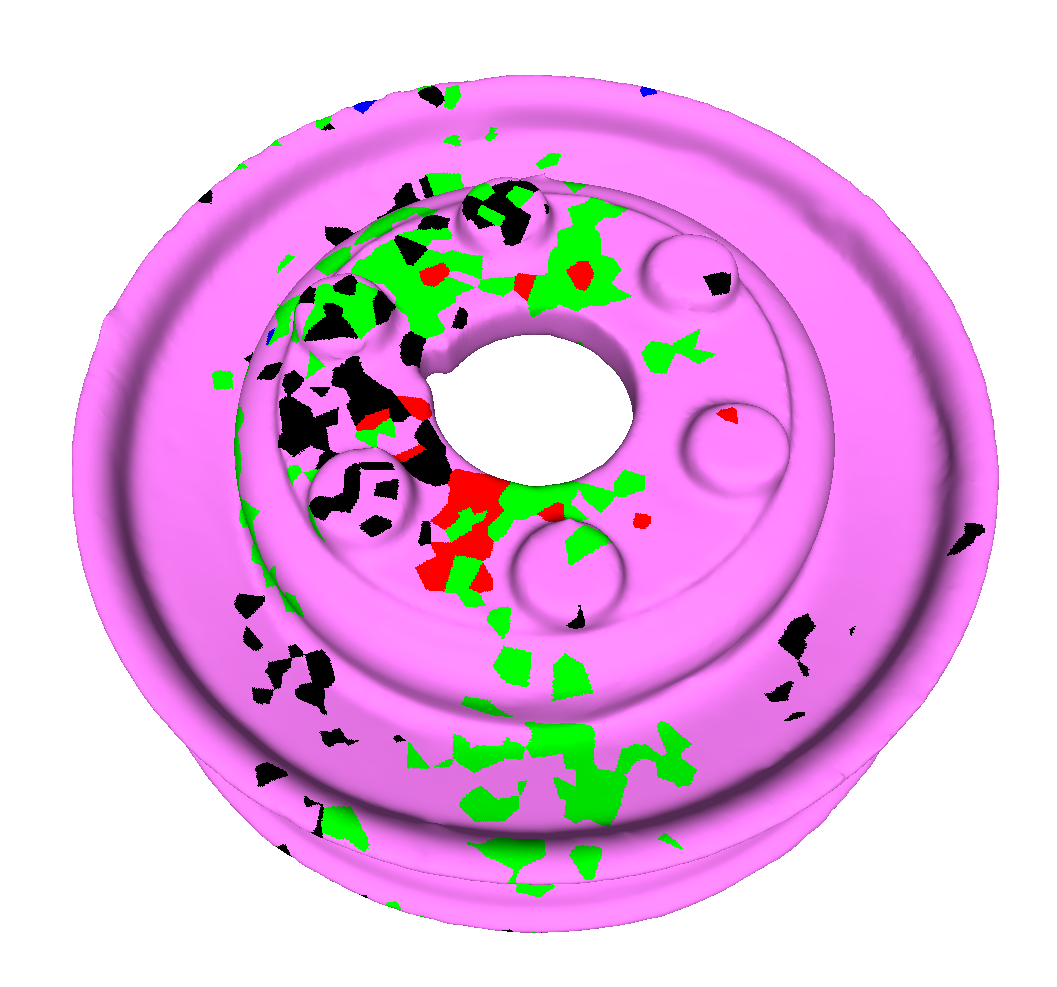}
  \end{tabular}
  &
  \begin{tabular}{l}
    \includegraphics[scale=0.085, trim={0.30cm 0.3cm 0.5cm 0.3cm}, clip]{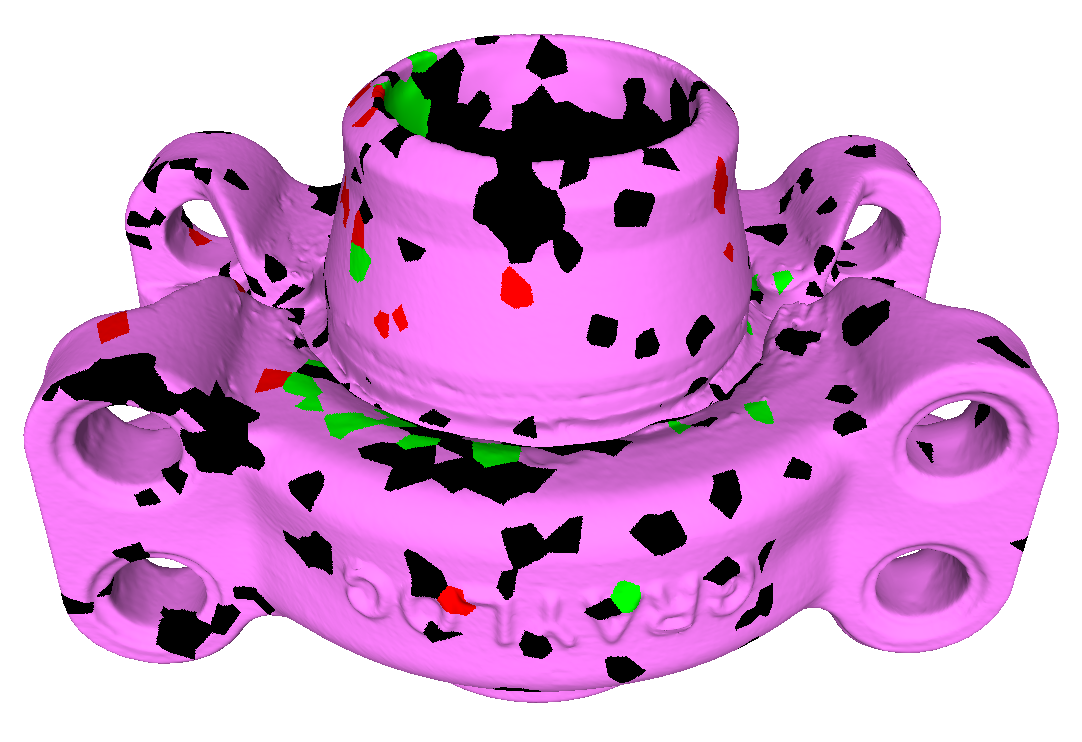}
  \end{tabular}
  \\
  
  \begin{tabular}{l}
  \centering
  \footnotesize
  Ours
  \end{tabular} &
  \begin{tabular}{l}
  \includegraphics[scale=0.30]{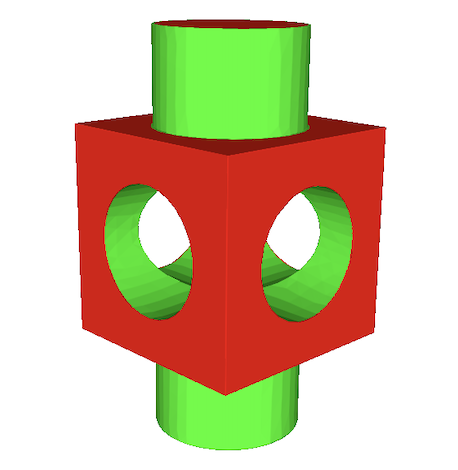}
  \end{tabular} 
  &
  \begin{tabular}{l}
  \includegraphics[scale=0.38, trim={0.25cm 0.5cm 0.25cm 0.25cm}, clip]{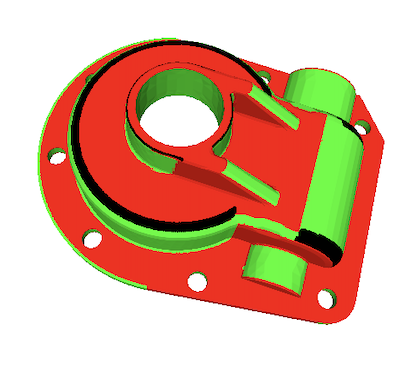}
  \end{tabular}
  &
  \begin{tabular}{l}
    \includegraphics[scale=0.42, trim={0.28cm 0.42cm 0.33cm 0.27cm}, clip]{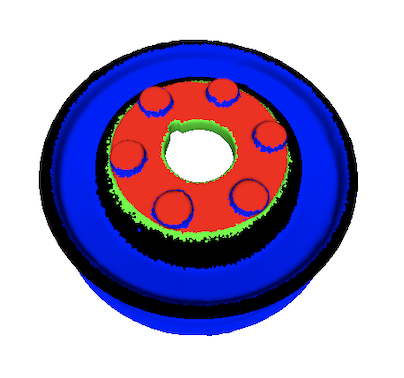}
  \end{tabular}
  &
  \begin{tabular}{l}
  \includegraphics[scale=0.38, trim={0.30cm 0.3cm 0.5cm 0.3cm}, clip]{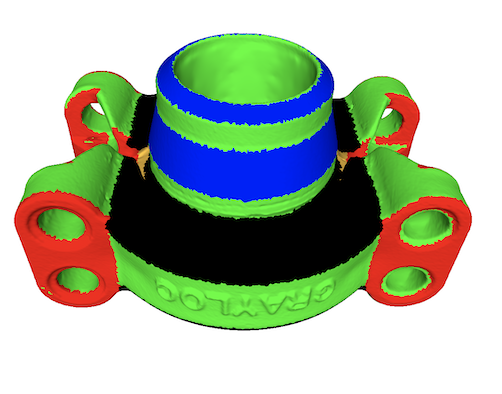}
  \end{tabular}
  \\
  
  & (a) & (b) & (c) & (d)\\
\end{tabular}
\caption{Primitive type recognition: a comparison between our approach, a RANSAC-based segmentation \cite{Schnabel2007}, and the method in \cite{Le2017}. Different colours correspond to different primitive types: planes (red), cylinders (green), cones (blue), tori (black), splines (pink), unsegmented (yellow). \label{fig:qualitative_comparison}}
\end{figure*}

The first analysis, proposed in Figure \ref{fig:qualitative_comparison}, is merely visual. Due to the lack of freely-downloadable implementations for some methods, it is indeed not possible to present this comparison other than qualitatively. The analysis consists of a comparison of our method with the RANSAC-based method introduced in \cite{Schnabel2007}, the patch aggregation approach in \cite{Le2017} and two recent deep learning architectures: ParSeNet \cite{Sharma:2020} and HPNet \cite{Yan:2021}. In this comparison, we have focused on models that solely present simple primitives, to show that even on these our approach gives a great performance; on the other hand, the capability to handle more complex primitives is undoubtedly an added value of our method. Similarly to \cite{Le2017}, we use different colours to represent different primitive typologies: red for planes, green for cylinders, blue for cones, black for tori, pink for (open and closed) B-splines and yellow for unsegmented parts. For a simple model like the block of Figure \ref{fig:qualitative_comparison}(a), all methods provide decent decompositions, although RANSAC misclassifies some primitives while the deep learning frameworks run into problems because of an inaccurate estimation of the surface normals: this assertion is supported by the location of misclassified points, which are largely limited to the sharp edges. In this regard, it is worth noting that -- considering that the networks were trained on clean data -- the more accurate the normals, the more precise  the final segmentation. In our case, the use of a voting procedure allows us to withstand possible perturbations. For the remaining more complex models, our method outperforms the competitors. In Figure \ref{fig:qualitative_comparison}(b), our approach is the only capable of correctly identifying portions of tori, misclassified by Le and Duan \cite{Le2017} and partly unsegmented by  RANSAC; ParSeNet and HPNet are able to reveal the presence of tori, but the resulting segmentation is unreliable along sharp edges -- again, mostly because of a deficiency in the normal estimates. Figures \ref{fig:qualitative_comparison}(c-d) show a RANSAC tendency to oversegment and misclassify complex models. While Le and Duan \cite{Le2017} obtain considerably improved results, their algorithm misses a thin cylinder (see Figure \ref{fig:qualitative_comparison}(c)) and all the tori in both models. Being these two objects acquired by low-quality scanners, the corresponding point clouds (and meshes) are affected by point cloud artifacts which, in turn, lead to an even greater error in the normal estimates. The resulting segmentations are unreliable and  mostly associated with spline segments. As previously mentioned, note that the two architectures were trained on the ABC dataset, which does not take into account the presence of noise or other types of perturbation in the data. Due to a lack of a sufficiently rich benchmark of scanned CAD objects, however, new training is not currently possible.
 
\begin{figure}[h!]
    \centering
    \begin{tabular}{c}
    \includegraphics[width=\columnwidth,trim={3.5cm 0 3cm 0cm}, clip]{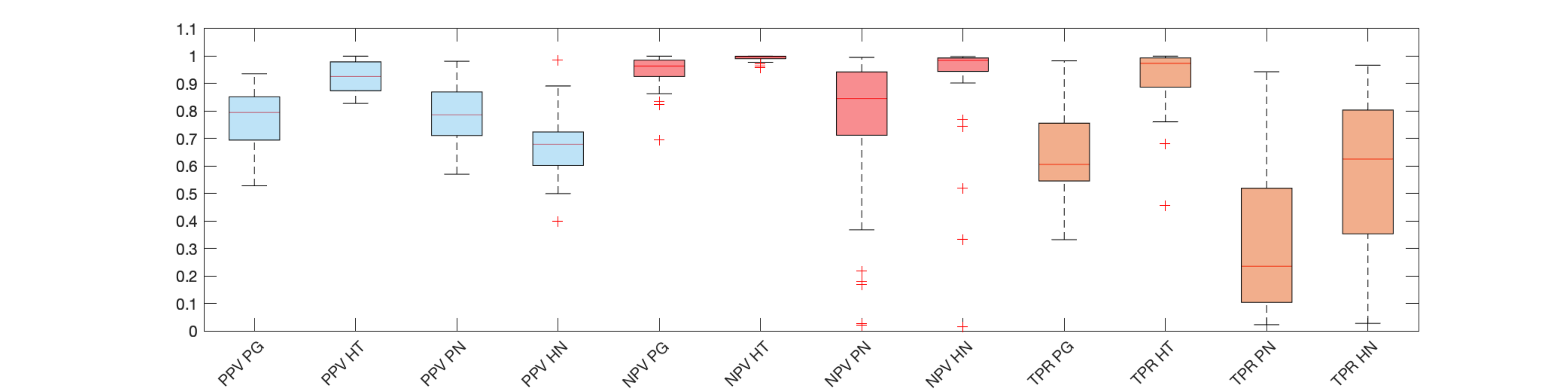}
    \\
    \includegraphics[width=\columnwidth,trim={3.5cm 0 3cm 0cm}, clip]{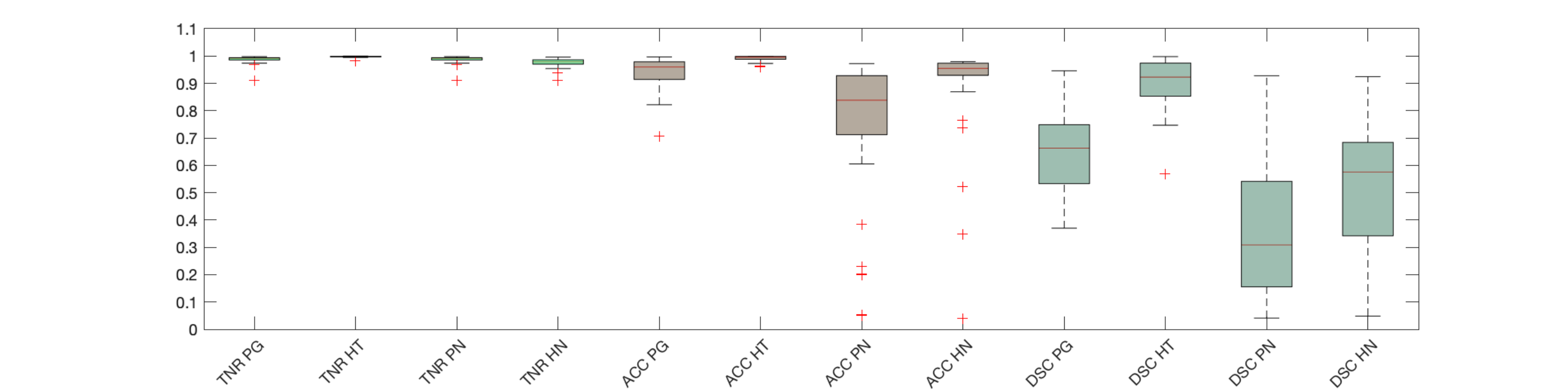}  
    \end{tabular}
    \caption{Comparison of our approach (HT) with other three methods: a primitive growing approach (PG), ParSeNet (PN) and HPNet (HN). The analysis is performed over the Fit4CAD benchmark \cite{Romanengo:2022}.}
    \label{fig:boxplots_Fit4CAD_whole}
\end{figure}

To offer a quantitative analysis of the robustness of our pipeline, we compare its performance with that of three state-of-the-art methods whose implementation was made freely available by the authors: a direct method approach on a primitive-growing (PG) framework \cite{Qie:2021}, adapted to handle point clouds as described in \cite{Romanengo:2022}, and the two learning-based methods from the previous comparison: ParSeNet (PN) and HPNet (HN). We conduct the study over the whole Fit4CAD benchmark \cite{Romanengo:2022}, which contains CAD point clouds defined by simple geometric primitives. In addition, Fit4CAD contains a selection of meaningful and validated models from \cite{Koch_2019_CVPR}, converted from meshes to point clouds. Figure \ref{fig:boxplots_Fit4CAD_whole} summarizes the performance of each method over the test set with respect to the following classification measures: Sørensen-Dice index (DSC), Positive Predicted Value (PPV), True Positive Rate (TPR), Negative Predicted Value (NPV), True Negative Rate (TNR) and accuracy (ACC). Given a point cloud, the benchmark defines these measures by comparing each point cloud segment in the ground truth to the most overlapping segment returned by a segmentation method, and then by averaging over all segments contained in that point cloud. Note that, given a point cloud segment in the ground truth, the points inside that segment are the positives while the points outside that segment are the negatives. We arrive at the following conclusions:
\begin{itemize}
    \item Learning-based methods show remarkably high TNRs -- said otherwise, the points they predict as being negative are almost always true negatives. On the other hand, they are more penalized by NPV: while it is ParSeNet that exhibits the highest variability and the lowest quartiles, both methods are seriously affected by outliers -- with some point clouds having this score below $0.2$.  A low NPV means that quite a few points predicted as negative are false negatives (e.g., common in heavily oversegmented models). In this case, these methods have been penalized by the low density of the point clouds, as well as from the possible presence of point cloud artifacts (e.g., missing data). 
    \item Our approach performs significantly better than the competitors in terms of TPR, i.e., it has similar proportions of correct (positive) predictions among positives. When it comes to the PPV, differences are more modest.
    \item In terms of accuracy, the four methods reach high scores, with the direct methods being the less prone to outliers: however, this metric is not completely reliable as this is a naturally unbalanced binary classification problem. A more reliable measure is provided by the Sørensen-Dice index (DSC). Our method  visibly outperforms the competitors in terms of the DSC. Intuitively, the higher the DSC, the more accurate segments returned by a segmentation method are -- with respect to the most overlapping segment in the  ground truth; to put it differently, DSC penalizes greatly both under- and oversegmentation.
\end{itemize}

When it comes to execution time, our Hough-based method and the primitive-growing approach were run on a desktop PC equipped with an Intel Core i9 processor (at 3.6 GHz) and a Windows 10 operating system. The average execution time, per model, are $286.0$ seconds for the PG-method and $50.7$ seconds for our pipeline. ParSeNet and HPNet were run on Google Colab pro equipped with NVIDIA-SMI 460.32.03 and CUDA  11.2. Regarding the average execution times, we have $5$ seconds for ParSeNet and $257.1$ seconds for HPNet (when the preprocessing of normals is applied).

\section{Conclusions}
To address the problem of recognition of simple and complex primitives in a point cloud, we have opportunely extended the family of geometric primitives to which the HT technique can be applied. The HT is naturally robust to noise and outliers and recognizes multiple instances of the same primitive. Thanks to an opportune preprocessing of the point cloud and its sub-parts, we are able to limit the number of parameters that are necessary to represent the primitive, thus reducing the computational complexity of the HT computation. In addition, the explicit extraction of position-independent primitive parameters and the use of a hierarchical clustering strategy permits us to identify maximal and compound primitives, thus reducing the output oversegmentation. Indeed, differently from spline-based primitives, our strategy is suited for primitive similarity reasoning and permits us to find maximal primitives.

Although learning methods have performed very well in recent years,
 when models present complex combinations of primitives (such as the examples in Figure \ref{fig:qualitative_comparison}(c,d)), direct methods perform even better, and our method is very competitive. In addition, our method has been validated on a whole benchmark and, compared with the others, it turns out to be the best.

\begin{figure*}[ht!]
\begin{tabular}{ccccc}
\hhline{~----}
& \multicolumn{1}{|c}{\cellcolor{ForestGreen!15} Segments (a)}
& \multicolumn{1}{|c}{\cellcolor{ForestGreen!15} Segments (b)}
& \multicolumn{1}{|c}{\cellcolor{ForestGreen!15} Segments (c)}
& \multicolumn{1}{|c|}{\cellcolor{ForestGreen!15} Segments (d)}
\\
\hline
\multicolumn{1}{|c}{\cellcolor{ForestGreen!15} }&\multicolumn{1}{|c}{ }& \multicolumn{1}{|c}{ } & \multicolumn{1}{|c}{ } & \multicolumn{1}{|c|}{ } \\
  \multicolumn{1}{|c}{\cellcolor{ForestGreen!15}\begin{tabular}{l}
  \centering
    {\footnotesize GT}
  \end{tabular}} & \multicolumn{1}{|c}{
  \begin{tabular}{l}
    \includegraphics[width=3.0cm,trim={1.0cm 1.0cm 0.75cm 0.5cm}, clip]{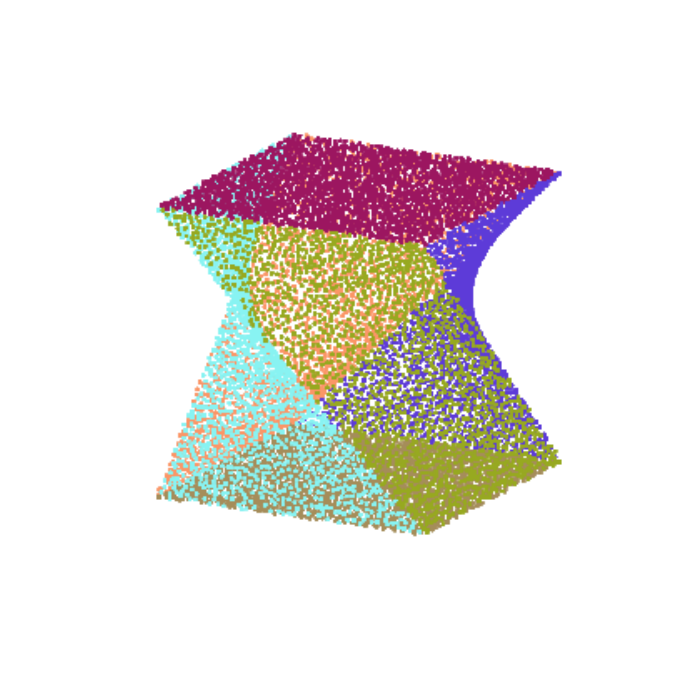}
  \end{tabular}
  }
  &
  \multicolumn{1}{|c}{
  \begin{tabular}{l}
    \includegraphics[width=2.5cm,trim={1.5cm 1.0cm 1.5cm 0.5cm}, clip]{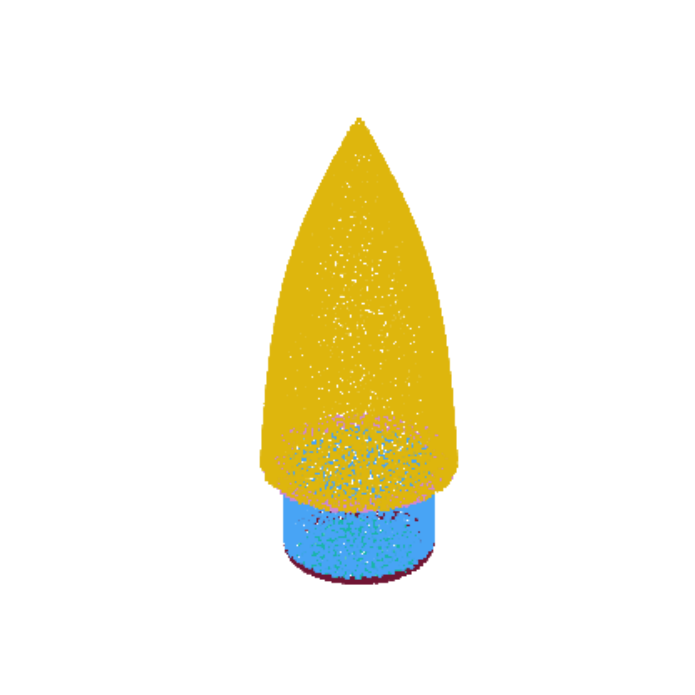}
  \end{tabular}}
   &
   \multicolumn{1}{|c}{
  \begin{tabular}{l}
    \includegraphics[width=2.75cm,trim={1.0cm 0.5cm 0.75cm 0.5cm}, clip]{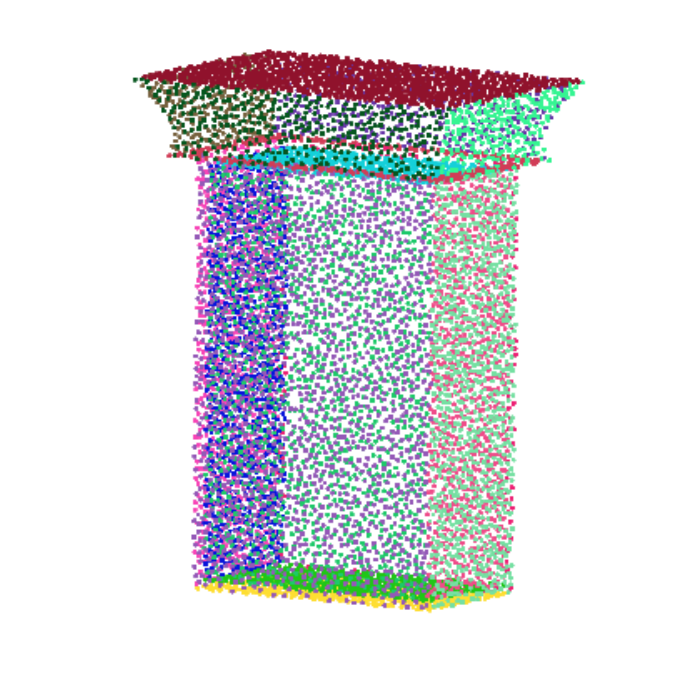}
  \end{tabular}
  }
   &
   \multicolumn{1}{|c|}{
  \begin{tabular}{l}
    \includegraphics[width=3.2cm,trim={1.0cm 1.0cm 0.75cm 0.5cm}, clip]{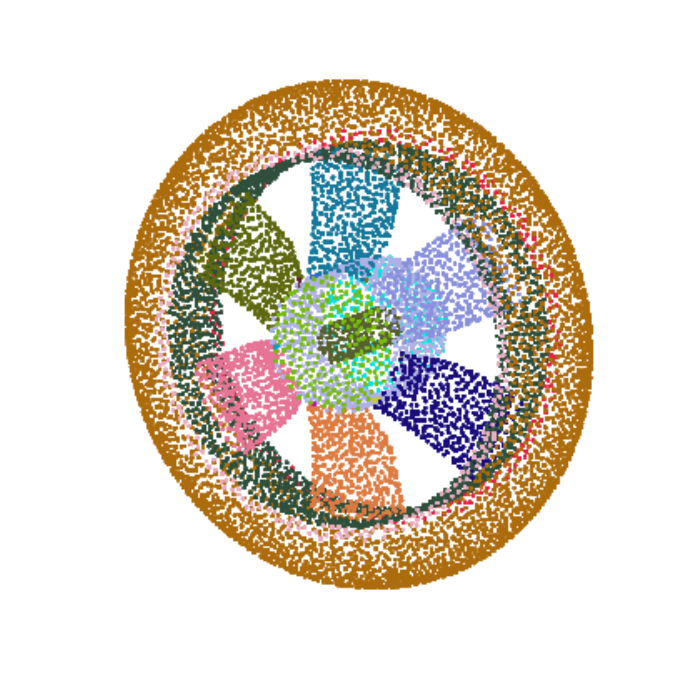}
  \end{tabular}}
  \\
  \multicolumn{1}{|c}{\cellcolor{ForestGreen!15} }&\multicolumn{1}{|c}{ }& \multicolumn{1}{|c}{ } & \multicolumn{1}{|c}{ } & \multicolumn{1}{|c|}{ } \\  
\hline
  \multicolumn{1}{|c}{\cellcolor{ForestGreen!15} }&\multicolumn{1}{|c}{ }& \multicolumn{1}{|c}{ } & \multicolumn{1}{|c}{ } & \multicolumn{1}{|c|}{ } \\ 
  \multicolumn{1}{|c}{\cellcolor{ForestGreen!15}\begin{tabular}{l}
  \centering
    {\footnotesize US}
  \end{tabular}} &
  \multicolumn{1}{|c}{\begin{tabular}{l}
    \includegraphics[width=3.0cm,trim={1.0cm 0.75cm 0.75cm 0.5cm}, clip]{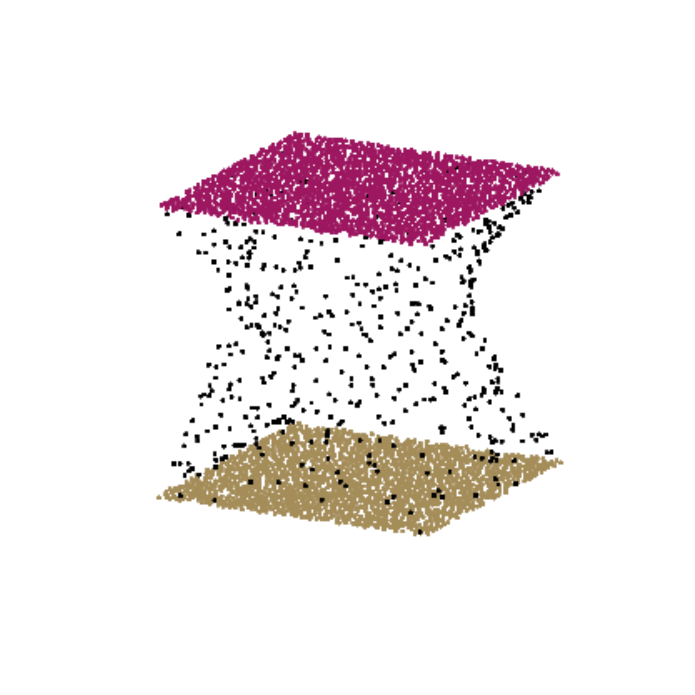}
  \end{tabular} }
  &
  \multicolumn{1}{|c}{\begin{tabular}{l}
    \includegraphics[width=2.5cm,trim={1.5cm 0.75cm 1.5cm 0.5cm}, clip]{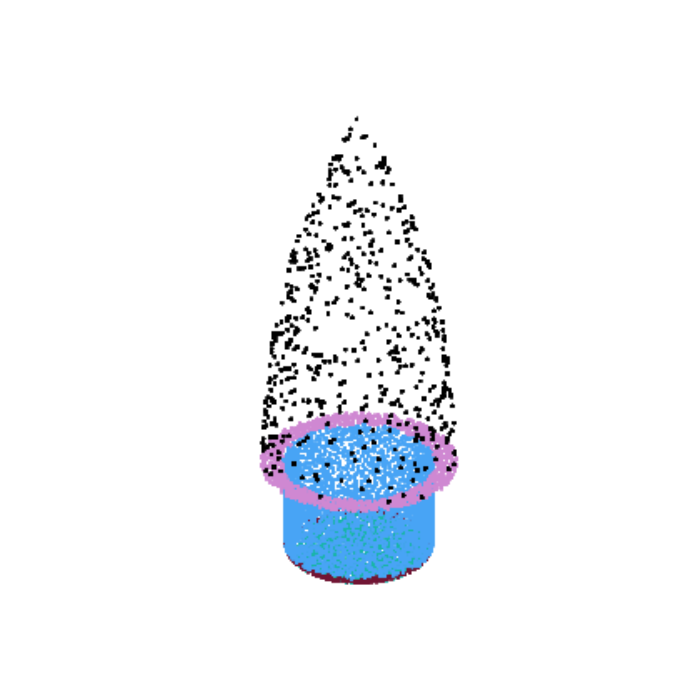}
  \end{tabular}}
  &
  \multicolumn{1}{|c}{\begin{tabular}{l}
    \includegraphics[width=2.75cm,trim={1.0cm 0.5cm 0.75cm 0.5cm}, clip]{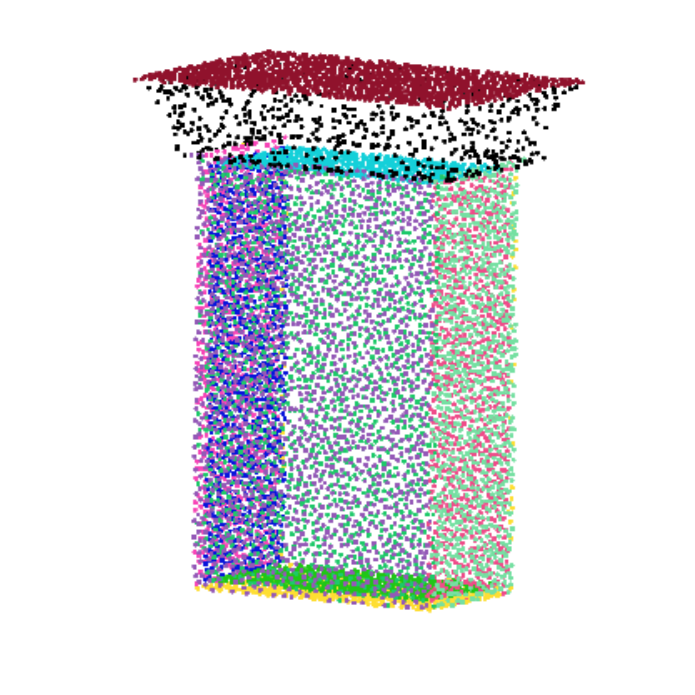}
  \end{tabular}}
  &
  \multicolumn{1}{|c|}{\begin{tabular}{l}
    \includegraphics[width=3.2cm,trim={1.0cm 1.0cm 0.75cm 0.5cm}, clip]{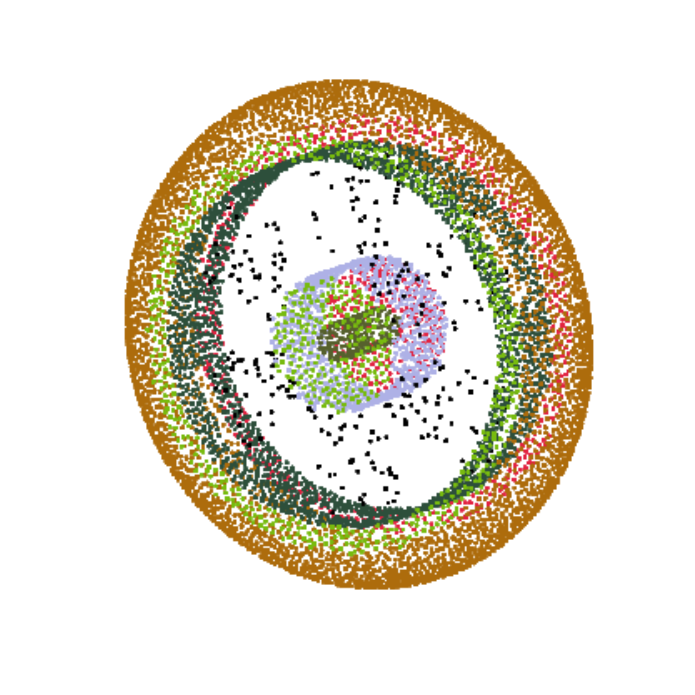}
  \end{tabular}}
  \\
\multicolumn{1}{|c}{\cellcolor{ForestGreen!15} }&\multicolumn{1}{|c}{ }& \multicolumn{1}{|c}{ } & \multicolumn{1}{|c}{ } & \multicolumn{1}{|c|}{ } \\
\hline
\end{tabular}
\caption{Four models from the ABC dataset \cite{Koch_2019_CVPR} containing spline segments\label{fig:ABC_splines}: comparison between the ground truth (GT) segmentation and our recognition of simple geometric primitives (US). To enhance the visibility of the primitives correctly recognized, the second row displays the segments classified as splines with a lower point density.}
\end{figure*}

The use of geometric primitives whose mathematical representations require a large number of parameters -- even in their standard forms -- is currently the main limitation of our pipeline; while theoretically possible, this would indeed increase the execution time exponentially, making the algorithm inapplicable in practice. An instructive example, shown in Figure \ref{fig:ABC_splines}, is given by point clouds that contain segments generated by spline surfaces: while our method cannot deal with spline surfaces, it does manage to recognize that parts of a point cloud do not originate from any primitive at our disposal. The first row of Figure \ref{fig:ABC_splines} shows the ground truth segmentation, while the second row shows the segmentation we achieve: the black dots correspond to points that are not recognized as simple geometric primitives, and that are labelled as ``unsegmented" because of the high mean fitting error; note that spline segments are undersampled for the sake of visibility, but this processing is realized only after segmenting the point clouds. 
On the contrary, methods that recognise splines tend to give any segment a label, as shown in Figure \ref{fig:qualitative_comparison}. 

As future work, we are thinking of other strategies to reduce the computational cost in the HT-based recognition step. Indeed, the parameter space can be further optimised by reducing its dimension through the use of a refinement strategy that discretizes the parameter space only in the relevant regions of interest. Another possible direction of investigation could concern the applicability of our method to non-orientable surfaces. Indeed we have only applied our method to objects with orientable surfaces since the datasets we have used exhibit only this type of surface. However, in theory, our approach could also apply to objects with non-orientable surfaces, as long as their equation exists in the literature and the system of equations is solvable with respect to the parameter (as in the case of M\"{o}bius strip). The challenge here would be to study a preprocessing technique to translate and/or rotate the point cloud into the standard form of that surface type.  For instance,  in the case of the M\"{o}bius strip, how to estimate the symmetry axis in order to place the point cloud so that we can fit it with a primitive in a standard form.

Regarding the 
automation of the process, we could take advantage of a pre-classification of the points in the input point cloud using a curvature-based characterization, for instance, the local surface variation proposed in \cite{Pauly2002} or the shape index and curvedness as used in \cite{Qie:2021} for ISO GPS segmentation. Once points are aggregated by their type, we could further split or aggregate these regions using our fitting method and possibly considering only a suitable subset of primitives, i.e., the primitives that are compatible with the recognized type of points, e.g., planar, cylindrical, spherical, etc.

\section*{Acknowledgements}
This work has been developed in the CNR research activities DIT.AD004.100, DIT.AD021.080.001 and DIT.AD021.125. 

The authors thank the VVS shape repository by the AIM@SHAPE consortium (\url{http://visionair.ge.imati.cnr.it}) and Dr. Alexander Leutgeb from RISC Software GmbH (Linz, Austria) for the models used in this paper.

\bibliographystyle{ieeetr}      

\end{document}